\RequirePackage[2024-06-01]{latexrelease}
\documentclass[aps,prc,superscriptaddress,nofootinbib,twocolumn,floatfix]{revtex4}
\usepackage{bm}
\usepackage{placeins}
\usepackage{float}
\usepackage{amsfonts}
\usepackage{bbm}
\usepackage{multirow}
\usepackage{graphicx,color,amsmath,amssymb}
\usepackage{pifont}
\usepackage[version=4]{mhchem}
\usepackage{hyperref}
\usepackage{color,soul}
\usepackage{tabularx}

\newcommand{\bea}{\begin{eqnarray}}
\newcommand{\eea}{\end{eqnarray}}
\newcommand{\eg}{\textit{e.g.}}

\newcommand{\ie}{\textit{i.e.}}

\newcommand{\lsim}{\lesssim}
\newcommand{\gsim}{\gtrsim}

\begin{document}

\title{Photon Emission from Nucleon-Nucleon Bremsstrahlung in Fermi-energy Heavy-Ion Collisions}

\author{Thomas Onyango}
\affiliation{Cyclotron Institute and Department of Physics and Astronomy, Texas A$\&$M University, College Station, TX 77843-3366, USA}
\affiliation{Lawrence Livermore National Laboratory, 7000 East Avenue, Livermore, CA 94550-9698, USA}
\author{Ralf Rapp}
\affiliation{Cyclotron Institute and Department of Physics and Astronomy, Texas A$\&$M University, College Station, TX 77843-3366, USA}

\date{\today}

\begin{abstract}
The emission of direct (hard and thermal) photons from nucleon-nucleon Bremsstrahlung in heavy-ion collisions at Fermi energies is analyzed. We utilize a photon emission rate based on a quantum-field-theoretical model together with nucleon distribution functions extracted from a coarse-graining method of transport model simulations.
The latter accounts for off-equilibrium effects during the early stages of nuclear collisions primarily occurring in the beam direction while the transverse-momentum distributions are amenable to a thermal description. With this setup, we quantify the contributions from first-chance nucleon-nucleon collisions and the subsequent transition to a thermal source to the photon energy spectrum measured from Ca-Ca collisions at 35~A$\cdot$MeV bombarding energy. We find that most of the hard photons are produced in the initial stages of heavy-ion collisions from primordial collisions where nucleons move with the initial collective nuclear motion while the emission from the later stages plays a sub-dominant role. We compare our calculations to experimental measurements of a differential photon-energy spectrum from a collision system of similar size and beam energy, thereby including acceptance cuts as applied in the detectors. 
\end{abstract}
%\pacs{13.75.Cs, 21.10.Gv, 21.65.+f}
%\PhySH{Low \& intermediate energy heavy-ion reactions, Heavy-ion reaction mechanisms, Nuclear Matter, Equations of state of nuclear matter, Nucleon distribution, Nucleon-nucleon interactions, photon production}
%\keywords{Heavy-Ion Collisions, Nuclear Matter, Thermal Photon Emission, Nucleon-Nucleon Bremsstrahlung}
\maketitle

%%%%%%%%%%%%%%%%%%%%%%%%
\section{Introduction}
\label{sec_intro} 
%%%%%%%%%%%%%%%%%%%%%%%%

Measurements of electromagnetic (EM) radiation, \ie, photons and dileptons, have a long history of investigating the thermodynamic and spectral properties of strongly interacting matter as formed in the aftermath of heavy-ion collisions (HICs)~\cite{Bauer:1986zz, Rapp:1999ej, Gale:2009gc, Bonasera:2006qb, Rapp:2014hha}. Real and virtual photons are continuously produced throughout the course of an HIC and thus can serve as a penetrating probe as they escape from the nuclear medium largely unaltered by final-state interactions while carrying information about their origin. At relativistic beam energies, the emission is dominated by the hot nuclear medium, including its resonance excitations, as well as partonic production from a quark-gluon plasma, while at nonrelativistic beam energies nucleon-nucleon ($NN$) interactions are the dominant source. Parametrically, the $NN$ collisions that produce photons are less likely than ones that do not by a factor of $\alpha_{EM} = e^2/4\pi\simeq1/137$ (the electromagnetic coupling constant). This greatly reduces the possibility of a photon being scattered or absorbed on its way out of the fireball and to the detector. Dileptons are the result of pair production whereby a virtual photon that is emitted subsequently decays into a lepton and its anti-lepton. This process involves an additional electromagnetic interaction, which suppresses its production probability by another factor of $\alpha_{EM}$ compared to real photons. The scarcity of dileptons encourages the use of photons as a probe into HICs, despite dileptons carrying the advantage of an additional variable represented by their virtuality, \ie, their invariant mass. 

In the present manuscript, we focus on heavy-ion collisions in the regime of the Fermi energy, $E_{\rm beam}\simeq40$~MeV, at nuclear saturation density, $\rho_N = \rho_0 \equiv$~0.16~fm$^{-3}$. Several calculations of photon emission from $NN$ Bremsstrahlung in this context have been carried out before. Early works, some of which were simulations running Boltzmann-Uehling-Uhlenbeck (BUU) models, were built on $NN$ scattering using a constant cross section with probabilistic occurrences of Bremsstrahlung according to classical radiation theory~\cite{Ko:1987zz}. By taking advantage of the $NN$ cross section being much larger than the cross section for photon emission, the latter can be treated as a perturbation to the $NN$ collisions.
The calculations were benchmarked with fits to data from $\ce{^{14}N}+\ce{^{208}Pb}$ at 40~MeV/A, but the measurements in $\ce{^{14}N}+\ce{^{58}Ni}$ at 35~MeV/A were underestimated by about a factor of 5, although the slopes of the energy spectra were found to agree with the data; this study included the effects of Pauli blocking and nuclear mean fields.
An equation from first principles for photon emission from protons freely scattering off neutrons is described in Ref.~\cite{Jackson:1998nia}. This cross section was extended by Cassing et al.~\cite{Cassing:1986aya} by distinguishing the initial and final momenta to ensure energy conservation by using 
\begin{equation}
   \frac{d^2\sigma_{el}}{dE_\gamma d\Omega_\gamma}=\alpha\frac{R^2}{12\pi}\frac{1}{E_\gamma}(2\beta_f^2+3\beta_i^2\sin^2{\theta_\gamma}) \ , 
\label{eq_photon_cross_section}
\end{equation}
where $\Vec{\beta}_{i(f)}$ is the initial (final) velocity of the nucleons in the $NN$ CM frame. The hard-sphere scattering radius, $R$, is the only adjustable parameter in this cross section and was set to 2.1~fm (corresponding to a total $NN$ cross section of about 139~mb) to reproduce the hard-photon data from Ref.~\cite{koehler.18.933}. Medium effects were later added to render the cross section more realistic~\cite{Neuhauser:1987tvf}. 
In Ref.~\cite{Bauer:1986zz}, it was acknowledged that the in-medium twice-differential cross section does not accurately reproduce the photons' energy spectrum nor their angular distributions, but it was judged sufficient for the purposes of quantifying the medium corrections to the photon cross section in HICs and determining a timescale for hard-photon emission, rather than attempting a detailed realistic theoretical description of the microscopic free emission process; a fair description of the fully integrated photon cross section measured in \ce{^{12}C}+\ce{^{12}C} collisions at 40~MeV/A was obtained. Bauer et al. also employed their elementary cross section to study the $pn\to pn\gamma$ process. By comparing to $p+d$ data \cite{EDGINGTON1966523}, they extracted a hard-sphere scattering radius of $\sqrt{3}$~fm. 
An alternative formulation for calculating Bremsstrahlung from the one-body channels (convection and magnetization currents of external legs) and the two-body channel (meson exchange) using a $G$-matrix was developed by Nakayama and Bertsch~\cite{Nakayama:1989zzb}. The Bremsstrahlung amplitudes they derived were then parameterized for faster and easier implementation~\cite{Nakayama:1989zza}. This formulation was compared to experimental data for \ce{^{12}C}+\ce{^{12}C} collisions at 84 MeV/A \cite{Grosse:1986aa}. Although the data were underestimated by the theoretical calculations by a factor of 2, the comparison demonstrated that the meson-exchange channel is the dominant source of photons with energies $E_\gamma\gtrsim 70$~MeV. The energy spectra of hard photons ($E_\gamma\gtrsim30$ MeV) calculated from this theoretical model are consistent with measurements from \ce{^{14}N}+\ce{^{208}Pb} collisions at 40~MeV/A from Ref.~\cite{Stevenson:1986zz} and data for \ce{^{12}C}+\ce{^{154}Sm} collisions at 10~MeV/A~\cite{Gossett:1988}. 
A relativistic photon cross section was derived by Sch\"afer et al.~\cite{Schaefer:1991rp} using parameterizations of $NN$ amplitudes from Refs.~\cite{Horowitz:1985tw,Murdock:1986fs} for the same nucleon channels as Nakayama and Bertsch's $G$-matrix. Experimental data for $p + d \to\gamma + X$ at 195 MeV/A \cite{koehler.18.933} were reproduced quite well. This calculation further supports the importance of the meson-exchange channel. More recently, the cross section developed by 
Bauer et al. has been implemented in isospin-dependent quantum molecular dynamics (IQMD)~\cite{Wang:2020zty}, extended quantum molecular dynamics (EQMD)~\cite{Shi:2021far}, and BUU~\cite{Deng:2016pqu} transport models to investigate how collective nuclear motion affects hard-photon emission. Comparisons of their calculations to experimental data~\cite{Stevenson:1986zz} from \ce{^{14}N}+\ce{^{12}C} collisions at 20, 30, 40~MeV/A are detailed in Refs.~\cite{Wang:2020bzn,Shi:2020ark}. Compared to the 20~MeV/A data, a slightly harder slope was found at 30 and 40~MeV/A, providing good agreement with the data. A comparison to a heavier system, \ce{^{14}N}+\ce{^{208}Pb} at 40~MeV/A in Ref.~\cite{Deng:2016pqu}, was also made; the yield and slope of photons with energies $20 < E_\gamma < 60$~MeV were approximately reproduced. For $E_\gamma > 60$~MeV, the calculation produced fewer photons than measured, and the discrepancy grows further with increasing photon energy. This indicates that a mechanism for producing highly energetic photons is missing from their model. In another series of works, photon emission from $NN$ collisions was evaluated with $\sigma$-meson exchange model \cite{Gan:1994zz,Yong:2007cq,Guo:2023nmm}, 
\begin{equation}
   \frac{d^2P_\gamma}{dE_\gamma d\Omega_\gamma}=(1.671\times10^{-7})\frac{(1-y^2)^\alpha}{y}\text{ [MeV]}^{-1},
\label{eq_prob_differential}
\end{equation}
implemented into an isospin- and momentum-dependent BUU (IBUU) transport. In Eq.~(\ref{eq_prob_differential}), $y=E_\gamma/E_{max}$, $\alpha=0.7319-0.5898\beta_i$, where $E_{max}$ is the total energy available in the center-of-mass frame of the nucleons and $\beta_i$ is the initial nucleon velocity. IBUU simulations of $\ce{^{208}\text{Pb}}+\ce{^{208}\text{Pb}}$ collisions at 45~MeV/A were utilized to determine what effect the thickness of the neutron skin has on photon emission.

In the present manuscript, we employ a somewhat different approach based on a coarse-graining method of transport calculations, which enables an 
extraction of nucleon distribution functions that subsequently can be combined with rigorous quantum-field-theoretical calculations of photon emission rates. 
This method has been applied earlier in the context of ultra-/relativistic heavy collisions~\cite{Huovinen:2002im,Santini:2011zw,Liu:2017vlx,Galatyuk:2015pkq,Endres:2015fna}; here, we present a first application to collisions at Fermi energies. In doing so, we implement thermal nucleon distribution 
functions with off-equilibrium effects arising from the initial beam momentum, thereby providing a seamless transition from the anisotropic initial conditions 
(where the hardest photons are expected to be produced) to the nearly thermalized conditions at the end of the evolution~\cite{Onyango:2021egp}. 
In addition, the use of Fermi distribution functions accounts for Pauli blocking, which is mandatory at Fermi energies and turned out to be essential 
for an accurate description of the output from quantum molecular dynamics simulations~\cite{Onyango:2021egp}. Specifically, we employ photon emission 
rates from nucleon-nucleon Bremsstrahlung and convolute them over the coarse-grained space-time evolution extracted from 
\ce{^{40}Ca}+\ce{^{40}Ca} collisions 35~MeV/A. To simplify the analysis by taking advantage of spatial symmetries, we focus on central collisions but
apply a suitable centrality rescaling as well as experimental acceptance cuts to enable a meaningful comparison to available experimental photon spectra for a central sample of \ce{^{36}Ar}+\ce{^{98}Mo} collisions at 37~MeV/A.

The structure of the paper is as follows. In Section~\ref{sec_collision_evo}, we briefly describe the transition of two cold, liquid nuclei into hot, thermal matter during HICs and how coarse graining allowed us to extract the thermodynamic and collective properties of the system. In Section~\ref{sec_gam-rate}, we discuss the derivation of the photon emission rate from nucleon-nucleon collisions. In Section~\ref{sec_emissivity}, we benchmark the photon rate developed in this study by comparing the emissivity to established literature. In Section~\ref{sec_photon_rate}, we elaborate the dependence of the photon rate on the properties of the nuclear matter produced in HICs. We discuss experimental acceptance cuts in measuring photon energy spectra in Section~\ref{sec_exp_constraints}, which we include in our calculation. In Section~\ref{sec_mom_spectra}, we fold the photon emission rate over the medium evolution and discuss the emission from $NN$ Bremsstrahlung during the various stages. In Section~\ref{sec_energy_spectrum}, our theoretical approach is compared to the experimental data of a comparable collision system. In Section~\ref{sec_other_calculations}, the results of this study are compared to other theoretical works. We conclude in Section~\ref{sec_concl} with a summary of our findings and future perspectives.

%%%%%%%%%%%%%%%%%%%%%%%%
\section{Evolution of Heavy-Ion Collisions at Fermi Energy}
\label{sec_collision_evo} 
%%%%%%%%%%%%%%%%%%%%%%%%
The method for constructing a space-time dependent evolution of nucleon distribution functions in the present paper is the coarse-graining technique, whereby a transport calculation is discretized into local spatial cells and timesteps. Let us start by briefly describing previous works that have employed this method.

The emission spectra of EM radiation using coarse gaining were first studied by Huovinen et al. in Pb+Pb collisions at ultra-relativistic beam energies of 158~GeV/A at the CERN-SPS, based on Ultra-relativistic Quantum Molecular Dynamics (UrQMD) simulations \cite{Huovinen:2002im}. They defined cells in a cylindrical coordinate system of a radius, a space-time rapidity, an azimuthal angle as well as proper time. By extracting the momentum, energy, and net baryon density of a cell, the cell velocity was determined. By comparing the pertinent energy and baryon density in the restframe of the cell with that of a hadron resonance gas with the same particle content as in the transport code, the corresponding temperatures and densities were extracted. The photon and dilepton spectra were then computed by convoluting thermal emission rates over the results from the coarse graining.  Fair agreement with experimental data and, moreover, with the results from a hydrodynamic evolution model were found. This suggests that coarse graining can give a rather robust representation of the equation of state of nuclear matter produced from ultra-relativistic HICs. 
Similar calculations were later done at relativistic collision energies~\cite{Galatyuk:2015pkq,Endres:2015fna}. Specifically, coarse graining was applied to Au+Au collisions at 1.23~GeV/A using a grid of 21\textsuperscript{3} cubic cells with individual volumes of 1 fm\textsuperscript{3} \cite{Galatyuk:2015pkq}. The momenta of the nucleons and pions passing through a cell were binned into momentum distributions, averaged over all events, and boosted 
event-by-event into the thermal restframe. The local baryon density was extracted, and the temperature was obtained from an inverse-slope fit to the transverse-mass distributions of pions. The temperature extracted from the calculated invariant-mass spectra of thermal dilepton emission was found to reflect the highest temperature of the thermodynamics extracted from the coarse-graining procedure.
Coarse graining has also been applied to analyze thermal properties of the fireball in heavy-ion collisions at lower energies, albeit with typically substantially larger cell volumes. In Ref.~\cite{Zhou:2012bd}, the bulk properties of Au+Au collisions at bombarding energies ranging from 50-200~MeV have been investigated using Quantum Molecular Dynamics averaged over a sphere of radius 5~fm around the center of the collision. Using the generalized hot Thomas-Fermi formalism, parameterized nucleon distributions in transverse and longitudinal directions were extracted, and pertinent values of the number, energy and entropy densities calculated, as well as the shear viscosity using an effective $NN$ crosss section; the time scales of the evolution were found to be significantly smaller than in Ref.~\cite{Onyango:2021egp} reviewed below.  
In a further study, the quadrupole momentum of nucleon transverse-momentum distributions has been extracted from BUU-van-der-Waals transport simulations to study the thermal and transport properties in Au-Au collisions at 100-1300 MeV/A bombarding energies, with a central cube of varying size using 6$^2$ and 10$^2$ fm$^2$ of transverse area.
Finally, in Ref.~\cite{Liu:2017vlx}, a similar study has been conducted for IQMD simulations of \ce{^{129}Xe}+\ce{^{119}Sn} collisions at energies between 15 and 100 MeV/A utilizing a central cube of volume (6~fm)\textsuperscript{3}. 
In the present study, we follow our previous work for coarse graining \ce{^{40}Ca}+\ce{^{40}Ca} collisions at 35 MeV/A using a significantly smaller cell size, with the aim of evaluating local photon emissivities in a uniform approach comprising both the early and later phases of the collision dynamics.

\begin{figure}[t!]
\centering
\includegraphics[width=0.8\linewidth]{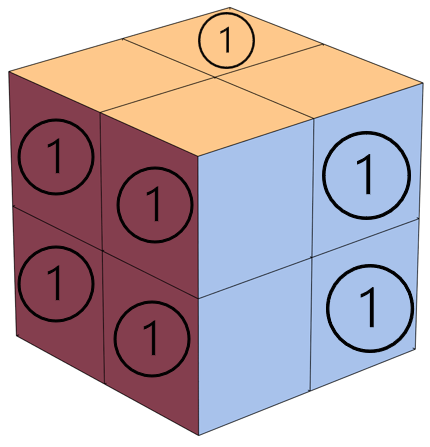}
\caption{Representation of the first (and innermost) layer of the coarse-graining grid.}
\label{fig_spatial_config1}
\end{figure}

\begin{figure}[htb!]
\centering
\includegraphics[width=0.95\linewidth]{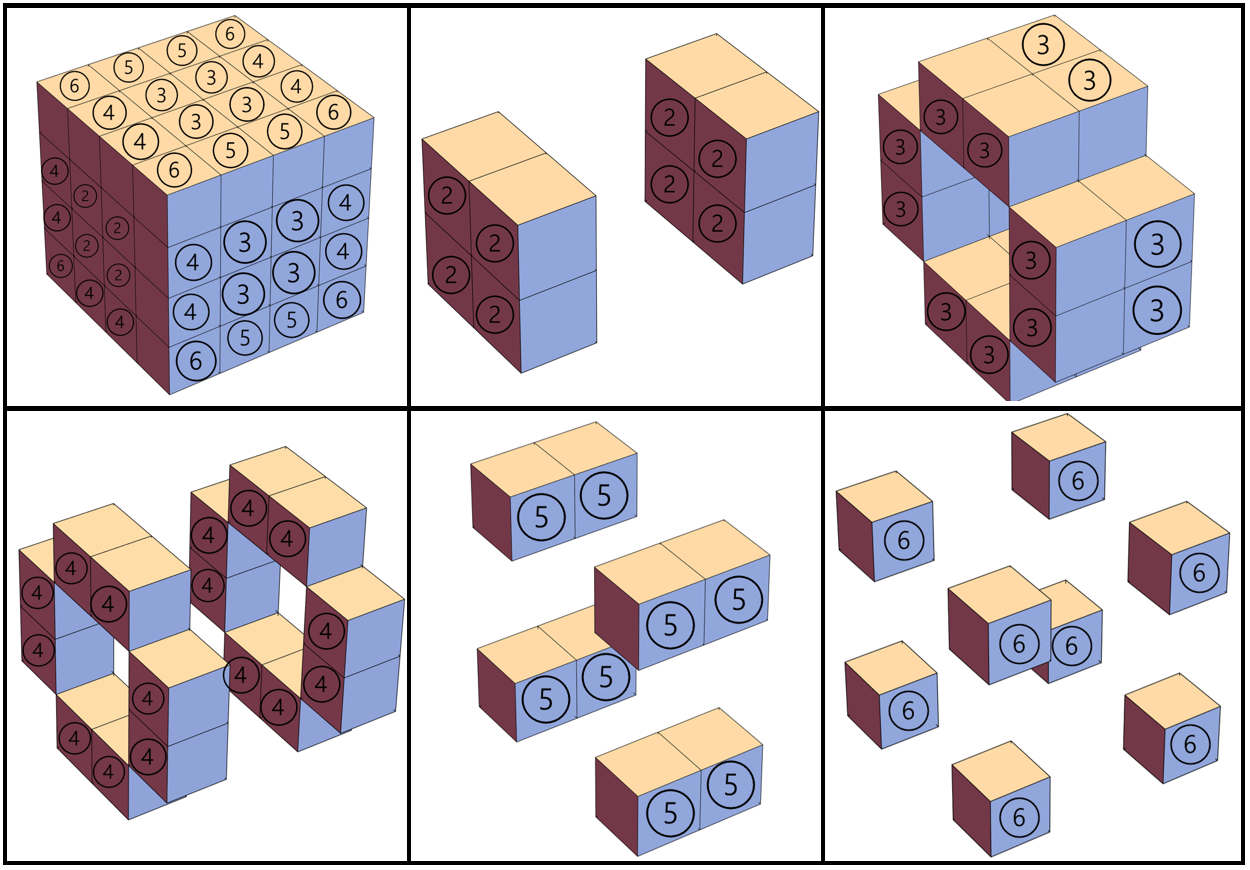}
\caption{Representation of the spatial configuration of the complete second layer (upper left panel) of the coarse-graining grid and the symmetry groups of the centers of the longitudinal faces (upper middle panel), the centers of the transverse faces (upper right panel), the edges of the longitudinal faces (lower left panel), the edges of the transverse faces (lower middle panel), and the corners of the second layer (lower right panel).}
\label{fig_spatial_config2}
\end{figure}

\begin{figure}[t!]
\centering
\includegraphics[width=0.9\linewidth]{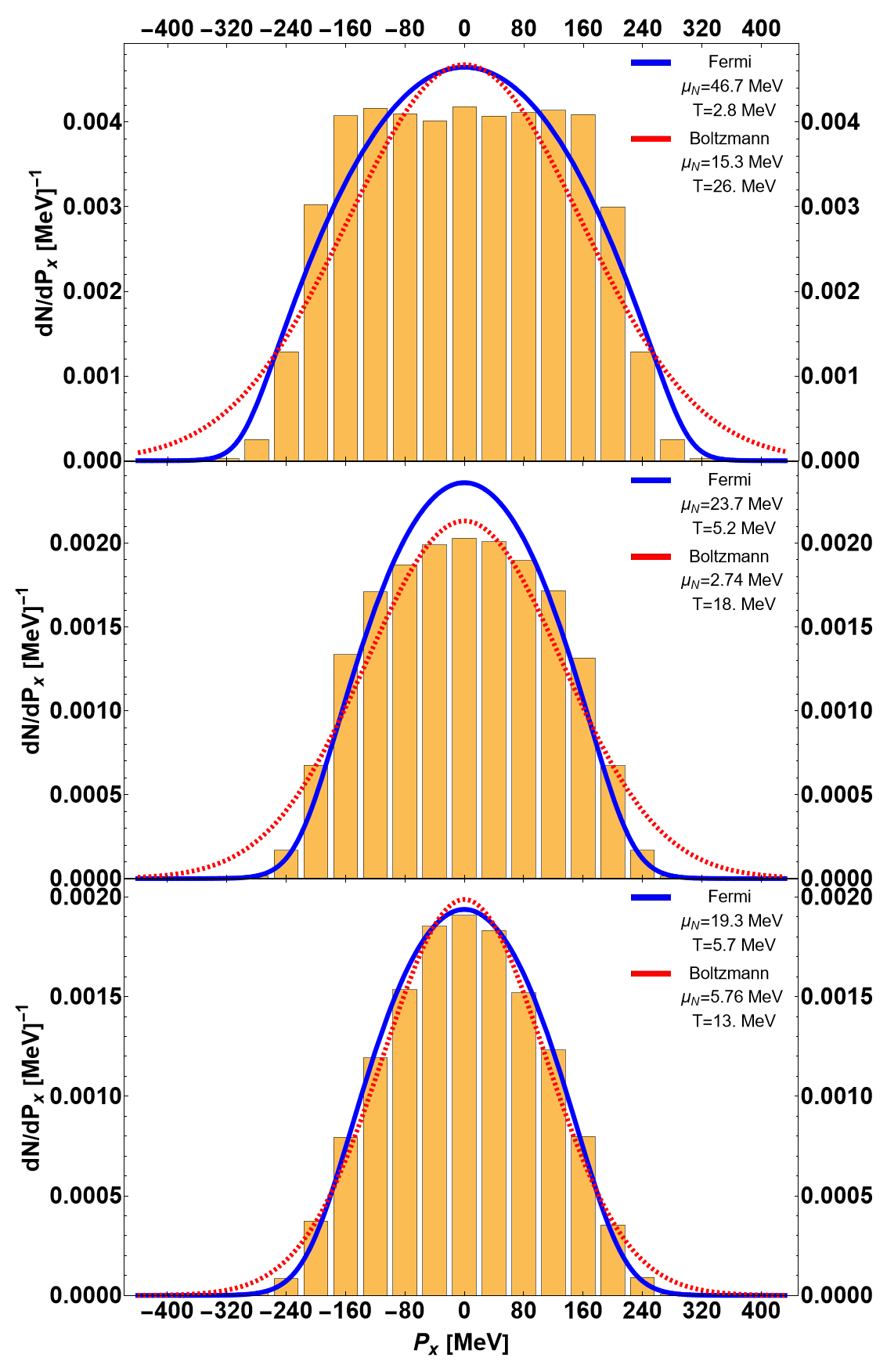}
\caption{Transverse-momentum distributions along the $x$-direction based on Fermi-Dirac (blue solid line) and Boltzmann (red dashed line) statistics compared to the CoMD transport outputs (orange histograms) for time snapshots at 65~fm/$c$ (top panel, corresponding to maximal density), 115~fm/$c$ (middle panel, near the start of the temperature plateau) and 175~fm/$c$ (bottom panel, near the onset of 3-D isotropy).}
\label{fig_trans_mom}
\end{figure}

Our coarse graining is deployed on the Constrained Molecular Dynamics (CoMD) model~\cite{Zheng:2014afa,Papa:2000ef,Papa:2005sp} using 24,000 simulated events of central collisions of \ce{^{40}Ca}+\ce{^{40}Ca} at 35 A$\cdot$MeV. The output of the code is individual positions and momenta of all nucleons. 
A grid of $8\times8\times8$ cells with volumes of (2 fm)$^3$ is constructed around the geometric center of the collision in its center-of-mass frame~\cite{Onyango:2021egp}. The central 8 cells are referred to as the first layer because they are all adjacent to the center of the collision, as portrayed in Fig.~\ref{fig_spatial_config1}. 
There are 56 cells that surround the first layer to form the second layer, portrayed in Fig.~\ref{fig_spatial_config2}. 
The cells in Figs.~\ref{fig_spatial_config1} and \ref{fig_spatial_config2} are categorized to denote cells which share a given symmetry with respect to the collision direction. The cells within a symmetry group can be analyzed together due to the cylindrical symmetry of the dynamics of central collisions. The second layer is surrounded by 152 cells that comprise the third layer. A fourth layer surrounds the third, and it contains 296 cells.
By tracking the nucleons at each timestep as they pass through the coarse-graining grid, the time dependence of local momentum distributions of nucleons in each cell of the grid can be determined. These distributions were averaged over all events. The distributions are corrected for the center-of-mass momentum so that they are in the thermal rest frame of the cell. The CM momentum is the first moment of the distribution and is calculated by summing the momenta of the nucleons in a cell across all events (for each timestep).
The distributions along the transverse direction (perpendicular to the beam axis) can be described reasonably well by a thermal ansatz, parameterized with fit functions
\begin{equation}
   \frac{dN}{dp_{x}}(t)=d\cdot V_{cell} \int{\frac{dp_{y} dp_z}{(2\pi)^3}f(E;\mu_N,T)},
\label{eq_xmom}
\end{equation}
and  
\begin{equation}
   \frac{dN}{dp_{y}}(t)=d\cdot V_{cell} \int{\frac{dp_x dp_z}{(2\pi)^3}f(E;\mu_N,T)},
\label{eq_ymom}
\end{equation}
where $E=(p_x^{\ 2}+p_y^{\ 2}+p_z^{\ 2})/2M$ is the nucleon kinetic energy with (average) nucleon mass $M$, $V_{\rm cell}$ the cell volume, $\mu_N$ the nucleon chemical potential, $T$ the temperature, $d=4$ the (spin-isospin) degeneracy, and $f$ the nucleon's thermal Fermi-Dirac distribution,
\begin{equation}
   f_N(E;\mu_N,T)=\big(1+e^{\frac{E-\mu_N}{T}}\big)^{-1} \ .
\label{eq_fermi_distribution}
\end{equation}
This allows for a straightforward extraction of chemical potential and temperature, and gives significantly better results than a Boltzmann ansatz which, in particular, does not capture the high-momentum tails well, which are the most relevant part for high-momentum photon production. An example of the parameterized fits in comparison to the transport output of the nucleon momentum spectra is displayed in Fig.~\ref{fig_trans_mom}. 
We also find that the distributions from the $y$-direction can be described using the same parameters extracted from the $x$-direction (within statistical fluctuations). 

\begin{figure}[thb!]
\centering
\includegraphics[scale=0.25]{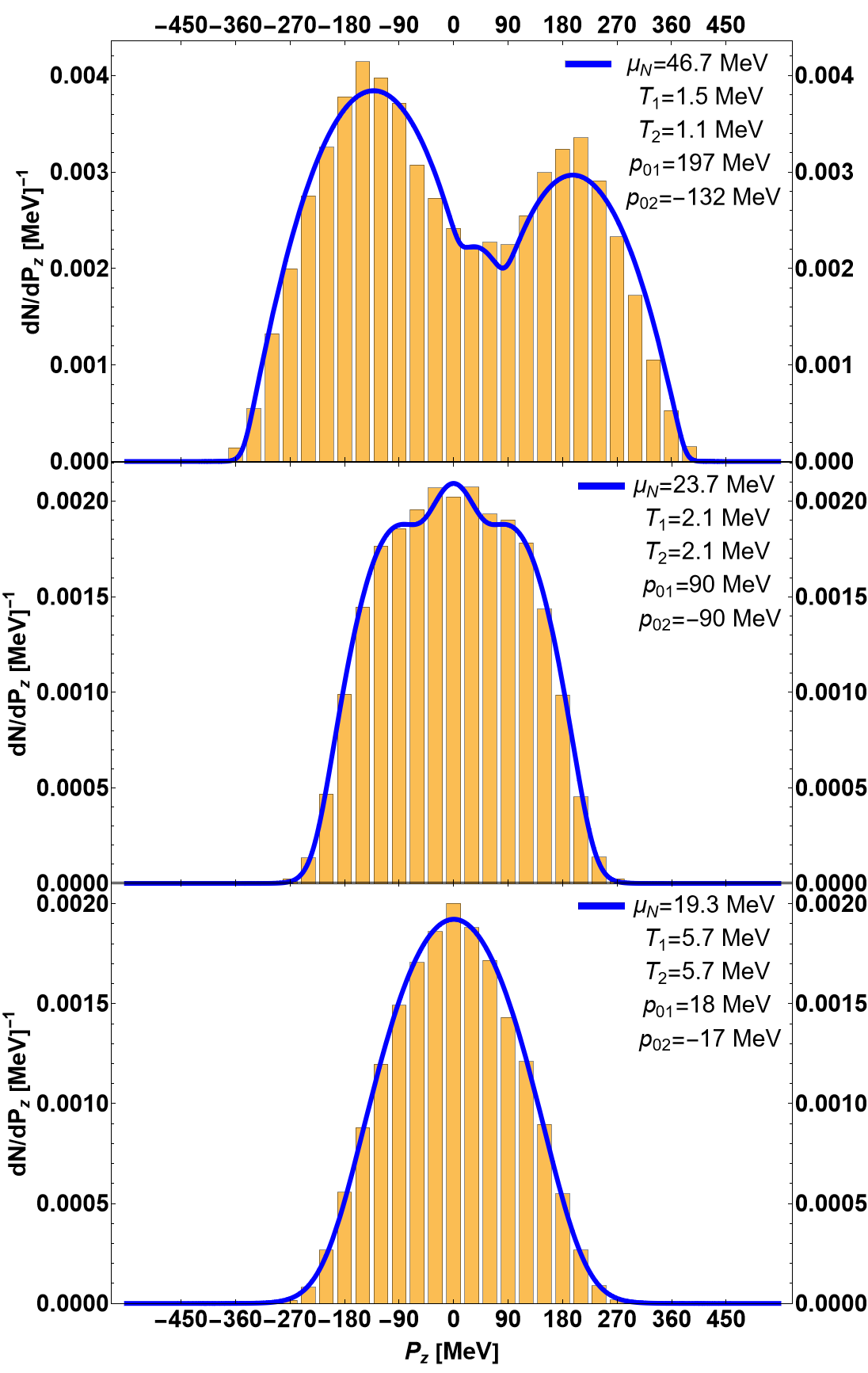}
\caption{Comparison of Fermi fit functions for the longitudinal momentum distributions (blue lines) to the transport distributions (orange histograms) for time snapshots at 65~fm/$c$ (top panel; maximal compression), 115~fm/$c$ (middle panel, onset of temperature plateau), and 175~fm/$c$ (bottom panel; onset of isotropy)}
\label{fig_long_mom}
\end{figure}

The longitudinal momentum distributions proved to be a more complicated case. At early times, they exhibit a two-hump structure reflecting the arrivals of the 2 nuclei within a given cell. An example of the longitudinal distribution from the center of the collision at different times is shown in Fig.~\ref{fig_long_mom}.
The fit function used for longitudinal distributions,
\begin{equation}    
\begin{split}    
f(\Vec{p};\mu_N,T,p_{01},&p_{02},\xi_1,\xi_2,w)=\\
&\frac{\sqrt{\xi_1}w}{1+\exp[\frac{p_\perp^2+\xi_1(p_z-p_{01})^2}{2mT}-\frac{\mu_N}{T}]}\\
+&\frac{\sqrt{\xi_2}(1-w)}{1+\exp[\frac{p_\perp^2+\xi_2(p_z-p_{02})^2}{2mT}-\frac{\mu_N}{T}]},      
\end{split}
\label{eq_p0}
\end{equation}
is a two-centroid distribution around momenta $p_{01,02}$ and includes a weight parameter $w$ to conserve baryon number. The widths of the longitudinal distributions (which are proportional to temperature) were found to be significantly narrower than those from the transverse direction during the compression and part of the expansion stages of the collision, requiring lower and slightly different temperatures from each other. This is accounted for with two thermal stretch parameters, $\xi_{1,2}$, to mimic an effective ``longitudinal" temperature. 

\begin{figure}[htb!]
\centering
\includegraphics[width=\linewidth]{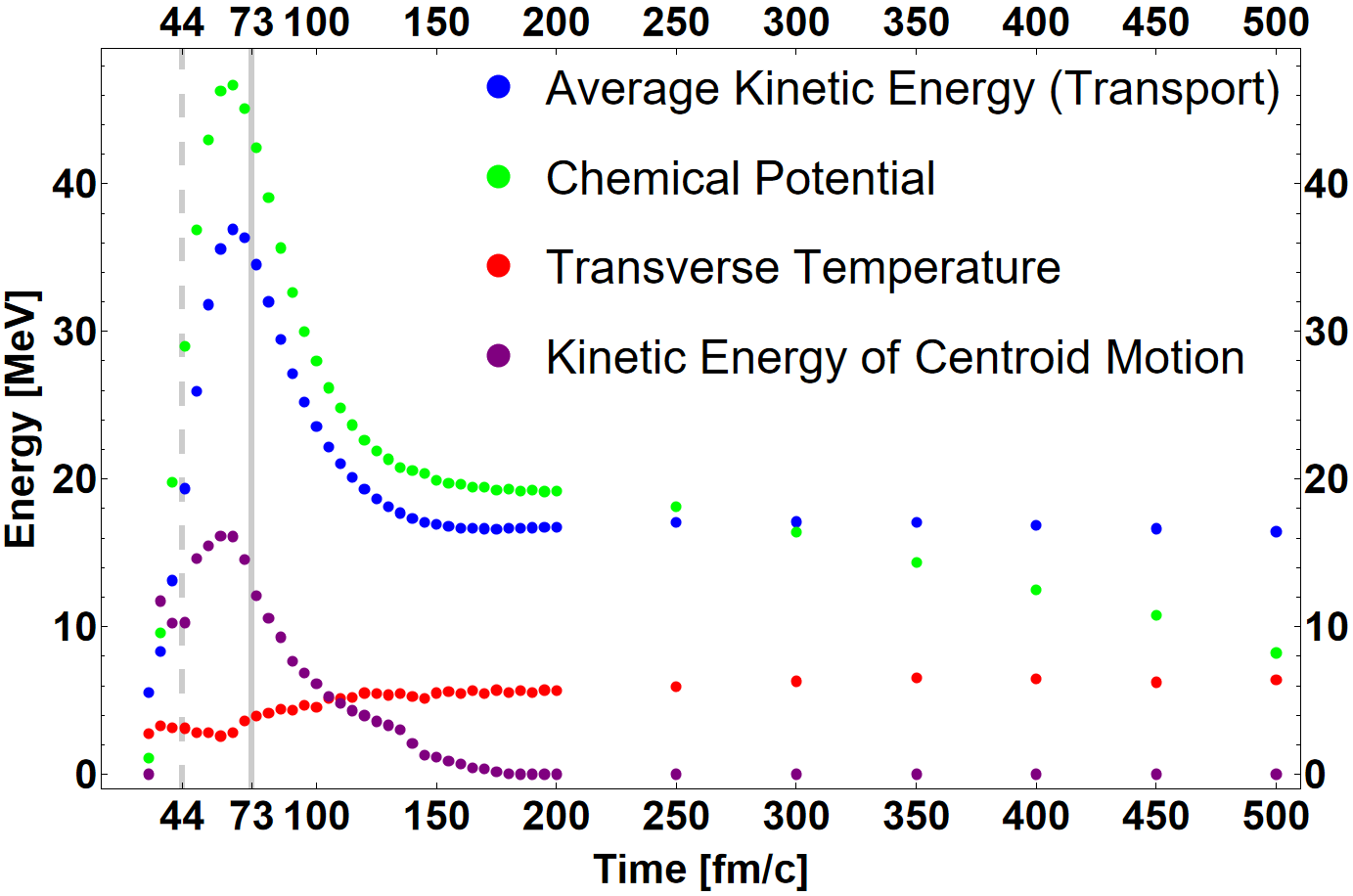}
\caption{Time evolution of the thermodynamic properties of nuclear matter extracted from the first layer of the coarse-graining grid of CoMD transport simulations, i.e., the average total kinetic energy (blue), nucleon chemical potential (green) and temperature (red) from fits of the transverse-momentum spectra, and the kinetic energy of the motion of the two centroids in the longitudinal direction (purple). The dashed line marks the approximate time when the nuclei would touch when treated as classical spheres, and the solid line indicates when the two nuclei would reach full geometric overlap if there were no interactions.}
\label{fig_kinetics_central}
\end{figure}
\begin{figure}[bht]
\begin{minipage}{\linewidth}
\centering
\includegraphics[width=\linewidth]{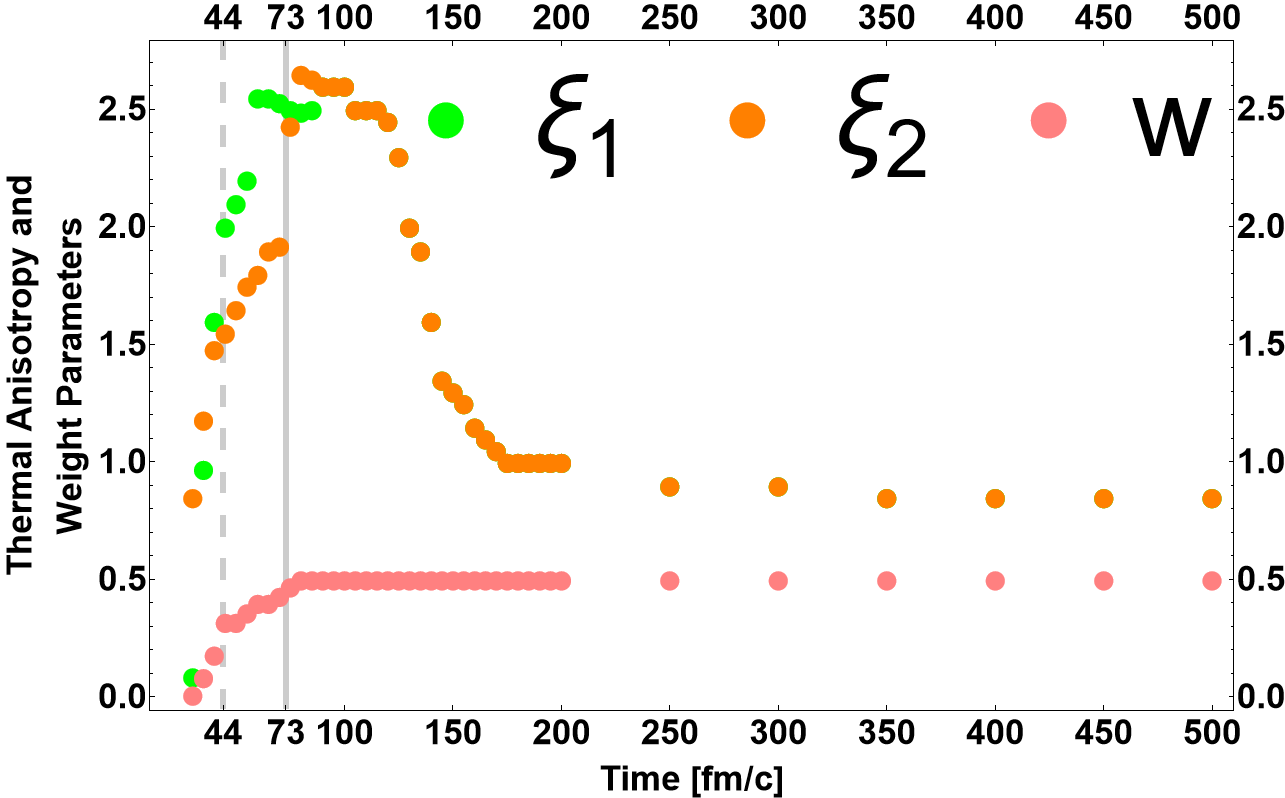}
\end{minipage}
\caption{Evolution of the weight $w$ (pink) and thermal stretch parameters $\xi_{1,2}$ (green, orange) for the effective longitudinal temperatures at the center of the collision. The dashed and solid lines are the same as in Fig.~\ref{fig_kinetics_central}.}
\label{fig_stretch}
\end{figure}
\begin{figure}[t!]
\centering
\includegraphics[width=\linewidth]{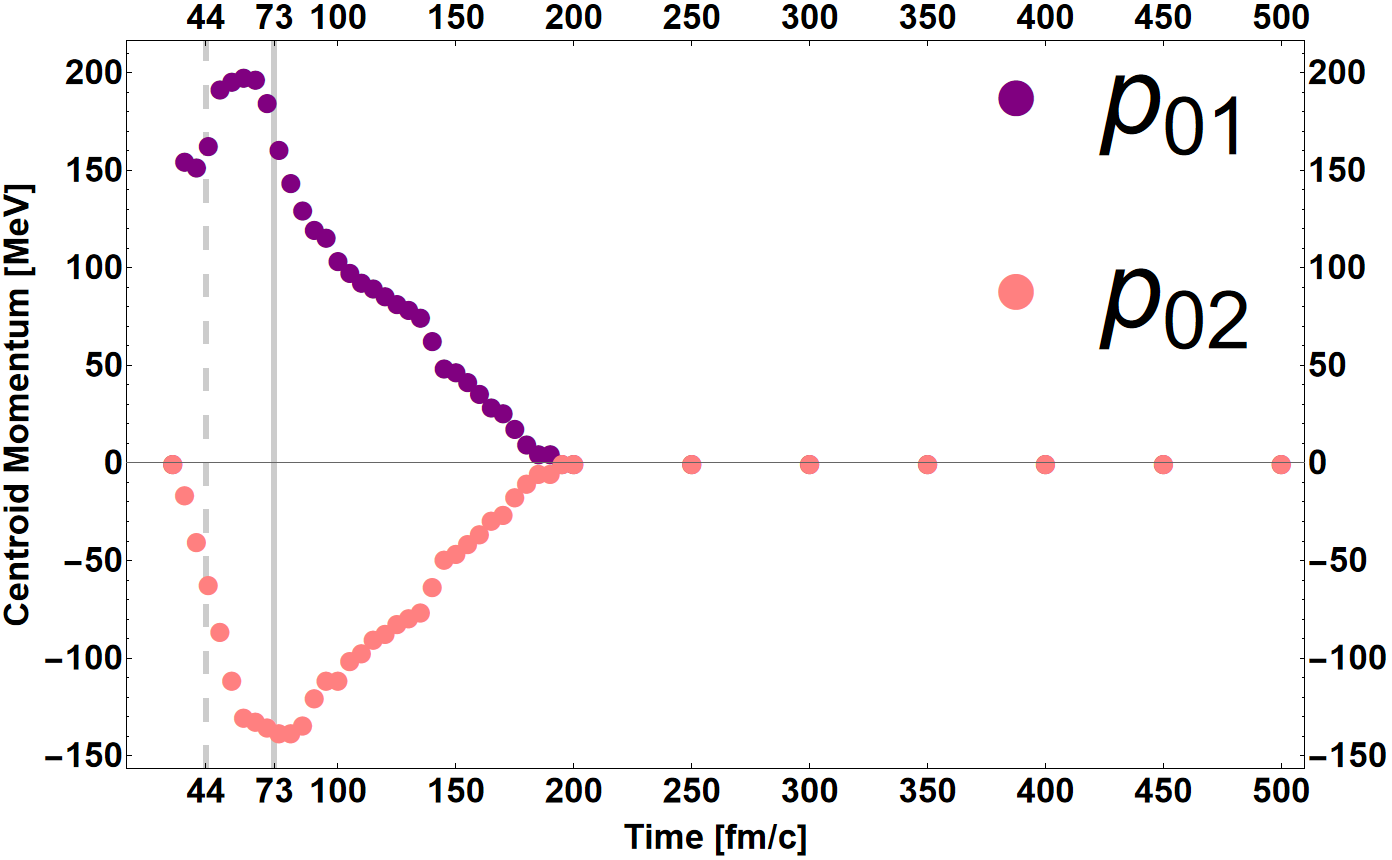}
\caption{Time evolution of the two centroid momentum parameters $p_{01,02}$ (purple, pink) extracted from the center of the collision. The dashed and solid lines are the same as in Fig.~\ref{fig_kinetics_central}.}
\label{fig_centroid_evo}
\end{figure}

The resulting time evolution of the thermodynamic properties of the nuclear matter at the center of the collision is displayed in Figs.~\ref{fig_kinetics_central}, \ref{fig_stretch}, and \ref{fig_centroid_evo}.
As a further piece of information, we also calculated the average kinetic energy per nucleon in a cell directly from the transport output, independent of the parameter extraction, 
\begin{equation}    
\begin{split}    
    E_k=\frac{<p_x^2>+<p_y^2>+<p_z^2>}{2M},     
\end{split}
\label{eq_kinetic_energy}
\end{equation}
where 
\begin{equation}    
\begin{split}    
   <p_x^2>=\frac{\int dp_x p_x^2 \frac{dN}{dp_x}}{\int dp_x \frac{dN}{dp_x}} \ , 
\end{split}
\label{eq_average_square_momentum}
\end{equation} 
and likewise for $<p_y^2>$ and $<p_z^2>$. The distributions used in Eq.~(\ref{eq_average_square_momentum}) are the orange histograms shown in Figs.~\ref{fig_trans_mom} and \ref{fig_long_mom}. 
In addition to displaying the values of the centroid momenta in Fig.~\ref{fig_centroid_evo}, we have plotted the kinetic energy related to centroid motion in Fig.~\ref{fig_kinetics_central}. This type of kinetic energy, defined as 
\begin{equation}    
\begin{split}    
    E_c=w\frac{p_{01}^2}{2M}+(1-w)\frac{p_{02}^2}{2M},     
\end{split}
\label{eq_centroid_energy}
\end{equation}
quantifies the amount of kinetic energy related to collective motion that has not (yet) been converted into random thermal motion. In a slight modification to Ref.~\cite{Onyango:2021egp}, where
\begin{equation}    
\begin{split}    
    E_c=\frac{p_{01}^2}{2m}+\frac{p_{02}^2}{2M} \   
\end{split}
\label{eq_wrong_centroid_energy}
\end{equation}
was used, we here include the weight parameter, which normalizes the value according to the baryon number of each nucleus.

%%%%%%%%%%%%%%%%%%%%%%%%
\section{Photon Emission Rate}
\label{sec_gam-rate} 
%%%%%%%%%%%%%%%%%%%%%%%%
With a description of the evolution of nuclear matter in heavy-ion collisions at Fermi energy, we proceed to calculating a local production rate of photons from nucleon-nucleon interactions. We first recapitulate the derivation of the thermal emission rate (Sec.~\ref{ssec_NN-brems}), detail how we determine and implement the pertinent $NN$ cross section (Sec.~\ref{ssec_cross-sec}), and give a brief discussion of the soft-photon approximation (Sec.~\ref{subsec_soft-gam}).

%%%%%%%%%%%%%%%%%%%%%%%%%%%%%%%%%
\subsection{Nucleon-Nucleon Bremsstrahlung}
\label{ssec_NN-brems}
\begin{figure}[ht!]
	\centering
	\includegraphics[width=\linewidth]{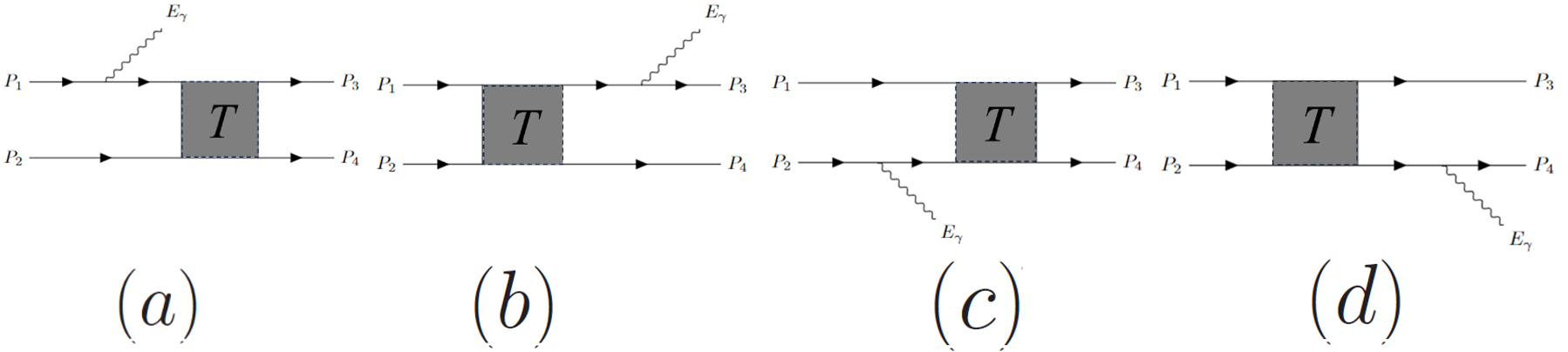}
	\caption{Diagrams of nucleon-nucleon bremsstrahlung where a photon is emitted from external nucleon lines.}
    \label{fig_ext_bremsstrahlung}
\end{figure}

\begin{figure}[tbh!]
	\centering
	\includegraphics[width=0.5\linewidth]{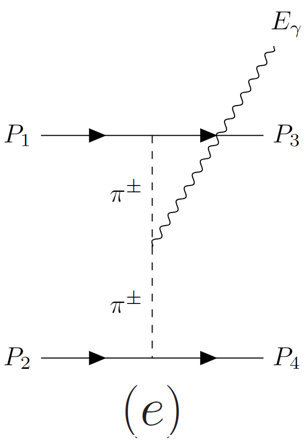}
	\caption{Diagram of nucleon-nucleon Bremsstrahlung where a photon is emitted from the internal process of exchanging a charged pion.}
    \label{fig_int_bremsstrahlung}
\end{figure}

%%%%%%%%%%%%%%%%%%%%%%%%%%%%%%%%
It was demonstrated almost 70 years ago how the emission of soft photons, \ie, those whose energies are much smaller than the energy scales involved in the collisions producing them, can be related to the elastic collisions of nucleons~\cite{Low:1958sn}, and subsequently applied in Ref.~\cite{Nyman:1968jro}.
The justification of the soft-photon approximation is based on the eikonal approximation \cite{Low:1958sn,Burnett:1967km}. We reproduce this argument for photon emission from the external legs of nucleon-nucleon scattering~\cite{Nyman:1968jro} here. Let us separate the matrix element for photon emission, 
$\mathcal{M}$, into two terms depending on their dependence on the photon's energy, $E_\gamma$,
\begin{equation}
    \mathcal{M}(K)=\mathcal{M}^{I}(K)+\mathcal{M}^{II}(K),    
\end{equation}
where $\mathcal{M}^{I}$ contains terms that are singular in $E_\gamma$ and 
\begin{equation}
    \mathcal{M}^{II}(K)=\mathcal{M}^{II}(0)+\mathcal{O}(E_\gamma) \ .    
\end{equation}
For nucleon-nucleon collisions, electric charge is conserved. This can be written through current conservation as
\begin{equation}
    K^\mu\mathcal{M}_\mu(K)=0,   
\end{equation}
which is the momentum space analogue of $ (\partial/\partial x^\mu)\mathcal{M}_\mu(x)=0$.
In quantum field theory, the emission rate of photons with energy $E_\gamma$ per 4-volume and 3-momentum produced from the external 
legs of the scattering nucleons which are close to their mass shell can be written as
\begin{equation}
\begin{split}
    E_\gamma\frac{d^7N_\gamma}{d^3kd^4x}=&\frac{E_\gamma}{2E_\gamma(2\pi)^3}\\
    &\int \frac{d^3p_1}{2E_1 (2\pi)^3}\frac{d^3p_2}{2E_2 (2\pi)^3} f_N(E_1) f_N(E_2)\\
    &\int d\Pi |\mathcal{M}|^2_{NN\gamma} 
\label{eq_photon_rate}
\end{split}
\end{equation}
where 
\begin{equation}
\begin{split}
    d\Pi=\frac{d^3p_3}{2E_3(2\pi)^3}\frac{d^3p_4}{2E_4(2\pi)^3}  (1-f_N(E_3))(1-f_N(E_4))\\
    \times(2\pi)^4\delta^4(P_1+P_2-P_3-P_4-K)    
\end{split}
\end{equation}
is the nucleon final-state phase space and
\begin{equation}        
    |\mathcal{M}|^2_{NN\gamma}=4\pi\alpha(\epsilon^\mu \Tilde{J}_\mu)^2 64\pi^2E_{CM}^2\frac{d\sigma_{NN}}{d\Omega_{CM}}    
    \label{eq_bremsstrahlung_matrix}
\end{equation}
is the scattering matrix element; 
$P_i=(E_i,\Vec{p}_i)$ are the nucleon four-momenta with $i=1,2,3,4$; $K=(E_\gamma,\Vec{k})$ is the photon's four-momentum; 
$\epsilon^\mu$ the photon polarization vector; $\frac{d\sigma_{NN}}{d\Omega_{CM}}$ the differential elastic $NN$ scattering cross section; and $\Tilde{J}_\mu=J_\mu$ or $L_\mu$ the nucleon electromagnetic currents~\cite{Rrapaj:2015wgs,Chang:2016ntp}. The photon coupling to the electric dipole current, 
\begin{equation}        
    J_\mu=\bigg(\frac{P_1}{P_1\cdot K}-\frac{P_3}{P_3\cdot K}\bigg)_\mu \ ,   
    \label{eq_dipole_current}
\end{equation}
is the leading-order term in the soft-photon energy expansion~\cite{Low:1958sn,Heller:1968cry,Nyman:1968jro,Rrapaj:2015wgs,Schaefer:1991rp}. This implies that the emission rate from collisions of protons on neutrons is the dominant contribution. The dipole current corresponds to two of the diagrams in  Fig.~\ref{fig_ext_bremsstrahlung},  $(a)$ with $(d)$ and $(c)$ with $(b)$ correspond to charge-exchange scattering. $(a)$ with $(b)$ and $(c)$ with $(d)$ represent the direct channels (neutral exchange).
For $pp$ collisions, the leading-order term is proportional to the quadrupole electric current. This causes the amplitudes for $pp\to pp\gamma$ to be suppressed by a factor of $v^2\propto T_{CM}/M$ where $v$ is the initial speed of the nucleons, $T_{CM}\equiv(\Vec{p_1}-\Vec{p_2})^2/4M$ is the 
total kinetic energy in the nucleon-nucleon center-of-mass frame. The pertinent rate can be obtained by inserting the quadrupole current, 
\begin{equation}        
    L_\mu=\bigg(\frac{P_1}{P_1\cdot K}+\frac{P_2}{P_2\cdot K}-\frac{P_3}{P_3\cdot K}-\frac{P_4}{P_4\cdot K}\bigg)_\mu,   
    \label{eq_quadrupole_current}
\end{equation}
and the cross section appropriate for those elastic collisions. The quadrupole current corresponds to considering all four of the diagrams from Fig.~\ref{fig_ext_bremsstrahlung}. 
Equation~(\ref{eq_bremsstrahlung_matrix}) specifies how $NN\to NN\gamma$ emission is coupled to $NN\to NN$ elastic scattering. The differential d$\Pi$ contains the Pauli blocking factors $(1-f_N(E_3))(1-f_N(E_4))$ which play a significant role in collisions producing highly degenerate nuclear matter. The soft photons are said to ``freely stream" out, so it is not necessary to include a $(1+f)$ Bose enhancement factor~\cite{Brinkmann:1988vi}. In our exploratory study here, we do not consider photon emission from internal Bremsstrahlung, an example of which is the charged-pion exchange shown in Fig.~\ref{fig_int_bremsstrahlung}. 
%This process corresponds to $\mathcal{M}^{II}$ and has a leading order term of $E_\gamma/T_{CM}$ in the soft-photon energy expansion. Internal bremsstrahlung processes must also conserve charge which can be characterized by 
%\begin{equation}
 %  K^\mu\mathcal{M}^{II}_\mu(K)=0 \ .   
%\end{equation}
%valid for any $K$. In the soft photon limit this relation becomes 
%\begin{equation}
 %   K^\mu\mathcal{M}^{II}_\mu(0)=0,   
%\end{equation}
%and further
%\begin{equation}
 %   \mathcal{M}^{II}_\mu(0)=0   
%\end{equation}
%because each component of $K$ is an independent variable.

After making use of the azimuthal symmetry of the nucleon distribution functions and enforcing conservation of momentum and energy, the photon emission rate for the $pn\to pn\gamma$ process, which contributes at the dipole order, simplifies to
\begin{equation}
\begin{split}
    E_\gamma\frac{d^7N_\gamma}{d^3kd^4x}=\frac{\alpha}{24\pi^7M^3 E_\gamma^2}\int{dp_\perp d\phi dp_z dq_\perp dq_z\ r_{np}^2 p_\perp}\\
  \times(p_\perp^2+p_z^2)q_\perp\ f_1f_2\int d\phi'dp'_z(1-f_3)
   (1-f_4)\\
   \times\bigg[1-\frac{p_\perp\sqrt{p_\perp^2+p_z^2-M E_\gamma-p_z'{ }^2}\cdot\cos{(\phi-\phi')}+p_z p_z'}{\sqrt{p_\perp^2+p_z^2}\sqrt{p_\perp^2+p_z^2-M E_\gamma}}\bigg] \ .
\end{split}
\label{eq_cylindrical_text}
\end{equation}
Here, $\Vec{p}$ $(\Vec{p'})$ denotes the incoming (outgoing) relative momentum, $\Vec{q}$ the total momentum, and $r_{np}$  the hard-sphere scattering radius for $pn$ scattering; its connection to the cross section is discussed in the next section. While Eq.~(\ref{eq_photon_rate}) was defined in the thermal rest frame of a coarse-graining cell, the integration variables of Eq.~(\ref{eq_cylindrical_text}) are defined in the nucleon-nucleon center-of-mass frame. The simplification from Eq.~(\ref{eq_photon_rate}) to Eq.~(\ref{eq_cylindrical_text}) can be found in App.~\ref{app_photon_rate}. 

%%%%%%%%%%%%%%%%%%%%%%%%
\subsection{Implementation of Cross Sections}
\label{ssec_cross-sec} 
\begin{figure}[htb!]
	\centering
	\includegraphics[width=\linewidth]{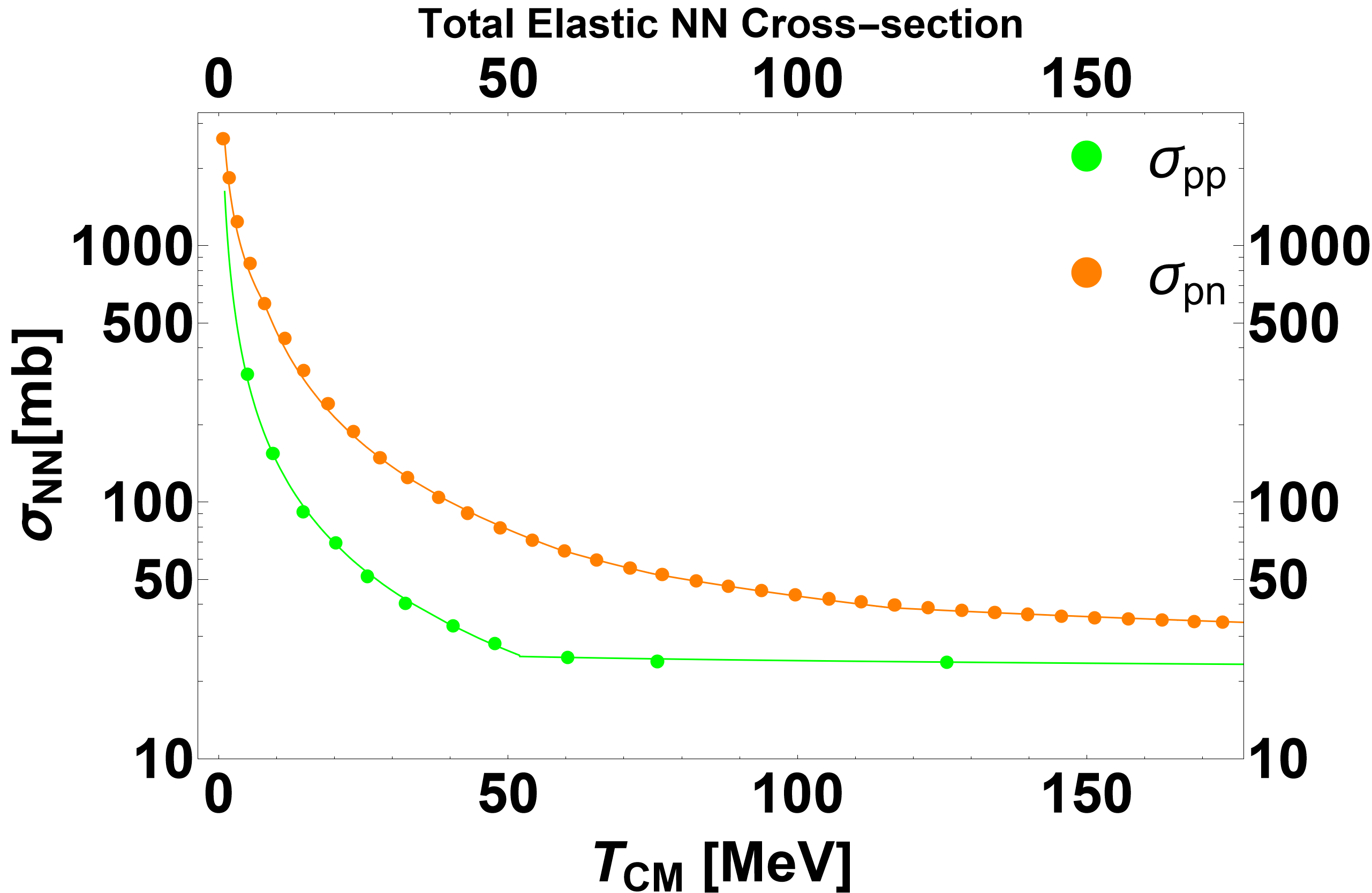}
	\caption{Experimental data for total $pn$ (green dots) and $pp$ (orange dots) cross sections from Refs.~\cite{Rrapaj:2015wgs} and \cite{RevModPhys.65.47} compared to their parameterizations (lines of the same colors).}
    \label{fig_parameterized_cs}
\end{figure}

\begin{figure}[thb!]
	\centering
	\includegraphics[width=0.85\linewidth]{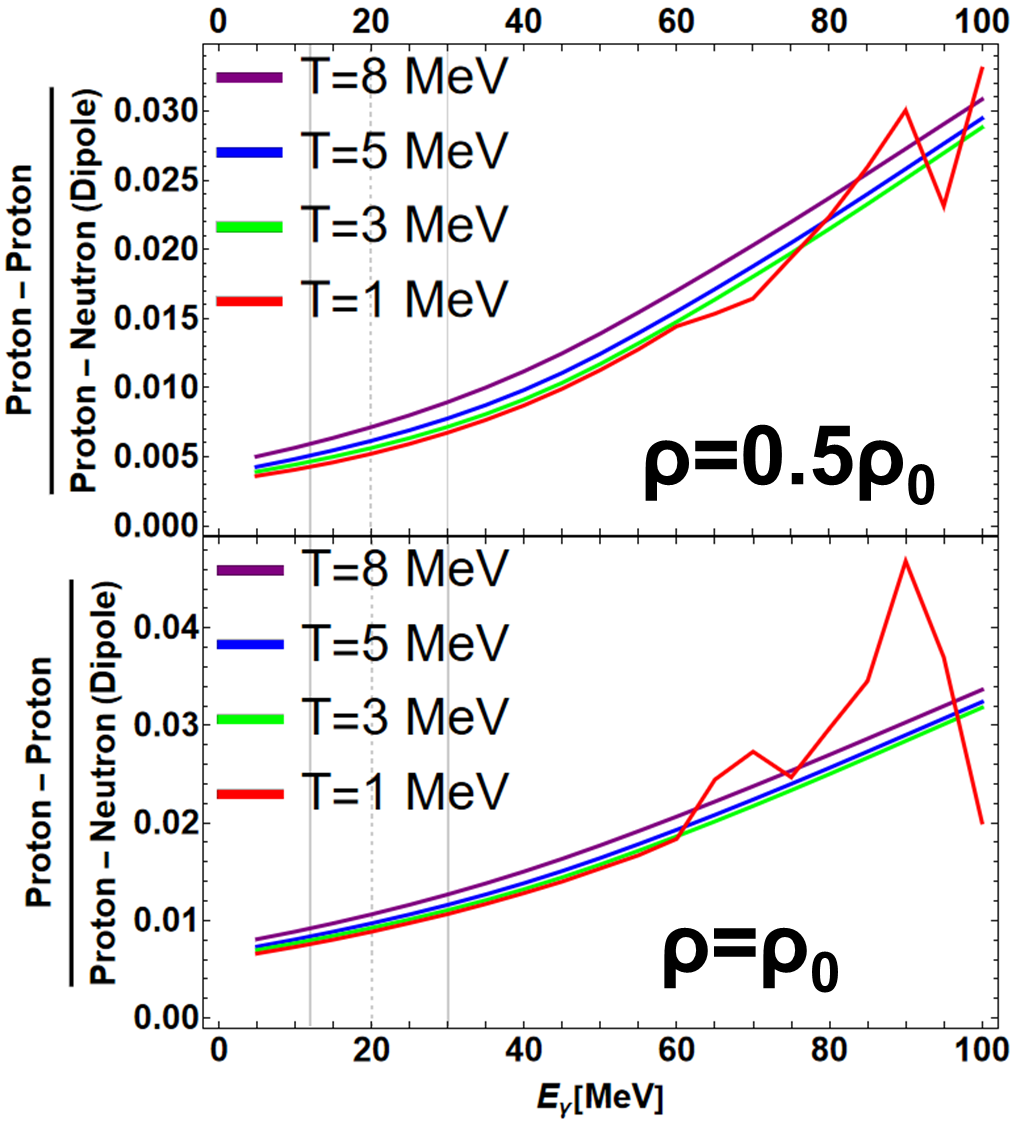}
	\caption{The ratio of photon rates calculated from $pp$ collisions to $pn$ collisions at temperatures $T=1$ (red), $3$ (green), $5$ (blue), and $8$ (purple) MeV at half (upper panel) and full (lower panel) saturation density. 
    %The vertical lines are the same as in Fig.~\ref{fig_ratio_spa}.
    }
	\label{fig_pp_photon_rate}
\end{figure}

%%%%%%%%%%%%%%%%%%%%%%%%
In the present work, we do not attempt to employ a microscopically calculated nucleon-nucleon cross section. We instead opt to parameterize experimental data via a cross section, following what has been done in Refs.~\cite{Zhou:2012bd,Hartnack:1989sd,Hartnack:1997ez,Bass:1994cq}, to represent the collision energy dependence of the total cross section. Specifically, we utilize a parameterization of the $np$ cross section data presented in Ref.~\cite{Rrapaj:2015wgs} (based on data of Ref.~\cite{Stoks:1993tb}); $S$-wave scattering is assumed to be the dominant contribution, so the cross section was parameterized as $\sigma_{np}=\pi r_{np}^2$ where $r_{np}$ is a hard-sphere scattering radius. The collision energy dependence of the cross section is encoded in a parameterization of  $r_{np}$ (in fm) as
\begin{equation}
    r_{np}(T_{CM})=\left\{ 
        \begin{array}{ll}
          9.04065\ T_{CM}^{\ -0.354978}  \quad 0<T_{CM}<8 \text{ MeV} \\
          13.5565\ T_{CM}^{\ -0.549805}  \quad 8\leq T_{CM}<59 \text{ MeV} \\
          8.91011\ T_{CM}^{\ -0.446881}  \quad 59\leq T_{CM} <75 \text{ MeV}\\
          5.8584\ T_{CM}^{\ -0.349762}  \quad 75\leq T_{CM} <116 \text{ MeV}\\
          2.32566\ T_{CM}^{\ -0.155409}  \quad 116 \text{ MeV }\leq T_{CM}  \ .
        \end{array}
    \right. \\
\end{equation}
A comparison between the parameterized $np$ cross section and experimental data is shown in Fig.~\ref{fig_parameterized_cs}.
This cross section is different from the one used in previous applications to photon emission, 
\begin{equation}
    \frac{d^2\sigma_{np\to np\gamma}}{dE_\gamma d\Omega_\gamma}=\alpha\frac{R^2}{12\pi}\frac{1}{E_\gamma}(2\beta_f^{\ 2}+3\beta_i^{\ 2}\sin^2\theta_\gamma),
\end{equation}
with a constant hard-sphere scattering radius of $R\approx2.1$~fm~\cite{Bauer:1986zz} and $R=\sqrt{3}$~fm~\cite{Cassing:1986aya} (with $\beta$: nucleon velocity, $\theta_\gamma$: angle between the photon and incoming nucleon).
Probabilistically, in a collision of isospin symmetric nuclei, one expects half of all $NN$ collisions to be $pn$ collisions. 
The cross section for $pp$ collisions is based on a parametrization of experimental data from Fig.~5 of Ref.~\cite{RevModPhys.65.47}, parameterized with its corresponding hard-sphere scattering radius (in fm) of 
\begin{equation}
    r_{pp}(T_{CM})=\left\{ 
        \begin{array}{ll}
          7.1873\ T_{CM}^{-0.5267}  \quad 0<T_{CM}<52\text{ MeV} \\
          0.999341\ T_{CM}^{-0.0286}  \quad 52\text{ MeV}\leq T_{CM} \ . 
        \end{array}
    \right. \\
\end{equation}
The cross sections are compared in Fig.~\ref{fig_parameterized_cs}, where it can be seen that the $pp$ cross section is $\sim$60-70$\%$ of the $pn$ cross section for $T_{CM}<100$~MeV. 
For the quadrupole contribution pertinent to $pp$ collisions, the photon emission rate can be written as 
\bea
  E_\gamma\frac{d^7N_\gamma}{d^3kd^4x}=\frac{26\alpha}{135\pi^5M^6 E_\gamma^2}\int dp d\theta dq\ \sin{\theta}\ r_{pp}^{\ 2}\ p^6
\nonumber\\
 \times\sqrt{p^2-ME_\gamma}q^2\ f_1f_2\int d\theta'\ \sin{\theta'}(1-f_3)(1-f_4)
\nonumber \\
   \times\bigg[1-\frac{1}{8}(3+\cos{2\theta}+3\cos{2\theta}\cos{2\theta'}+\cos{\theta'})\bigg] \ .
 \label{eq_quadrupole_spherical}
\eea
A comparison between the photon rates from $pp$ and $pn$ collisions under nominal thermodynamic conditions for the HICs analyzed in this study is shown in Fig.~\ref{fig_pp_photon_rate}; $pp$ Bremsstrahlung contributes at $<4\%$ of the dipole component in $pn$ Bremsstrahlung.

Photon emission from neutron-neutron scattering is also possible despite neutrons being electrically neutral. They possess a magnetic moment due to their intrinsic spin and their constituents being electrically charged. An argument can be made that the cross section for $nn$ collisions is approximately the same as for $pp$ collisions~\cite{Charagi:1990zz} when ignoring Coulomb interactions (otherwise, the $pp$ cross section would be suppressed compared to the $nn$ cross section). It has also been reasoned that the cross sections for $pn$ (with total isospin $I$=1) and $nn$ are the same since there is no Coulomb interaction in either scenario. The production of photons from $nn$ collisions is expected to be suppressed compared to the $pp$ case by $(\mu_n/\mu_p)^2$~\cite{Low:1958sn,Herrmann:1991tu,Rrapaj:2015wgs}. The coupling of photons to neutrons is similar to the coupling in Eq.~(\ref{eq_bremsstrahlung_matrix}) except that the nucleon quadrupole current would include the anomalous magnetic moment in lieu of the electric charge \cite{Nyman:1968jro}. The photon yield from $nn$ Bremsstrahlung is further reduced by a kinematic factor of $E_\gamma/p=(E_\gamma/T_{CM})v$ for the leading-order term for magnetic ``quadrupole" contributions compared to the electric dipole~\cite{Low:1958sn,Herrmann:1991tu}. This amounts to a factor of $\sim0.6$ under conditions relevant to this study, \eg, $E_\gamma\lesssim80$ MeV, $T_{CM}=18$ MeV, and $v=0.136c$.

%%%%%%%%%%%%%%%%%%%%%%%%
\subsection{Soft-Photon Approximation}
\label{subsec_soft-gam} 
\begin{figure}[ht!]
	\centering
	\includegraphics[width=0.95\linewidth]{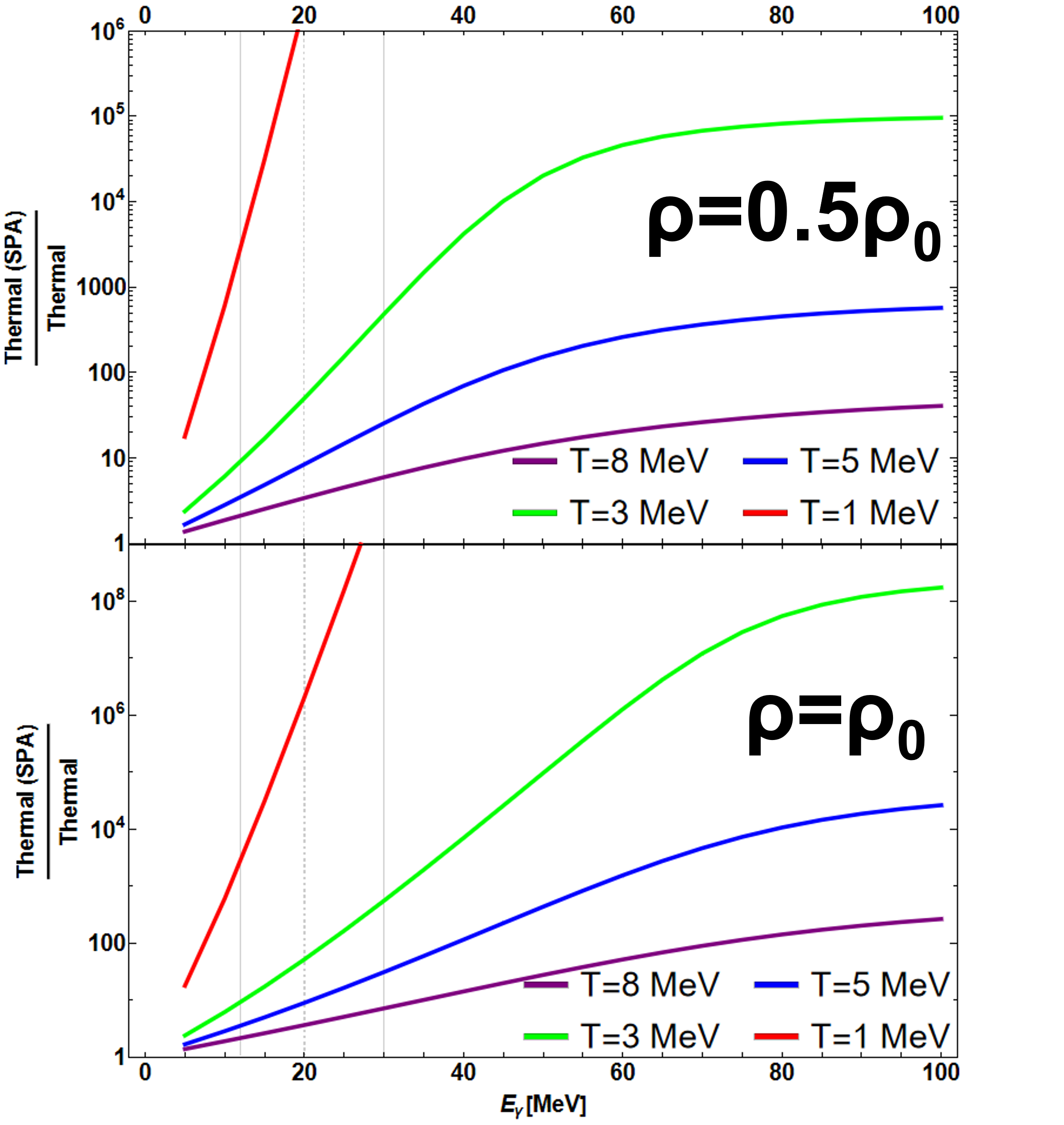}
	\caption{Ratio of thermal photon rates with SPA over thermal photon rates that consider the energy of the emitted photon during final state calculations at half (upper panel) and full (lower panel) saturation density ($\rho_0=0.16$ fm$^{-3}$) at temperatures $T=8$ (purple), $5$ (blue), $3$ (green), $1$ (red) MeV. 
    %The vertical solid lines denote the approximate bounds of the component of the photon energy spectrum 
 %(see Fig.~\ref{fig_anatomy_of_photon_spectrum}) 
 dominated by radiation from thermal sources.}
    \label{fig_ratio_spa}
\end{figure}

%%%%%%%%%%%%%%%%%%%%%%%%
The soft-photon approximation (SPA) usually entails neglecting the energy of the photon relative to that of the nucleons (since the photon energy is equal to its momentum, the latter is always well below the nucleon's momentum, $p=\sqrt{2ME}$, in the nonrelativistic regime). This approximation can become problematic if large degeneracies in the nucleon distributions are present. 
At the temperatures and densities expected in HICs around Fermi energies (cf. Fig.~\ref{fig_kinetics_central}), degeneracy effects are large, causing substantial final-state Pauli blocking especially at very low temperatures, that can be expected to be exacerbated when the final-state nucleon-energies are reduced due to photon emission. This is illustrated in Fig.~\ref{fig_ratio_spa}, where the suppression caused by including the photon energy in the energy conservation of the emission process can lead to a reduction of the rate by several orders of magnitude for temperatures and photon energies of interest for our study.
Consequently, we will not adopt the SPA for the photon energies in our calculations. On the other hand, the photon 3-momentum remains reasonably small compared to the typical nucleon momenta involved, and thus we will neglect it in the four-momentum-conserving $\delta$-function.
%\RR{Thomas, is this correct? If not, please modify.}
%The explanation is correct.

%%%%%%%%%%%%%%%%%%%%%%%%
\section{Benchmark using Emissivity}
\label{sec_emissivity} 
\begin{figure}[ht!]
	\centering
	\includegraphics[width=\linewidth]{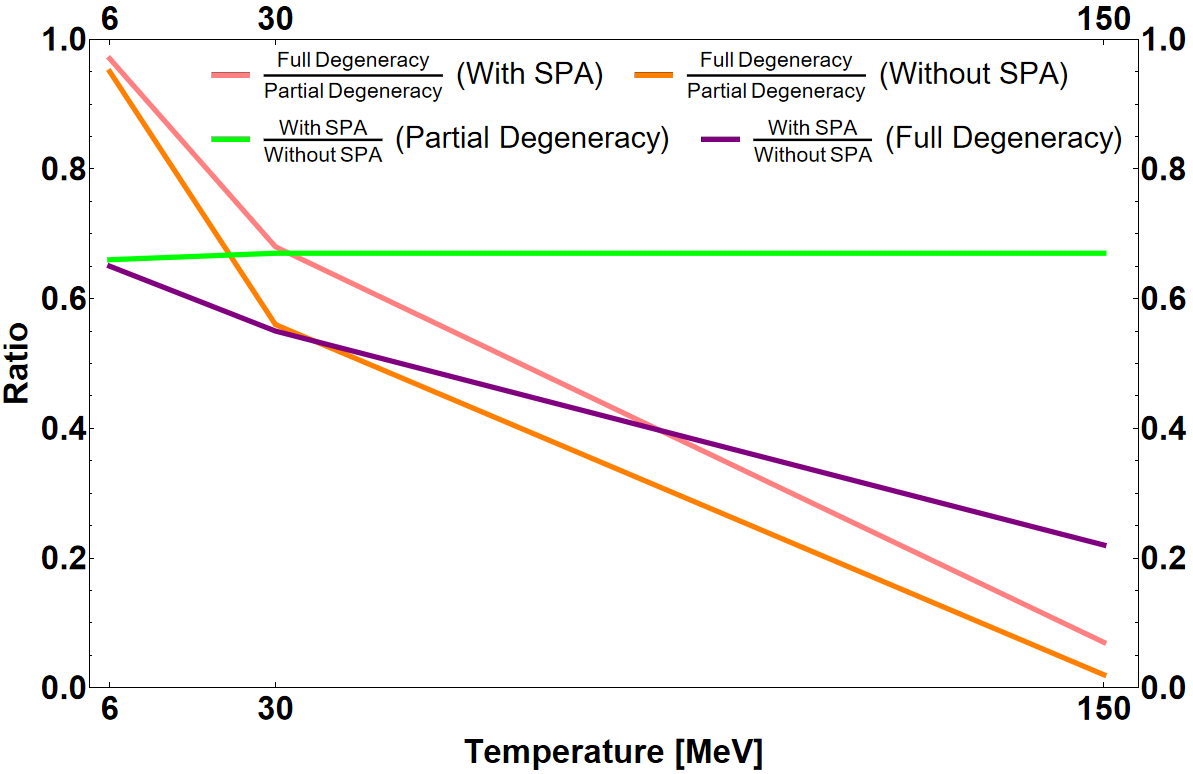}
	\caption{Ratios of emissivities of a partially degenerate description to a fully degenerate description of the nucleons with the SPA (pink), a partially degenerate description to a fully degenerate description of the nucleons without the SPA (orange),  SPA to no-SPA while treating the nucleons as partially degenerate (green),  and SPA to no-SPA while treating nucleons as fully degenerate (purple).}
    \label{fig_emissivity_treatment}
\end{figure}

%%%%%%%%%%%%%%%%%%%%%%%%
To benchmark our photon rate, we compare the emissivity, \ie, the  integrated photon rate,
\begin{equation}
    \Dot{\epsilon}=\int{d^3k\ E_\gamma\frac{d^7N_\gamma}{d^3kd^4x}} \ ,
\end{equation}
using different levels of approximations for quantum effects and energy conservation to determine the significance of those treatments. We have calculated photon emission using Boltzmann and Fermi functions for the nucleon distribution functions while also including or excluding Pauli blocking and the SPA. To directly compare our results to Fig.~4 (left panel) of Ref.~\cite{Rrapaj:2015wgs}, the emissivities are calculated at saturation density ($\rho_0=0.16$~fm$^{-3}$) and $T=30$~MeV (thermodynamic values representative of a typical supernova) and cataloged in Tab.~\ref{tab_emissivity} alongside a temperature typical of Fermi-energy HICs and a high temperature where quantum effects in the distribution functions do not make a significant difference. 
%A fiducial value calculated at $\rho_0$ and $T=30$ MeV from  is included.
\begin{table}[thb!]
	\centering
	\begin{tabular}{|m{5em}|m{2cm}|m{2cm}|m{2cm}|}
		\hline
		\multicolumn{4}{|c|}{Emissivity [MeV$\cdot$fm$^{-4}$] ($\rho=\rho_0$)} \\
		\hline
		Conditions & $T=150$ MeV & $T=30$ MeV & $T=6$ MeV \\
		  \hline
		Reddy \cite{Rrapaj:2015wgs} & 1.58$\times10^{-3}$ & 5.71$\times10^{-5}$ & 4.92$\times10^{-6}$\\ 
		\hline
		Boltzmann & 1.59$\times10^{-3}$ & 5.70$\times10^{-5}$ & 4.92$\times10^{-6}$\\
            \hline
		Partially degenerate (with SPA) & 1.64$\times10^{-3}$ & 7.40$\times10^{-5}$ & 1.99$\times10^{-5}$\\
		\hline
		Partially degenerate & 1.09$\times10^{-3}$ & 4.94$\times10^{-5}$ & 1.33$\times10^{-5}$\\ 
		\hline
		Degenerate (with SPA) & 1.59$\times10^{-3}$ & 5.01$\times10^{-5}$ &  1.35$\times10^{-6}$\\ 
		\hline
		Degenerate & 1.04$\times10^{-3}$ & 2.78$\times10^{-5}$ & 3.01$\times10^{-7}$\\ 
		\hline
	\end{tabular}
	\caption{Table of photon emissivities at various temperatures (second line) calculated by  Reddy et al. (third line), using Boltzmann distributions (fourth line), partially degenerate matter (blocking factors excluded) with the SPA (fifth line), partially degenerate matter (blocking factors excluded) without the SPA (sixth line), degenerate matter (blocking factors included) with the SPA (seventh line), and degenerate matter (blocking factors included) without the SPA (eighth line) at high temperature (second column), a fiducial temperature of supernovae (third column), and a temperature typical of Fermi-energy HICs (fourth column).}
\label{tab_emissivity}
\end{table}
The table is meant to be read from top to bottom with the calculations typically becoming more accurate. It starts by using Boltzmann distributions for nucleons, where our results are found to agree with those of Ref.~\cite{Rrapaj:2015wgs}. Subsequently, the nucleons are treated as fermions emitting photons whose energies are assumed to be negligible relative to the nucleon energies involved in the nuclear medium, \ie, in the SPA. Then the SPA is relaxed, and furthermore, Pauli blocking is accounted for, which constitutes the full calculation. Finally, the SPA is reinstated for a degenerate medium. The significance of the different treatments of photon emission and nuclear matter is illustrated by the amount of reduction in photon emission given as percentages in Fig.~\ref{fig_emissivity_treatment}.
For the case of partially degenerate matter, the amount of reduction in photon emission due to relaxing the SPA does not change with temperature. The amount of reduction does change with temperature for degenerate matter due to the phase space, which is already dramatically reduced by Pauli blocking, being further limited by emitting a photon. We can see that, for the thermodynamic conditions representative of HICs at Fermi energy, Pauli blocking is exceedingly consequential, causing a more than $90\%$ reduction in the photon emissivity. In regards to relaxing the SPA, its effect on highly degenerate matter is considerable causing a $78\%$ reduction in production compared to when the SPA is implemented. For the thermodynamic conditions relevant for the present study, the emissivity calculated in the approximations used by Reddy et al.~\cite{Rrapaj:2015wgs} and the emissivity based on Eq.~(\ref{eq_photon_rate}) (with fully degenerate matter and without SPA) differ by more than an order of magnitude, reiterating the importance of final-state Pauli blocking and the need to go beyond the SPA.

%%%%%%%%%%%%%%%%%%%%%%%%%%%%%%%%%%%%%%%%%%%%%%%%%%%%%%
\section{Properties of Photon Emission Rate}
\label{sec_photon_rate}
\begin{figure}[t!]
	\centering
	\includegraphics[width=\linewidth]{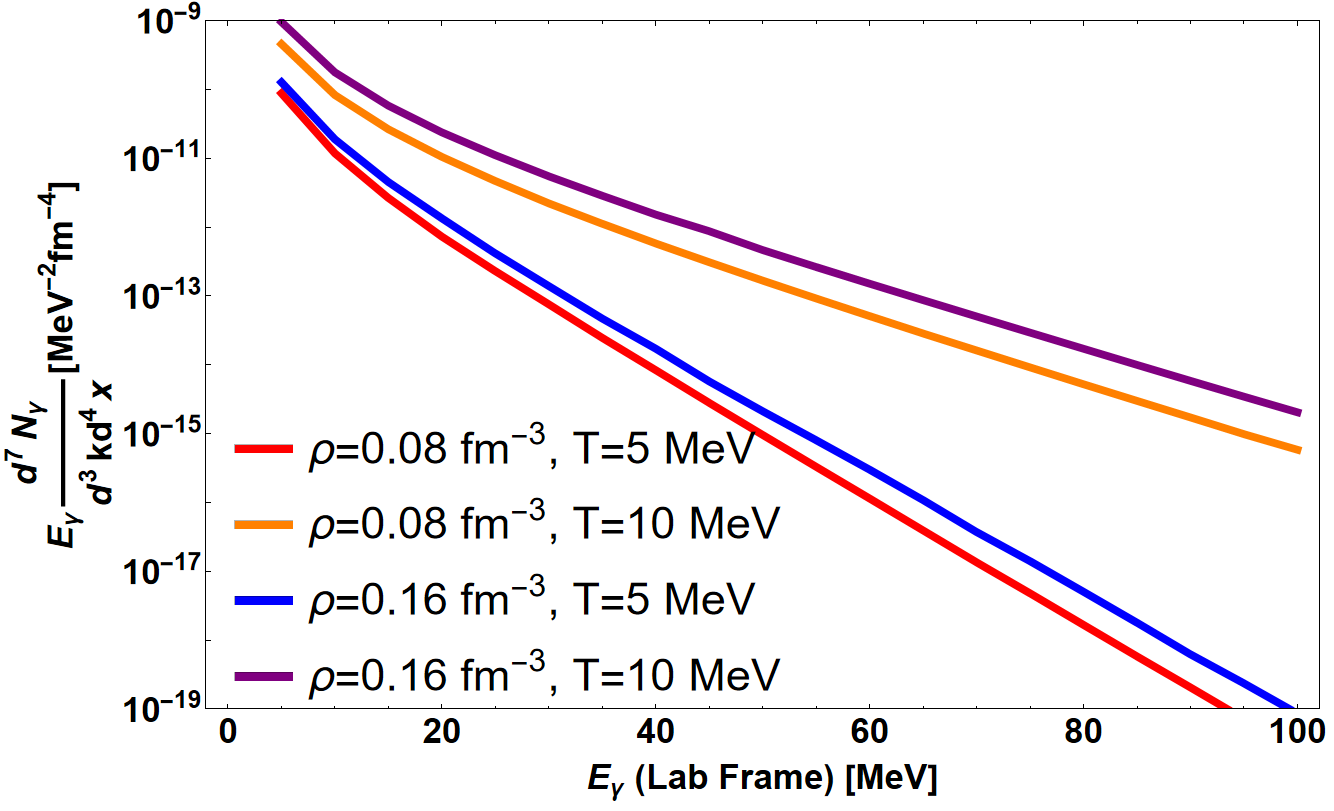}
	\caption{Comparison of photon emission rates in thermal nuclear matter at full- and half-saturation densities for temperatures of $T=5$ and 10\,MeV.}
    \label{fig_thermal_rate}
\end{figure}

%%%%%%%%%%%%%%%%%%%%%%%%%%%%%%%%%%%%%%%%%%%%%%%%%%%%%% 
We start by plotting the photon production rate for thermal (isotropic) nuclear matter (\ie, without any off-equilibrium effects) in Fig.~\ref{fig_thermal_rate} for a combination of 2 temperatures and densities resembling the thermodynamics of Fermi energy collisions. One recognizes a much stronger dependence on temperature than on nuclear density. 

The next step is to include the effect of the centroid motion of the two incoming nuclei for which we employ the distribution functions of Eq.~(\ref{eq_p0}) with a symmetric set-up, \ie, $p_{01,02}\to\pm p_0$ where $p_0\equiv\frac{|p_{01}-p_{02}|}{2}$. The photon rate turns out to be rather sensitive to this effect: with a centroid momentum of $p_0$=200\,MeV, which is near the maximum value found in our coarse-graining analysis, the rates are increased by several orders of magnitude with a weak residual dependence on temperature. An important feature of the distribution functions with collective motion is that the colliding nucleons can drop to very low momenta (even rest) while encountering little blocking due to the dip in the two-hump distribution, recall Fig.~\ref{fig_long_mom}.
\begin{figure}[ht!]
	\centering
	\includegraphics[width =\linewidth]{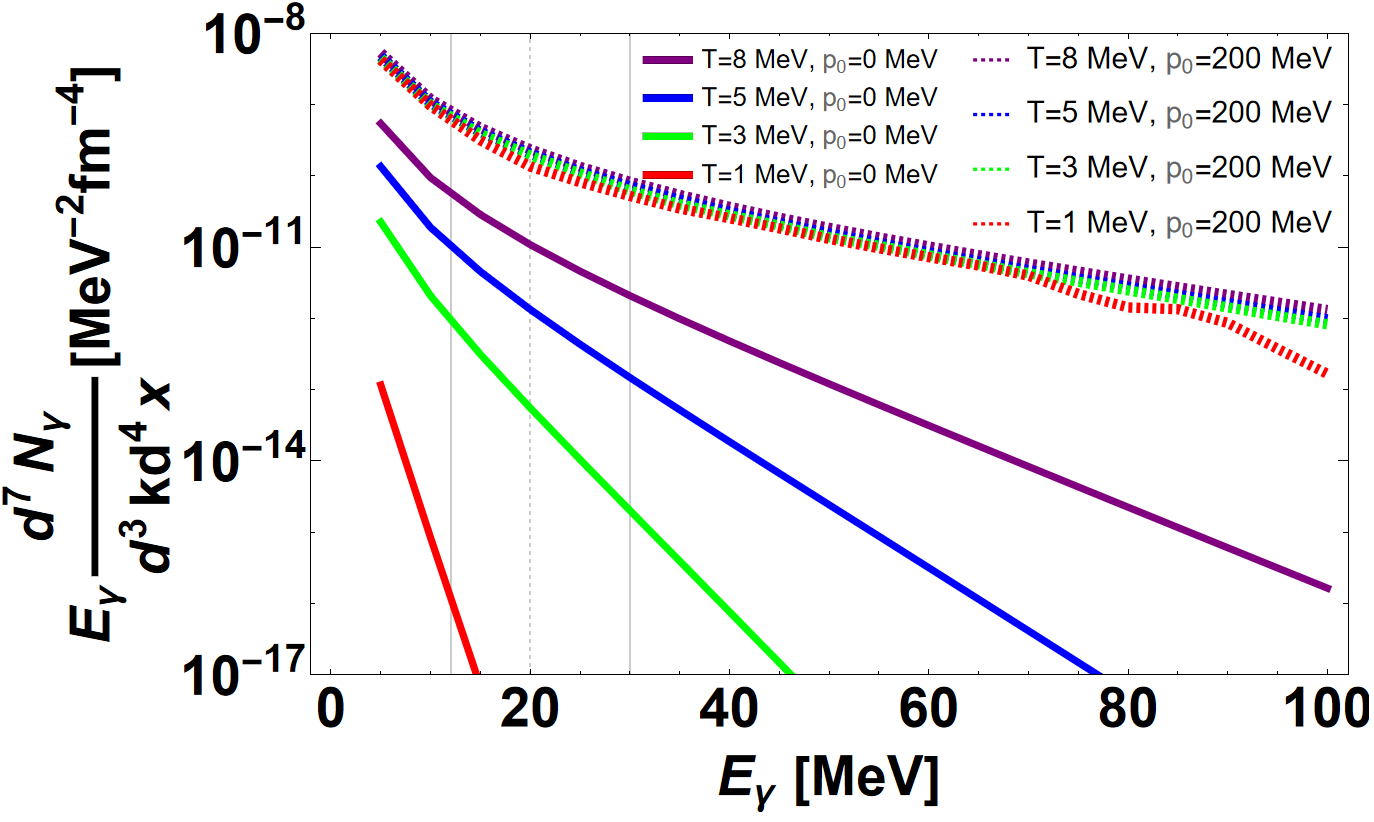}
	\caption{Comparison of photon emission rates within nuclear matter at saturation density without (solid lines) and with (dashed lines) centroid momentum in the distribution functions at temperatures $T=8$ (purple), $5$ (blue), $3$ (green), $1$ (red) MeV. 
    %The vertical lines are the same as in Fig.~\ref{fig_ratio_spa}.
    }
	\label{fig_centroidal_rate}
\end{figure}

\begin{figure}[ht!]
	\centering
	\includegraphics[width=\linewidth]{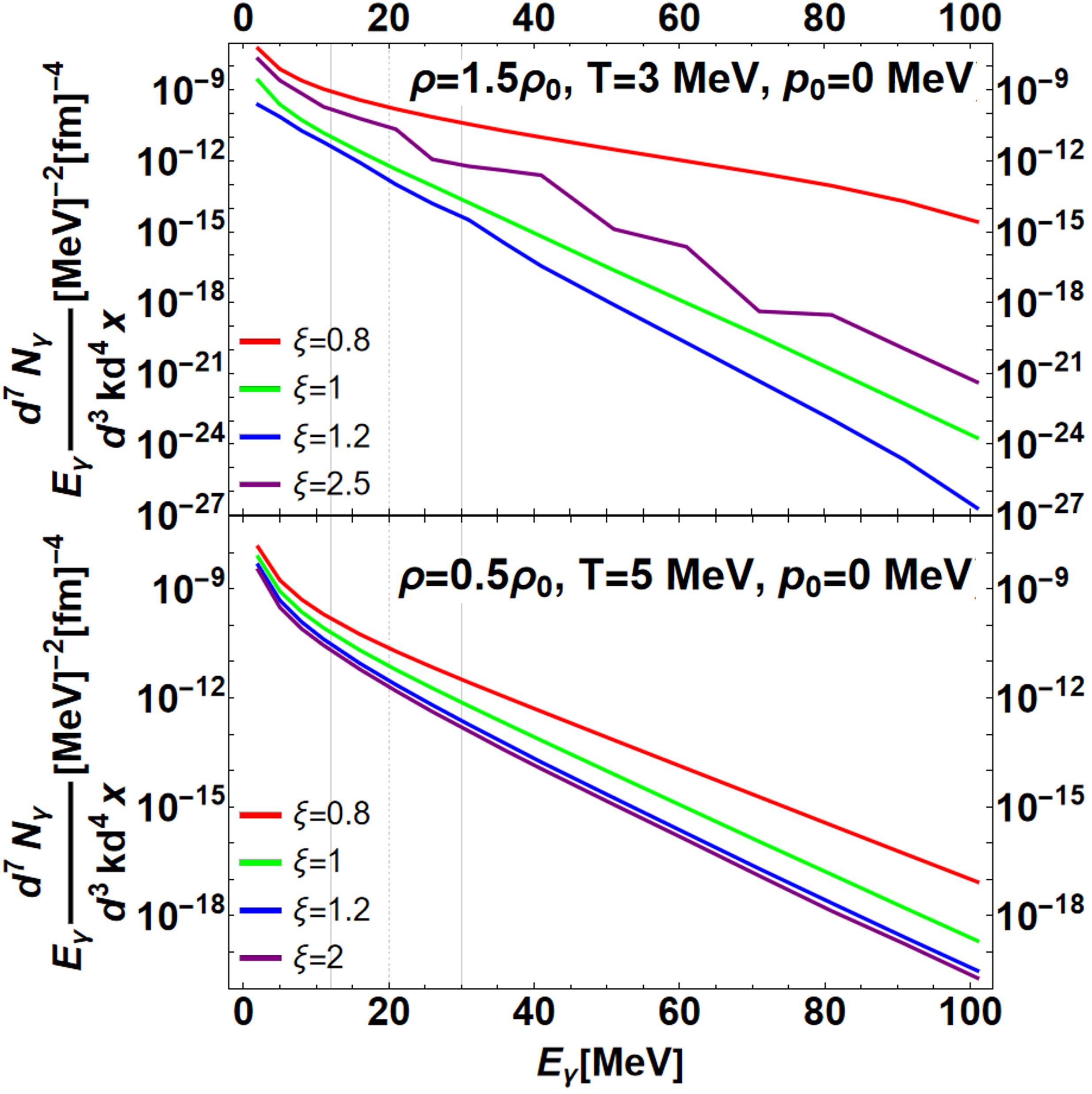}
	\caption{Photon emission rate at various values of $\xi$ at $1.5\times$ saturation density and $T=3$ MeV in the upper panel and $0.5\times$ saturation density and $T=5$ MeV in the lower panel. All curves are calculated without centroid motion. 
    %The vertical lines are the same as in Fig.~\ref{fig_ratio_spa}.
    }
    \label{fig_anisotropic_photon_rate}
\end{figure}

\begin{figure}[ht!]
	\centering
	\includegraphics[width=0.75\linewidth]{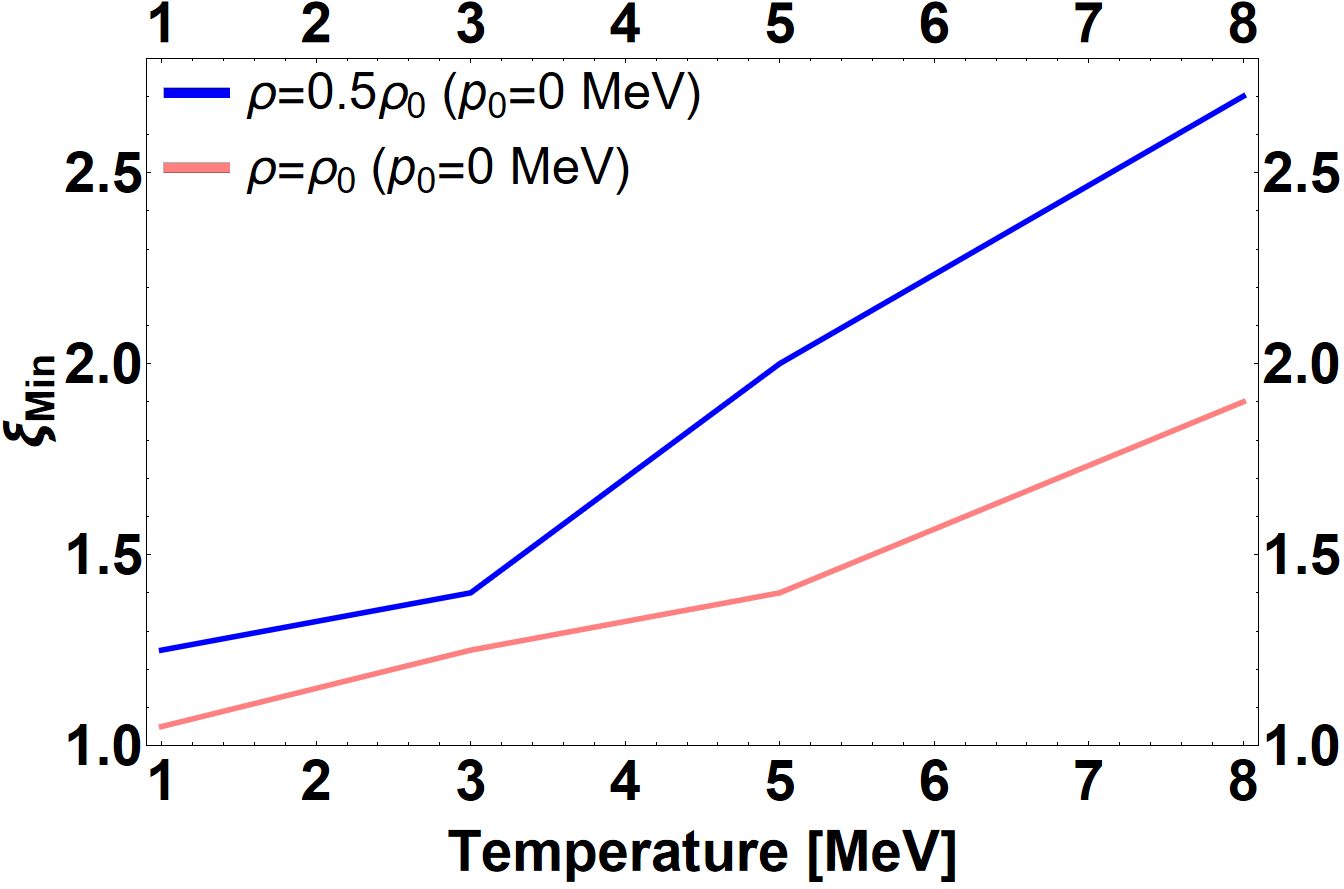}
	\caption{Temperature dependence of the thermal stretch parameter corresponding to the minimum photon rate under the conditions of half (pink) and full (blue) saturation density and temperatures $T=8$, $5$, $3$, $1$~MeV.}
    \label{fig_min}
\end{figure}

\begin{figure}[ht!]
	\centering
	\includegraphics[width=0.9\linewidth]{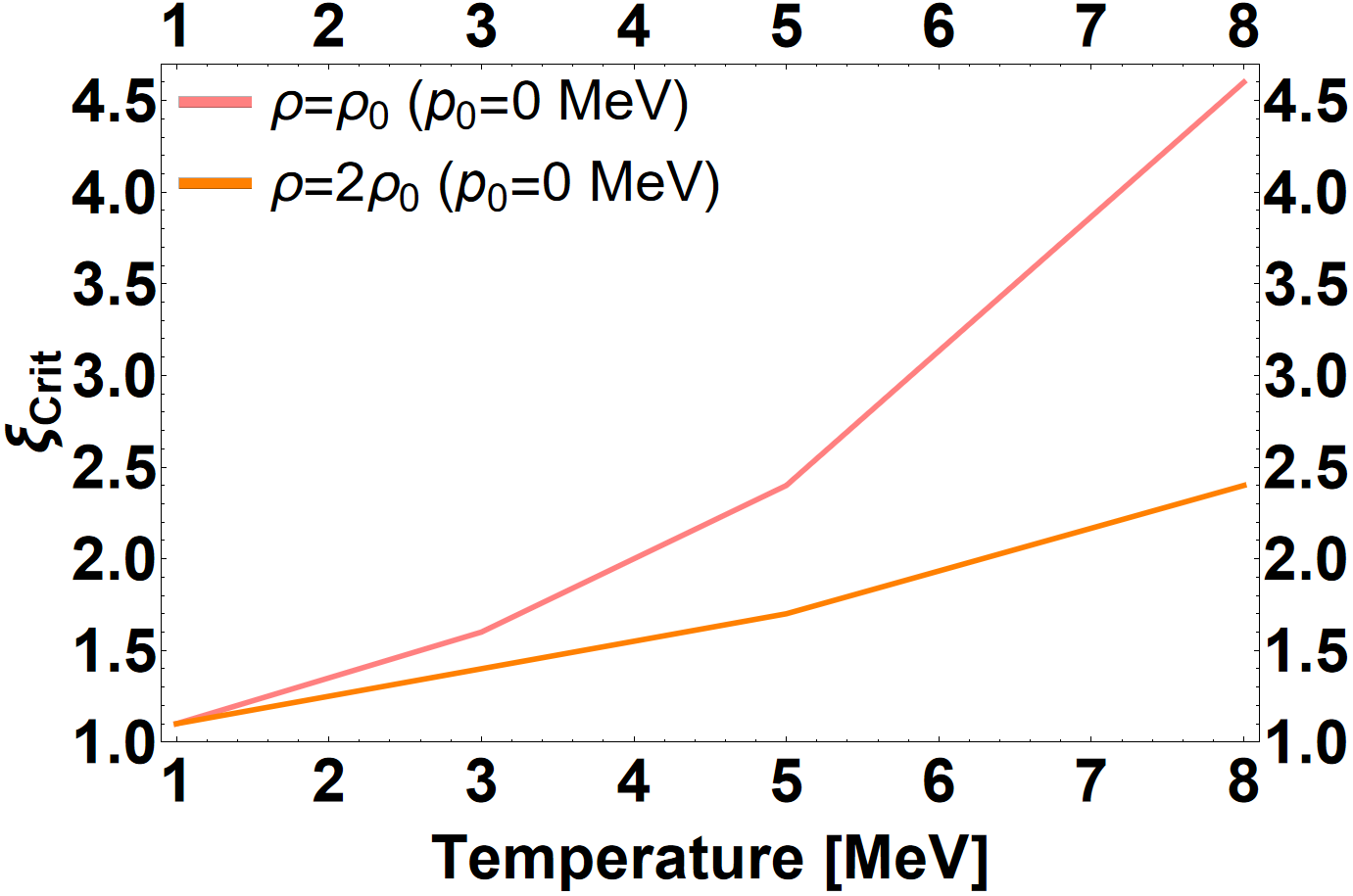}
	\caption{Temperature dependence of the thermal stretch parameter at which the photon rate is approximately equal to the photon rate without anisotropy ($\xi=1$) at full (blue line) and twice (orange line) saturation densities .}
    \label{fig_crit}
\end{figure}

\begin{figure}[thb!]
	\centering
	\includegraphics[width=\linewidth]{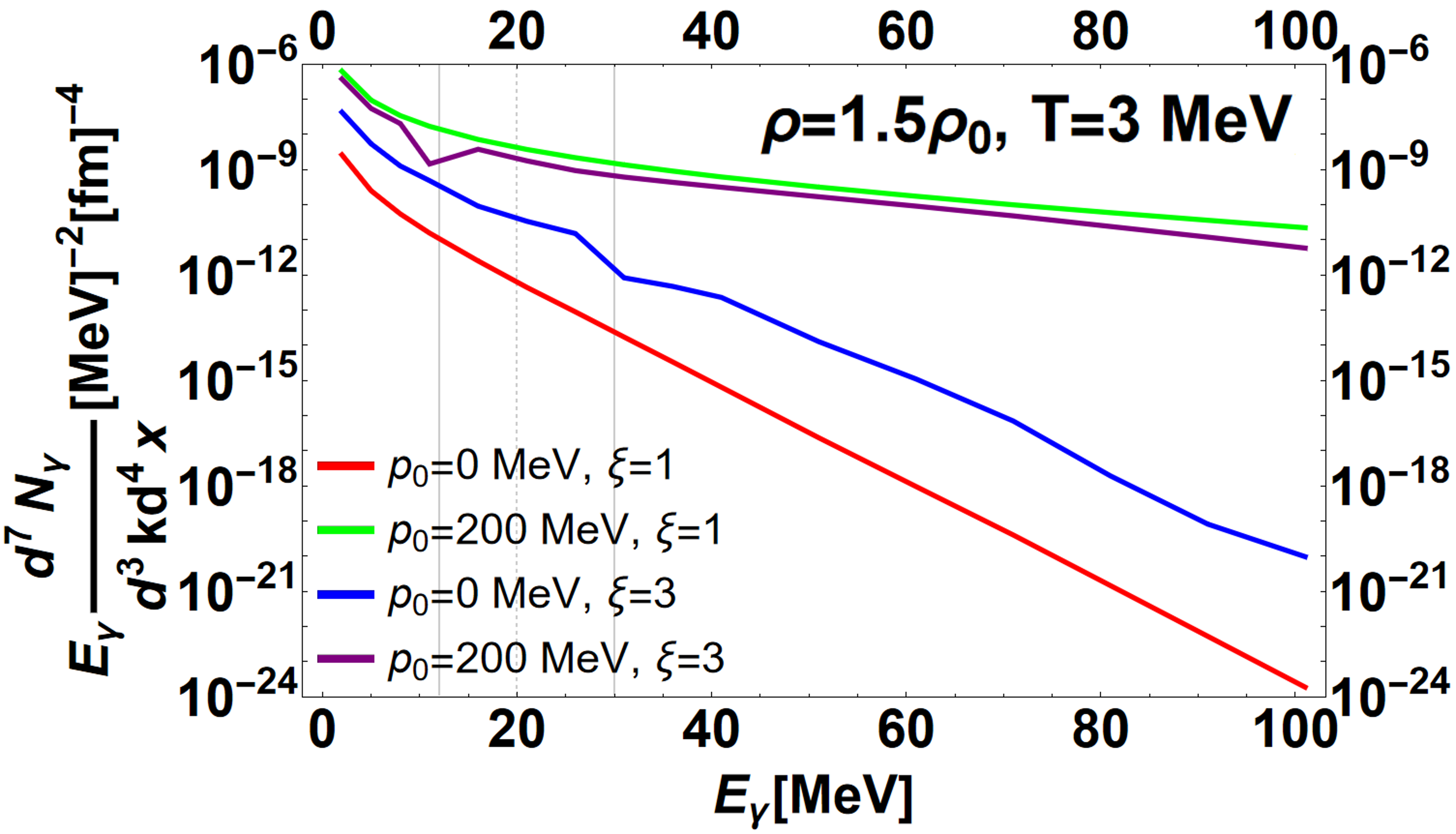}
	\caption{Photon emission rates at 1.5$\rho_0$ and $T=3$ MeV: without centroid momentum nor thermal anisotropy (red), with centroid momentum but without thermal anisotropy (green), without centroid momentum but with thermal anisotropy (blue), and with centroid momentum and thermal anisotropy (purple). 
    %The vertical lines are the same as in Fig.~\ref{fig_ratio_spa}.
    }
    \label{fig_compare_centroid_anisotropy}
\end{figure}

The thermal stretch parameters $\xi_{1,2}$ quantify the discrepancy between the temperatures extracted from the transverse and longitudinal directions. They also have an appreciable effect on the magnitude and slope of the photon rate, see Fig.~\ref{fig_anisotropic_photon_rate}.
Similar to $p_{01}$ and $p_{02}$, we approximated $\xi_{1,2}\to\xi\equiv(\xi_{1}+\xi_{2})/2$ for the analysis of the effects of thermal anisotropy in this work. For $\xi<1$, the photon rate increases due to the increase in the ``longitudinal" temperature ( $T_Z^{eff}=T/\xi$). For $1<\xi<\xi_{Min}$, the rate expectedly decreases, but increases again for $\xi_{Min}<\xi$ to approximately its value for equilibrium conditions at a quasi-equilibrium value $\xi=\xi_{QE}\gg\xi_{Min}$ and even more with further temperature anisotropy. This is in line with the argument made about vacant low-momentum states in regards to nucleon-nucleon collisions with high centroid momentum. At low photon energies, the effect is small. Meanwhile, at high energies, the photon rate changes significantly. The effect from varying the temperature anisotropy is more dramatic for situations of high density and low temperature than for low density and high temperature. We illustrate this by plotting the temperature dependence of $\xi_{Min}$ at half and full saturation density in Fig.~\ref{fig_min}. 
and $\xi_{QE}$ at full and twice saturation density in Fig.~\ref{fig_crit}.
The effect of large centroid motion opening low-momentum states that are normally blocked can be further enhanced when the stretch parameters are sufficiently high. Plotting the photon rate at the limits of the parameters extracted from Sec.~\ref{sec_collision_evo}, we see that it is more sensitive to centroid motion than temperature anisotropy, cf.~Fig.~\ref{fig_compare_centroid_anisotropy}. 
Similar to the case of isotropic temperatures, the photon rate has little sensitivity to temperature anisotropy when there is a significant amount of centroid motion in the system. We note that the thermal fit function, defined in Eq.~(\ref{eq_p0}), has the possibility of exceeding 1 under conditions when $\sqrt{\xi_{1(2)}}>1+\exp[\frac{p_\perp^2+\xi_{1(2)}(p_z-p_{01(02)})^2-2m\mu_N}{2mT}]$, which occurs during the initial stages of HICs when there is significant temperature anisotropy and high density. This poses an issue when calculating the Pauli blocking factors. Since negative blocking factors are unphysical, we manually exclude these configurations via theta functions as $\Theta(1-f_3)$ and $\Theta(1-f_4)$ on the blocking factors. 

%%%%%%%%%%%%%%%%%%%%%%%%%%%%%%%%%
\section{Experimental Constraints \label{sec_exp_constraints}}
%%%%%%%%%%%%%%%%%%%%%%%%%%%%%%%%%
\begin{figure*}[ht!]
	\centering
	\includegraphics[width=\linewidth]{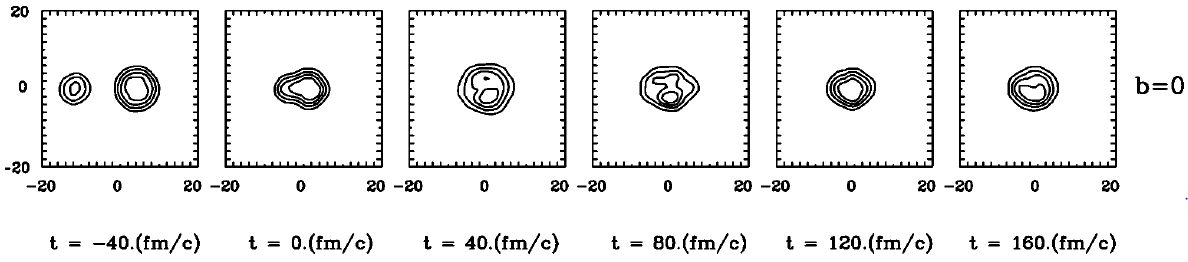}
	\caption{Contour plots of the spatial density as a function of time in the reaction plane for impact parameter $b$=0~fm from Ref.~\cite{Santonocito:2002hy}.}
    \label{fig_time_evo_central}
\end{figure*}
The collision system closest in energy and size to $\ce{^{40}\text{Ca}}+\ce{^{40}\text{Ca}}$ at  35A$\cdot$MeV (as considered in this work) for which photon spectra have been measured are $\ce{^{36}\text{Ar}}+\ce{^{98}\text{Mo}}$ collisions at 37A$\cdot$MeV~\cite{Piattelli:1996ppx,Santonocito:2002hy}. 
For a realistic comparison, we therefore need to account for an appropriate centrality selection and also implement the experimental angular acceptance cuts on the photons in the lab frame. The latter, in particular, requires boosting our emission spectra from the local restframe back into the laboratory system.  
In this section, we lay out how we rescale our calculations for central  $\ce{^{40}\text{Ca}}+\ce{^{40}\text{Ca}}$ collisions to results appropriate for centrality classes in the $\ce{^{36}\text{Ar}}+\ce{^{98}\text{Mo}}$ system (Sec.~\ref{subsec_scaling}), describe the procedure for the Lorentz boost from the local restframe of the individual coarse-graining cells to the lab frame (Sec.~\ref{subsec_boost}), which, in turn, enables us to apply the angular acceptance cuts in the lab frame reflective of the placement of detector units  (Sec.~\ref{subsec_detector}). 

%%%%%%%%%%%%%%%%%%%%%%%%%%%%%%%
\subsection{Collision Scaling}
\label{subsec_scaling}
\begin{figure}[ht!]
	\centering
	\includegraphics[width=0.75\linewidth]{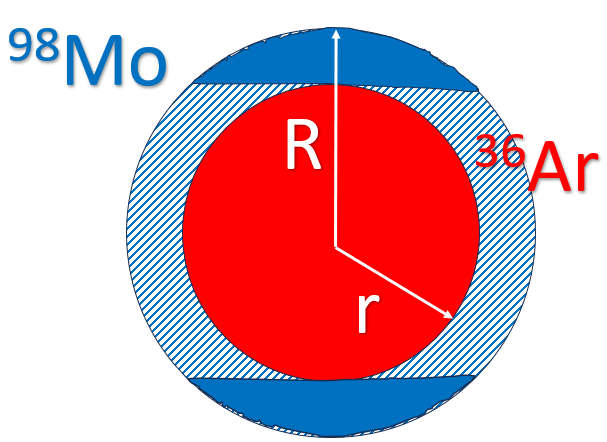}
	\caption{Diagram of $\ce{^{36}\text{Ar}}$ boring a tunnel through $\ce{^{98}\text{Mo}}$ during a central collision.}
    \label{fig_tunnel_schematic}
\end{figure}

%%%%%%%%%%%%%%%%%%%%%%%%%%%%%%%
In HICs with non-zero impact parameter, one usually distinguishes the spectator nucleons from the projectile and target, which are outside the reaction (or overlap) zone, and the participant nucleons in the reaction zone where the nuclei strongly interact. At beam energies around the Fermi energy, these regions are not as kinematically distinct as compared to high-energy HICs. Thus, spectator regions interact with the participant zone significantly. Our task here is to rescale our photon spectra from $\ce{^{40}\text{Ca}}+\ce{^{40}\text{Ca}}$ collisions, which in this work were simulated at zero impact parameter, to those measured in $\ce{^{36}\text{Ar}}+\ce{^{98}\text{Mo}}$ at 37A$\cdot$MeV~\cite{Piattelli:1996ppx,Santonocito:2002hy} which is of comparable size and only slightly higher in beam energy.
%and rescaled to show how impact parameter can affect photon production. 
However, the mass difference between the target and projectile nuclei is rather large. For central collisions, a sketch of the reaction's time evolution is shown in Fig.~\ref{fig_time_evo_central} taken from Ref.~\cite{Santonocito:2002hy}.
The argon nucleus (radius $r$) is expected to bore a hole through the molybdenum nucleus (radius $R$), as schematically depicted in Fig.~\ref{fig_tunnel_schematic}. 
To determine how many nucleons participate in the reaction zone, we calculate the volume of the ``tunnel". For the nuclear radii, we use $R_A = r_0A^{1/3}$ (with $r_0$= 1.14~fm) for nuclear mass number $A$, so that
 \begin{equation}
   \rho_0\frac{4}{3}\pi (r_0 A^{1/3})^3=A \ .
\label{eq_spherical_cap}
\end{equation}
With the spherical cap volume of
\begin{equation}
    V_{cap}=\frac{\pi h^2}{3}(3R-h),
\label{eq_spherical_cap_volume}
\end{equation}
($h=R-r$), the volume of the tunnel bored by the argon nucleus becomes
\begin{equation}
    V_{\rm tunnel}=V_{Mo}-2V_{\rm cap} \ .
\label{eq_tunnel}
\end{equation}
The total number of participants is the sum of the argon nucleus and the nucleons in the tunnel,
\bea
  A_{\rm part} &=& A_{\rm Ar}+A_{\rm tunnel} 
\nonumber \\
    &=& A_{\rm Ar}+\rho_0\bigg[\frac{4}{3}\pi R^3-2\frac{\pi(R-r)^2}{3}(2R+r)\bigg]
\nonumber\\
&=& 36+86.7
\nonumber\\
    &\approx& 123
\label{eq_central_number}
\eea
The maximum number of nucleons from the calcium nuclei captured in the coarse-graining grid is 77~\cite{Onyango:2021egp}.
Thus, the calculated results for central collisions of $\ce{^{40}\text{Ca}}+\ce{^{40}\text{Ca}}$ should be multiplied by a factor of 
$\big(\frac{123}{77}\big)^{4/3}$ to compare to central $\ce{^{36}\text{Ar}}+\ce{^{98}\text{Mo}}$ collisions. The exponent or 4/3 is based on the expectation that the dominant contribution arises from the early phase of the collisions, amounting to a $NN$ collision scaling (nucleons of the projectile typically interact with a row of target nucleons which leads to one extra dimension, \ie,  $A\cdot A^{1/3}=A^{4/3}$).

\begin{figure*}[ht!]
	\centering
	\includegraphics[width=\linewidth]{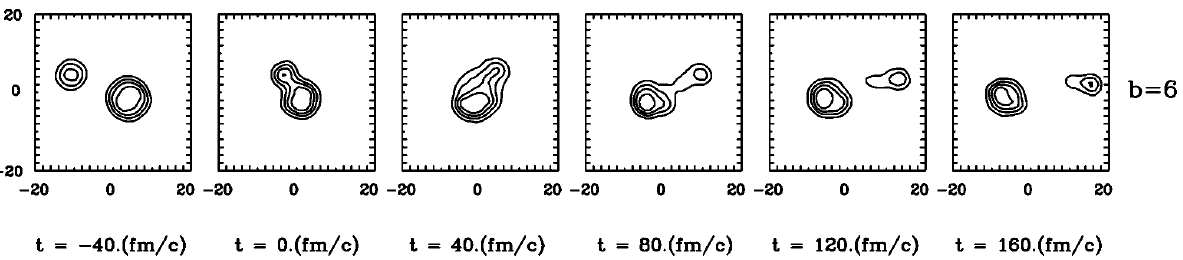}
	\caption{Contour plots of the spatial density as a function of time in the reaction plane for impact parameter $b$=6~fm, which is  $\approx0.63$b$_{max}$ from Ref.~\cite{Santonocito:2002hy}.}
    \label{fig_time_evo_peripheral}
\end{figure*}
\begin{figure}[ht!]
	\centering
	\includegraphics[width=0.75\linewidth]{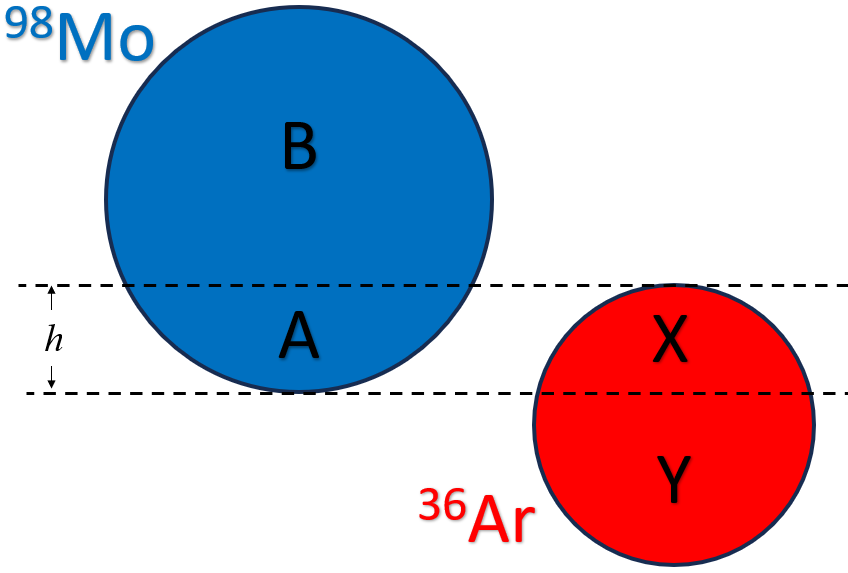}
	\caption{Illustration of the geometry of $\ce{^{36}\text{Ar}}$ and $\ce{^{98}\text{Mo}}$ prior to a peripheral collision.}
    \label{fig_transfer_schematic}
\end{figure}

At relatively large impact parameters $b$, with incomplete nuclear overlap, the main mechanism for reactions at intermediate beam energies is believed to be 
incomplete fusion~\cite{Nifenecker:1986gny}, especially for significantly asymmetric collision systems. This is a phenomenon where nucleons are 
transferred from the smaller nucleus to the larger one. An illustration of such a process at an impact parameter of 
$\approx\!0.63\,b_{\rm max}$ with $b_{\rm max}=1.2(A_{proj}^{1/3}+A_{targ}^{1/3})$ is shown in Fig.~\ref{fig_time_evo_peripheral}. 
Figure~\ref{fig_transfer_schematic} shows a schematic of the nuclei before colliding at a finite impact parameter. The radii of the argon and molybdenum nuclei are still $r=r_0(A_{\rm Ar})^{1/3}=3.8$~fm and $R=r_0(A_{\rm Mo})^{1/3}=5.3$~fm, respectively. 
%with $r_0=1.2$ fm.   why would we change that here compared to the above value of 1.14?
%
In Ref.~\cite{Piattelli:1996ppx}, the mass of the pre-equilibrium compound nucleus $A_{CN}$, which is the sum of the number $N_A$ of reactants from the projectile, number $N_X$ of reactants from the target, and the number $N_B$ of spectators from the target, is calculated to be 110 nucleons on average 
($N_Y$ is the number of spectators from the projectile), 
\begin{equation}
    A_{CN}=N_A+N_B+N_X
\label{eq_compound_nucleus}
\end{equation}
with
\bea
    N_Y&=&A_{\rm Ar}+A_{\rm Mo}-A_{CN}=24
\eea
and
\bea
    N_X&=&A_{\rm Ar}-N_Y=12 \ .
\eea
Since 1/3 of the nucleons of the argon nucleus are involved in the reaction, one has $V_X=V_{Ar}/3$, \ie,
\begin{equation}
    \frac{\pi h^2}{3}(3r-h)=\frac{1}{3}\frac{4\pi}{3}r^3
    \label{eq_peripheral_proj_volume}
\end{equation}
where $h$ is still the height of the spherical caps but is no longer equal to the difference in the radii of the nuclei. 
Solving Eq.~(\ref{eq_peripheral_proj_volume}) gives a height of 
\begin{equation}
    h\approx0.774r\approx3.07 \text{ fm} \ .
\end{equation}
The volume of the spherical cap of the molybdenum nucleus is
\begin{equation}
    V_A=\frac{\pi h^2}{3}(3R-h)=133.2 \text{ fm}^3 \ .
\end{equation} 
The number of nucleons from the molybdenum nucleus participating in the reaction becomes 
\begin{equation}
    N_A=\frac{V_A}{V_{Mo}}A_{Mo}\approx18 \ . 
\end{equation}
yielding the total number of nucleons participating in the reaction as $N_A+N_X=30$. Consequently, a spectrum from peripheral collisions should be multiplied by a factor of $\big(\frac{123}{30}\big)^{4/3}$ before being compared to the spectrum from central collisions. 
%Other experiments measuring the photon energy spectrum produced in HICs near the Fermi energy have been conducted and are mentioned in Sec.~\ref{sec_future_heavier_lower}.  

%%%%%%%%%%%%%%%%%%%%%%%%%%%%%%%%%%%%%
\subsection{Lorentz Boosts to the Laboratory Frame} 
\label{subsec_boost}
%%%%%%%%%%%%%%%%%%%%%%%%%%%%%%%%%%%%
\begin{figure}[th!]
	\centering
	\includegraphics[width=\linewidth]{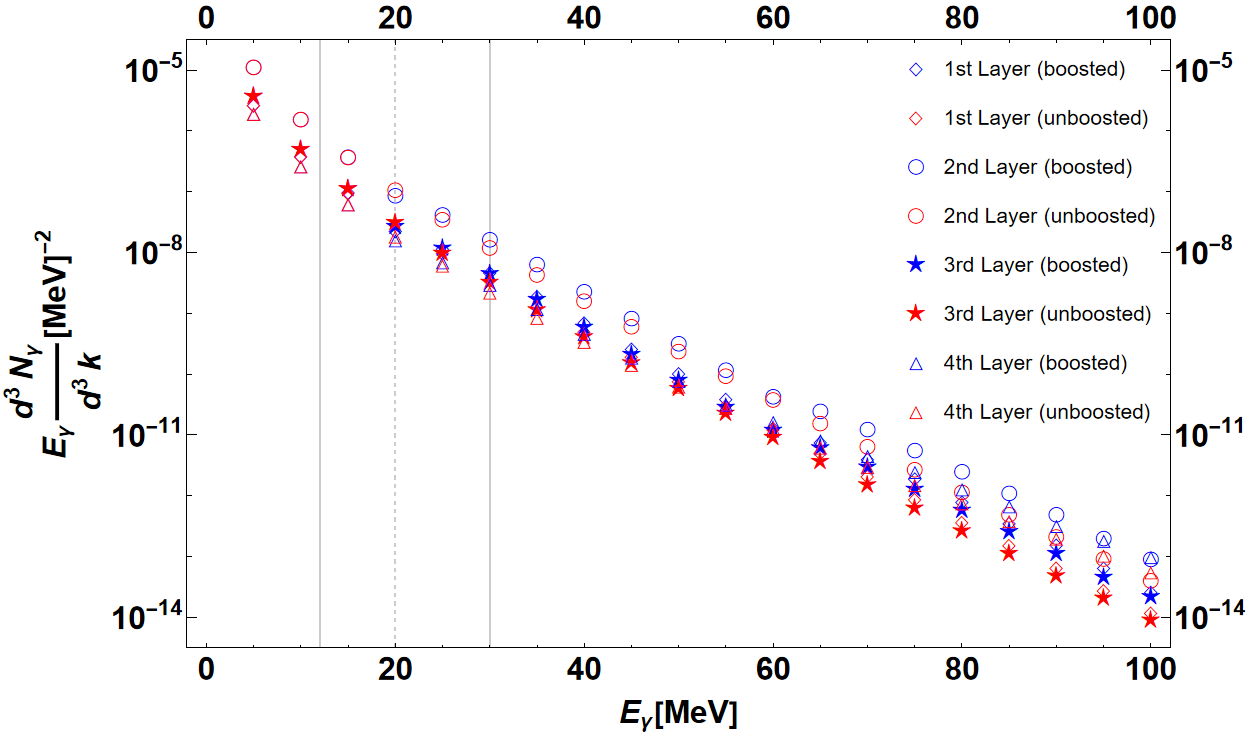}
	\caption{Comparison between the calculated momentum-differential spectrum of photons produced from the thermodynamic properties extracted from the transverse directions of central $\ce{^{40}\text{Ca}}+\ce{^{40}\text{Ca}}$ at 35\,A$\cdot$MeV with (blue symbols) and without (red symbols) the boost to the lab frame in the first (diamonds), second (circles), third (stars), and fourth layers (triangles) of the coarse graining grid. Centroid motion and thermal anisotropy are not included.
    %The vertical lines are the same as in Fig.~\ref{fig_ratio_spa}.
    }
	\label{fig_unboost_boost}
\end{figure}
The photon rates have been calculated in the thermal rest frame of each cell of the coarse-graining grid while the data were measured in the lab frame. 
The emission of the photons from the the space-time volume of the cells, therefore, needs to account for the collective velocity of each cell. This velocity receives two contributions: (i) the center-of-mass (cm) velocity of each cell, defined in the restframe of the collision system, which has been determined by the total momentum of each cell as $\vec v_{\rm cell}  = \vec P_{\rm cell} /M_{\rm cell}$, averaged over the total sample of simulated Ca-Ca collisions; (ii) the transformation from the cm system into the lab system which, for a fixed target, is moving relative to the former at the negative velocity of the projectile, \ie, at
  $v_{\rm lab} = \sqrt{2E_{\rm lab}/M}/2$ in negative $z$-direction (choosing the incoming beam direction in the positive $z$-direction). With $E_{\rm lab}$=35\,MeV, this amounts to $v_{\rm lab}$=0.1365$c$.  In the restframe of each cell, the emission is assumed to be isotropic and amounts to a differential momentum spectrum,
\begin{equation}
   E_\gamma\frac{d^3N_\gamma}{d^3k}=\int{\Delta V\Delta t\ E_\gamma\frac{d^7N_\gamma}{d^3kd^4x}} \ ,
   \label{eq_mom_spectrum}
\end{equation}
prior to the Lorentz boosts. Here, $E_\gamma\frac{d^3N_\gamma}{d^3k}$ has the physical meaning of representing how many photons have a momentum within the the phase space cell of $d^3k$ around $\vec k$. To determine how the isotropic angular distribution of emitted photons would look after a transformation, 20 photons in random directions at different energies were generated, and the boosts were applied relativistically to their vectors in a single step using the Galilean sum of the 2 velocities described above. 
% The photons are then rebinned according to their energies after the boosting process (and the spectrum is normalized by dividing by 20). \RR{Not sure what the following sentence means and why it would be needed; please clarify or drop} 
The photons calculated from the rate using the thermodynamic and collective properties extracted through the model in this study can be rebinned according to how the random photons were transformed. 
To compare photon production in different regions of the coarse-graining grid for different periods of time, it is necessary to use the photon momentum-differential spectrum because it has its space-time dependence integrated out. 
The results of this boosting process on a layer-by-layer comparison are shown in Fig.~\ref{fig_unboost_boost}. 
This boost had a rather modest but not insignificant effect on the spectra, slightly increasing the number of high-energy photons while decreasing the number of low-energy photons because the total number of photons is conserved.

%%%%%%%%%%%%%%%%%%%%%%%%%%%%%%%%%%%%%%%%
\subsection{Detector Placement and Acceptance}
 \label{subsec_detector}
%%%%%%%%%%%%%%%%%%%%%%%%%%%%%%%%%%%%%%%%

\begin{figure}[hbt!]
	\centering
	\includegraphics[width=\linewidth]{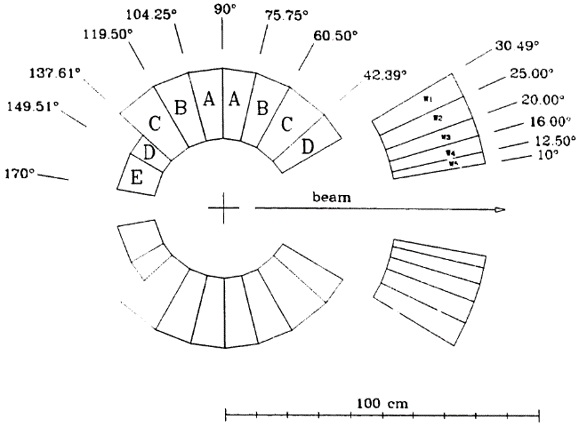}
	\caption{MEDEA detector arrangement from Ref.~\cite{MIGNECO199231}.}
	\label{fig_detector}
\end{figure}

The MEDEA detector at GANIL has been used to measure the energy spectra of emitted photons from multiple experiments \cite{Suomijarvi:1996zz,Piattelli:1999muy}, including the Ar+Mo system of interest here. Its experimental setup is shown in Fig.~\ref{fig_detector} and further detailed in Ref.~\cite{MIGNECO199231}.
The experimental data were recorded over the full azimuth and at a polar angle segment of 90\textdegree$\pm$14\textdegree \cite{Piattelli:1999muy}. 
The data were then renormalized to 4$\pi$ by dividing the data by the range of angular acceptance ($\approx$30\textdegree\ in the polar direction and 2$\pi$ in the azimuthal direction) and multiplying by 4$\pi$ \cite{Santonocito:2023jkl}. These cuts have been readily applied to our spectra in the lab system, referring back to the random generated photons as described in the preceding section. If a photon is found to be headed in a direction outside of the acceptance 
window of the MEDEA detector, that photon is deemed not detected. A momentum-differential spectrum composed of ``detected" is calculated. 
This spectrum is scaled by a factor of 
\begin{equation}
    \frac{4\pi}{2\pi(\cos{(76^{\circ}\cdot\frac{\pi}{180^{\circ}})}-\cos{(104^{\circ}\cdot\frac{\pi}{180^{\circ}})})}
\end{equation} 
to mimic how the experimental data were renormalized to represent a 4$\pi$ detector.

%%%%%%%%%%%%%%%%%%%%%%%%%%%%%%%
\section{Rate Implementation into Coarse-Grained Evolution}
\label{sec_mom_spectra}
%%%%%%%%%%%%%%%%%%%%%%%%%%%%%%%

\begin{figure}[htb!]
	\centering
	\includegraphics[width=\linewidth]{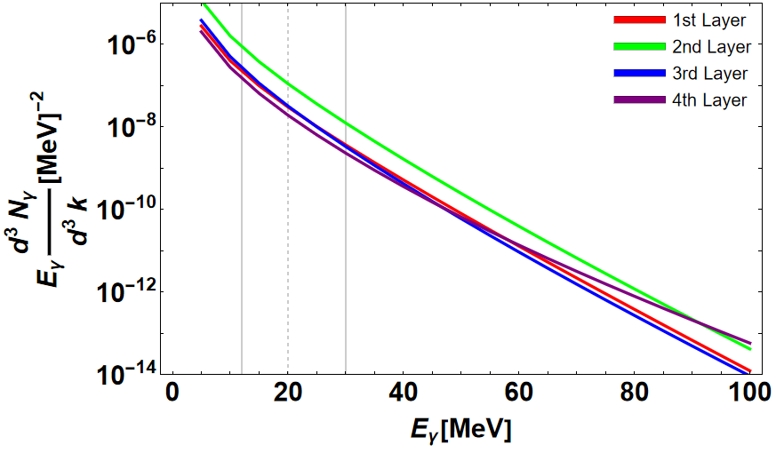}
	\caption{Photon momentum-differential spectra calculated using thermodynamic properties from the transverse direction for the first (red line), second (green line), third (blue line), and fourth (purple line) spatial layers of the coarse-graining grid. 
    %The vertical lines are the same as in Fig.~\ref{fig_ratio_spa}.
    }
    \label{fig_thermal_layer_decomp}
\end{figure}

After analyzing the properties of the emission rates in Sec.~\ref{sec_photon_rate} and describing the experimental conditions under which photons are measured in Sec.~\ref{sec_exp_constraints}, we now study the space-time profile of the photon emission spectra. We have recorded collision data for central Ca+Ca over the time duration of $40\lesssim t<550$~fm/$c$ and the different layers introduced in Sec.~\ref{sec_collision_evo}.
Employing our convention for setting the initial ``touching" time of the incoming nuclei at $\sim$40~fm/$c$, 
one can use classical kinematics to estimate the onset of emission from the different layers at approximately $t\sim40, 55, 70,$ and $85$~fm/$c$ in the first, second, third, and fourth layers, respectively. 
Based on these constraints, we define pertinent initial time intervals, followed by later intervals of typically 100~fm/$c$ in duration and a final one of 
150~fm/$c$. 
The time snapshots of the spectra can then be added up to study photon production layer by layer, which is displayed in 
Fig.~\ref{fig_thermal_layer_decomp} for the case of using isotropic thermal emission rates without centroid motion. 
In this scenario, the second layer produces the dominant contribution, which is mostly due to the relatively high temperatures in the late stages of the evolution.  

\begin{figure*}[bth!]
	\centering
	\includegraphics[width=\linewidth]{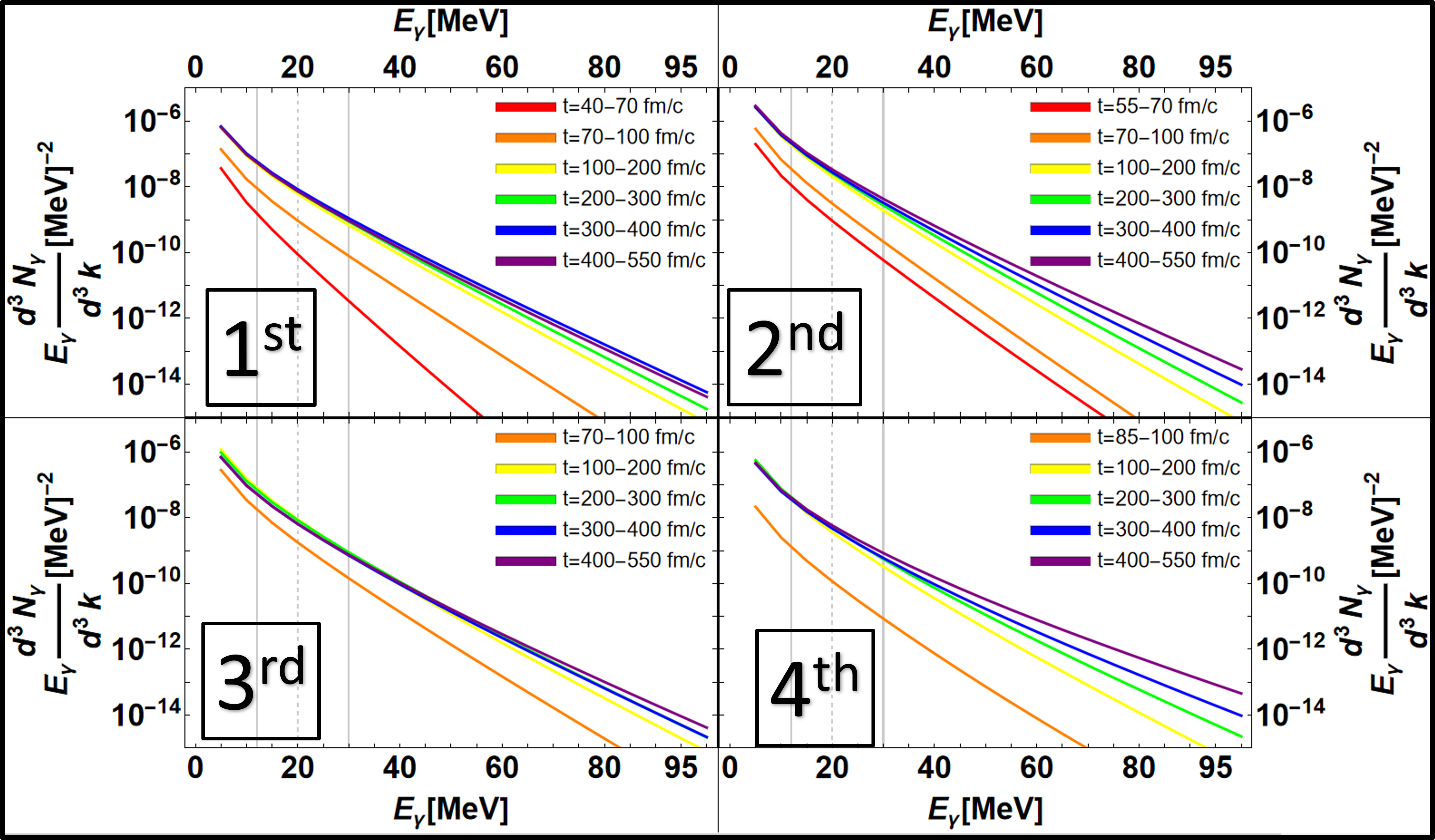}
	\caption{Photon momentum-differential spectra calculated using thermodynamic properties extracted from the transverse direction for different time slices for the first (top left panel), second (top right panel), third (bottom left panel), and fourth (bottom right panel) layers of the coarse-graining grid. The differences in the starting time for the first period of each panel are a consequence of initiating the calculation of photon production in any particular coarse-graining layer synchronously with the classical timing of when both nuclei moving with a velocity of $0.136c$ would overlap within that particular layer. 
    %The vertical lines are the same as in Fig.~\ref{fig_ratio_spa}.
    }
    \label{fig_thermal_time_window}
\end{figure*}

This can be seen more explicitly through a decomposition of each layer into 5-6 time windows as indicated above, which is shown in Fig.~\ref{fig_thermal_time_window}. 
Although high temperatures exist in the 3\textsuperscript{rd} and 4\textsuperscript{th} layers, they are too dilute to produce as many photons as the 1\textsuperscript{st} and 2\textsuperscript{nd} layers. Despite the photon rate's large sensitivity to higher temperatures leading to harder rates, density still plays a role, albeit a much smaller one than temperature. The monotonic increase in temperature for the duration of the collision causes late periods to produce the most thermal photons.  

\begin{figure}[htb!]
	\centering
	\includegraphics[width=0.95\linewidth]{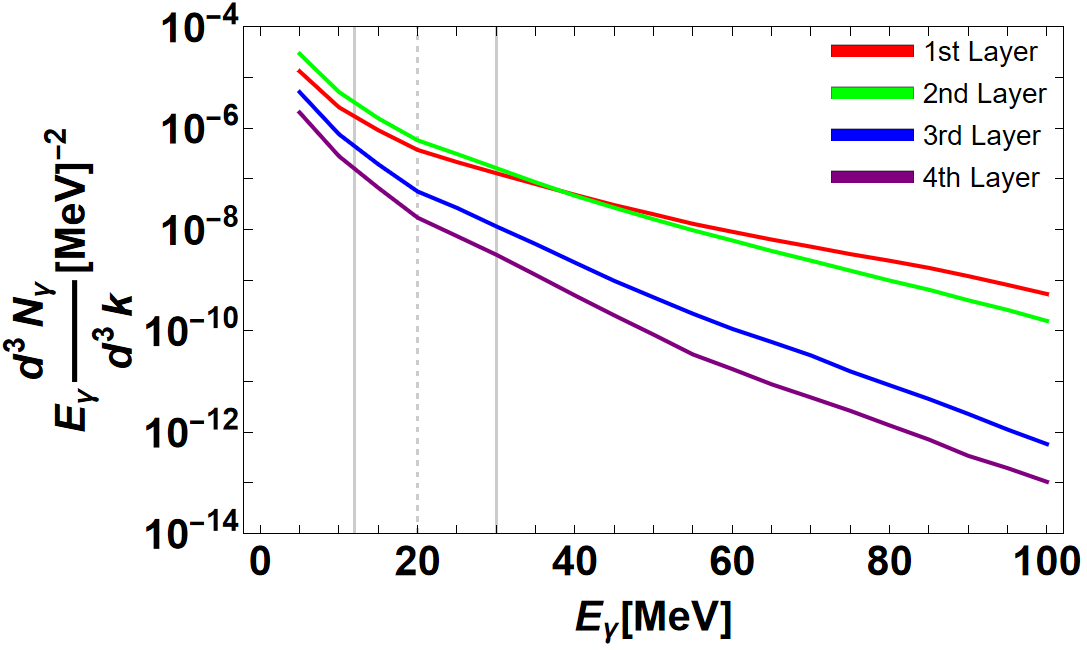}
	\caption{Comparison of photon momentum-differential spectra accounting for the presence of the centroid motion for the first (red), second (green), third (blue), and fourth (purple) layers of the coarse-graining grid. 
    %The vertical lines are the same as in Fig.~\ref{fig_ratio_spa}.
    }
    \label{fig_centroidal_layer_decomp}
\end{figure}

\begin{figure*}[thb!]
	\centering
	\includegraphics[width=\linewidth]{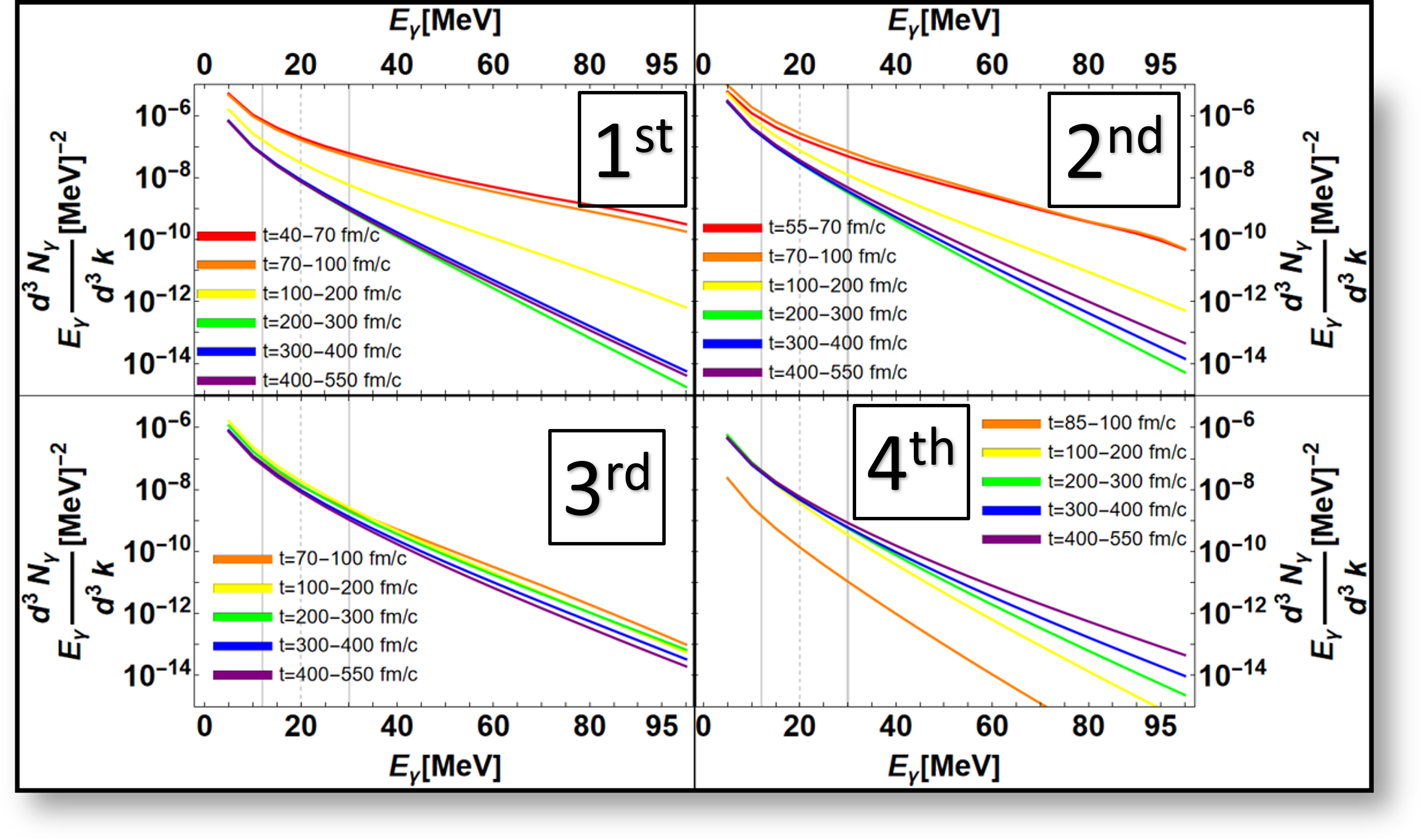}
	\caption{Photon momentum-differential spectra accounting for the presence of centroid motion for different time slices for the first (top left panel), second (top right panel), third (bottom left panel), and fourth (bottom right panel) layers of the coarse-graining grid. The differences in the starting time for the first period of each panel have the same cause as Fig.~\ref{fig_thermal_time_window}. 
    %The vertical lines are the same as in Fig.~\ref{fig_ratio_spa}.
    }
	\label{fig_centroidal_time_window}
\end{figure*}

\begin{figure}[t!]
	\centering
	\includegraphics[width=\linewidth]{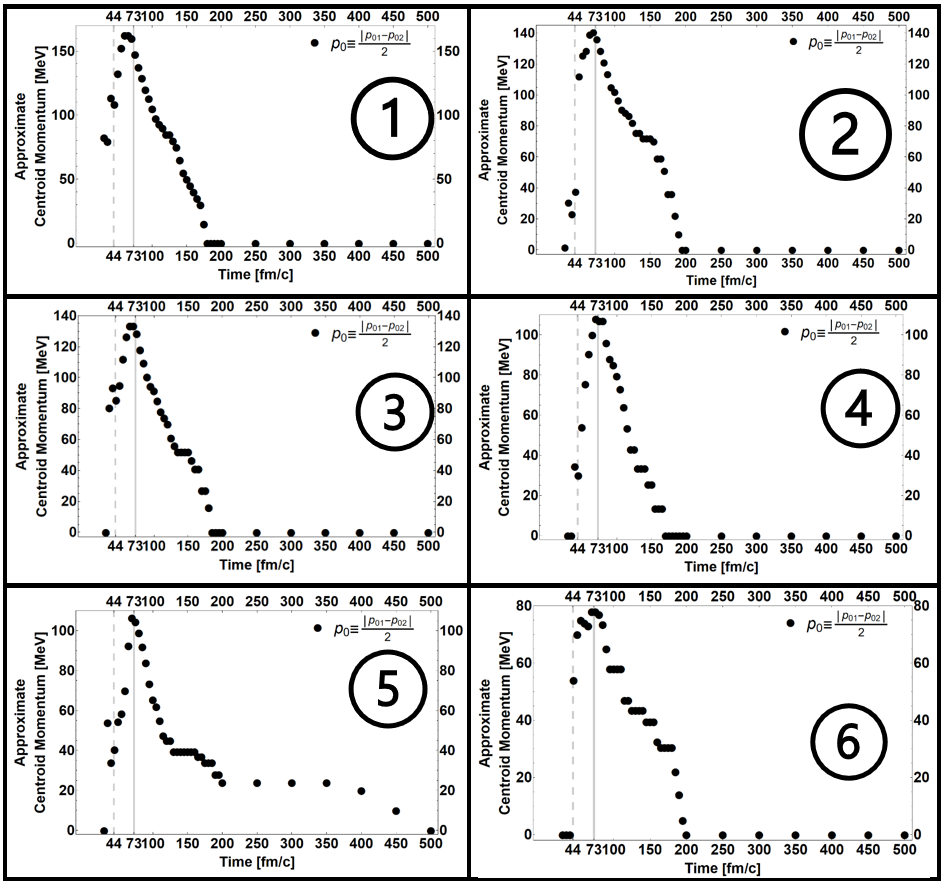}
	\caption{Time evolution of approximate centroid momentum distinguished by the symmetry groups corresponding to the first (group 1, recall Fig.~\ref{fig_spatial_config1}) and second layers (groups 2-6, recall Fig.~\ref{fig_spatial_config2}). 
    %The vertical dashed and solid lines are the same as in Fig.~\ref{fig_kinetics_central}.
    }
    \label{fig_effective_centroid_mom_evo}
\end{figure}

When accounting for centroid motion in the nucleon distribution functions, we find a dramatic increase in high-energy photon production, especially in the first two layers, see  Fig.~\ref{fig_centroidal_layer_decomp} (recall that the 3\textsuperscript{rd} and 4\textsuperscript{th} layers are empty upon initial nuclear impact).
The pertinent time decomposition confirms that this effect is almost entirely from the first two (relatively small) time windows, cf.~Fig.~\ref{fig_centroidal_time_window}.  
These two time windows, from the first two layers, largely dominate the total yields. Once the centroid motion has dissipated, the subsequent near-thermal emission sets in but turns out to be sub-dominant.
(We note that we approximated the centroid motion by a symmetric ansatz, $p_{01,02}\to\pm p_0$ where $p_0\equiv\frac{|p_{01}-p_{02}|}{2}$, just as we did in Sec.~\ref{sec_photon_rate}). 
%This approximation is moderately accurate during the compression phase of the collision at differences in momenta at ~and becomes more accurate throughout the expansion phase and later as shown in Fig.~\ref{fig_centroid_mom_approx_evo}. \RR{quantify or drop altogether}).
In the third layer, there is still some centroid motion, so the hierarchy of the time periods changes in a way similar to the first and second layers, but quantitatively, the enhancement over the thermal case is overall much less pronounced. The only cells in the fourth layer that have centroid momentum were the 8 cells in the centers of the faces along the beam axis. This means that the photon rate of the fourth layer was driven primarily by temperature, which is why it increases with time.
%\begin{figure}[t!]
%	\centering
%	\includegraphics[width=\linewidth]{graphic/Centroid_Momentum_Approximation(Ensemble).PNG}
%	\caption{Time evolution of the absolute value of the sum of the centroid momenta. 
%    %The vertical dashed and solid lines are the same as in Fig.~\ref{fig_kinetics_central}. \RR{I %don;t see any new information here and suggest to drop this figure}
%    }
%    \label{fig_centroid_mom_approx_evo}
%\end{figure}
To better illustrate the space-time structure of the centroid motion, the time evolution of the centroid momenta of the different symmetry groups of the 1\textsuperscript{st} and 2\textsuperscript{nd} layers is plotted in Fig.~\ref{fig_effective_centroid_mom_evo}. 
One notices a gradual decrease of the maximal centroid momentum as one moves from group 1 (first layer) to groups 2+3 (parallel attached to the first layer), 4+5 (single-diagonally attached to first layer), and group 6 (double-diagonally attached to the first layer).

\begin{figure}[thb!]
	\centering
	\includegraphics[width=\linewidth]{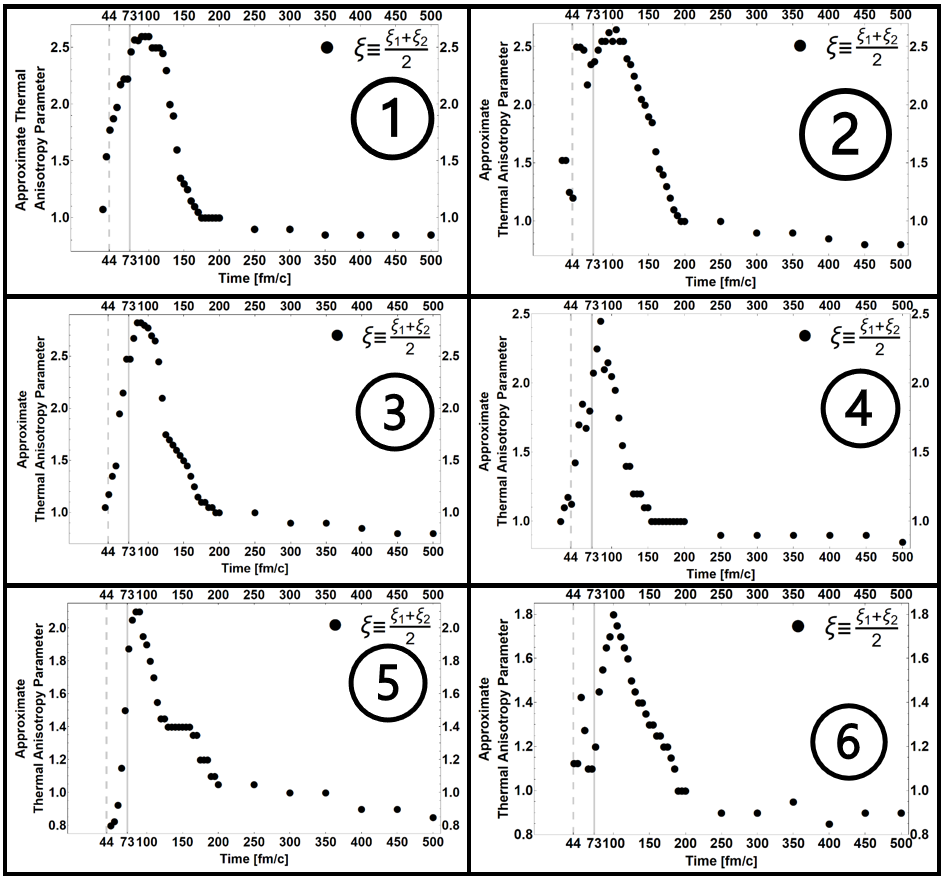}
	\caption{Time evolution of approximate thermal anisotropy parameter. 
    %The vertical dashed and solid lines are the same as in Fig.~\ref{fig_kinetics_central}.
    }
    \label{fig_effective_stretch_evo}
\end{figure}

\begin{figure}[thb!]
	\centering
	\includegraphics[width=\linewidth]{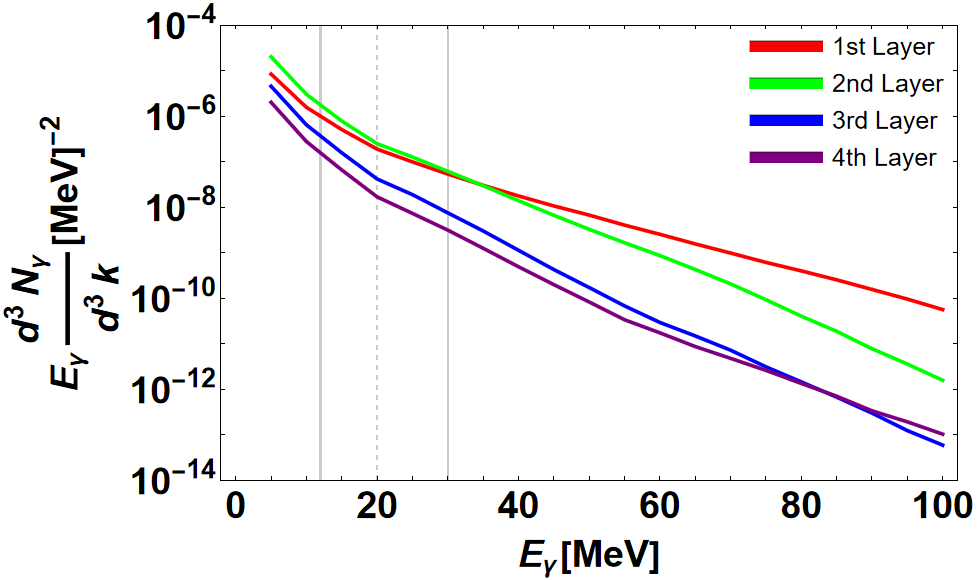}
	\caption{Comparison of photon momentum-differential spectra accounting for the presence of centroid motion and thermal anisotropy for the first (red), second (green), third (blue), and fourth (purple) layers of the coarse-graining grid. 
    %The vertical lines are the same as in Fig.~\ref{fig_ratio_spa}.
    }
    \label{fig_thermocentroidal_layer_decomp}
\end{figure}

\begin{figure*}[!t]
	\centering
	\includegraphics[width=\linewidth]{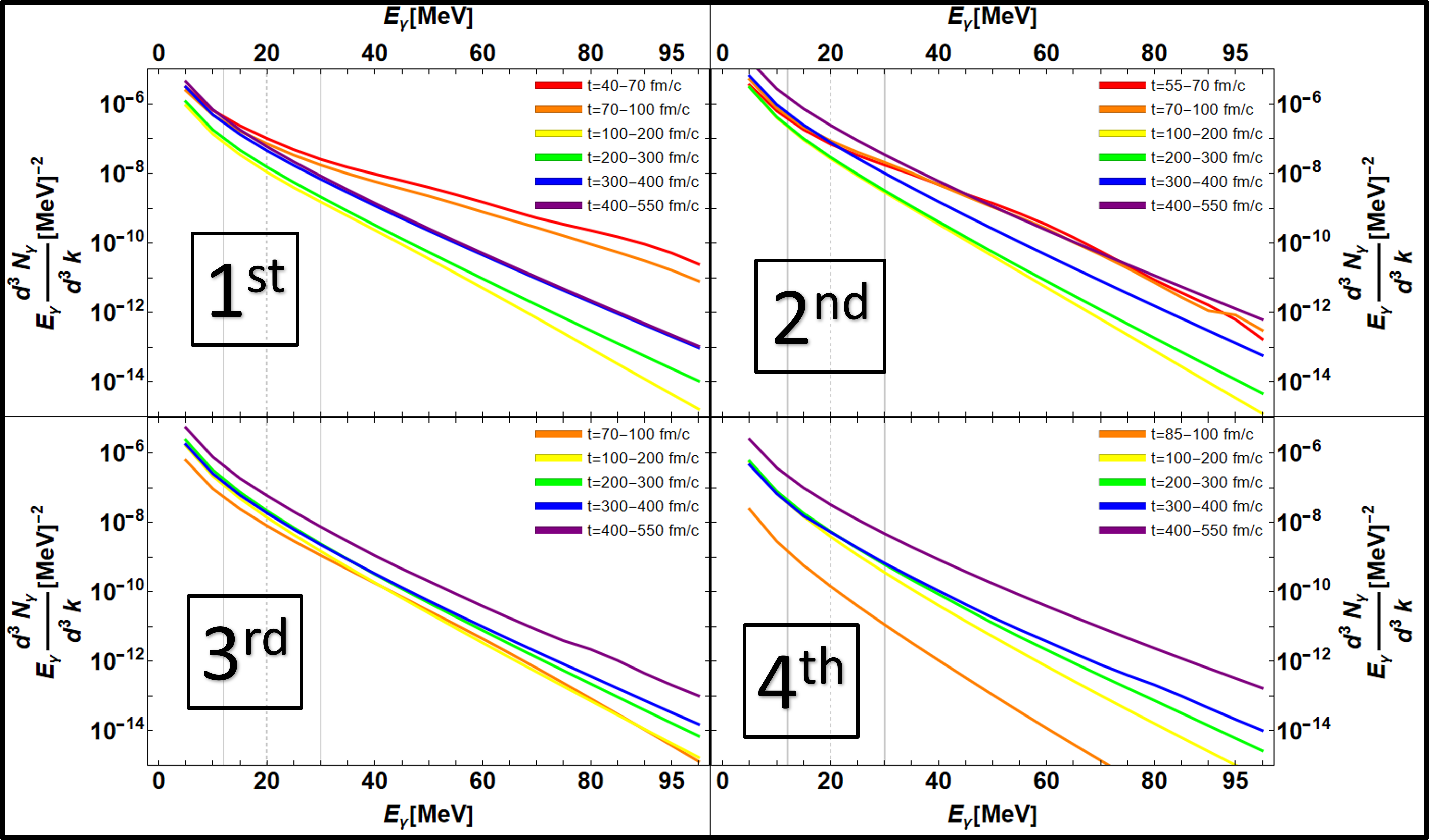}
	\caption{Photon momentum-differential spectrum calculated including thermal anisotropy, centroid motion, and thermodynamic properties extracted from the transverse direction for different time slices for the first (upper left panel), second (upper right panel), third (lower left panel), and fourth (lower right panel) layers of the coarse-graining grid. The differences in the starting time for the first period of each panel have the same cause as Fig.~\ref{fig_thermal_time_window}. %The vertical lines are the same as in Fig.~\ref{fig_ratio_spa}.
    }
    \label{fig_thermocentroidal_time_window}
\end{figure*}

We finally turn to the most realistic case with the full nucleon distribution functions which additionally include the reduction of the anisotropies in the ``longitudinal  temperatures" through the time-dependent stretch factors, $\xi_{1,2}$. Similar to the centroid momenta, $p_{01}$ and $p_{02}$, we approximate $\xi_{1,2}\to\xi\equiv(\xi_{1}+\xi_{2})/2$, whose time evolution is plotted in Fig.~\ref{fig_effective_stretch_evo}. 
%This approximation does not alter the accuracy of our framework for calculating photon emission.
%This approximation does not retain information on which nucleons have undergone more interpenetration by setting both stretch parameters to the same values. 
%This approximation is not very accurate during early times, especially for the cells in the centers of the longitudinal faces of the second layer (symmetry group \ding{193}), but after $t\sim100$ fm/$c$, the approximation becomes rather reliable, see Fig.~\ref{fig_stretch_approx_evo}. 
%\RR{does this, and if so how, impact the interpretation of the results below?}
%Thomas: come back to
%
%\begin{figure}[thb!]
%	\centering	\includegraphics[width=\linewidth]%{graphic/Stretch_Parameter_Approximation(Ensemble).PNG}
%	\caption{Time evolution of the absolute value of the difference between thermal anisotropy parameters. 
%    %The vertical dashed and solid lines are the same as in Fig.~\ref{fig_kinetics_central}.
%    }
%    \label{fig_stretch_approx_evo}
%\end{figure}
%
The temperature anisotropy does not change the hierarchy of the layers, as shown in Fig.~\ref{fig_thermocentroidal_layer_decomp}, but a significant reduction in the high-energy photon yields is found (directly reflecting the behavior of the pertinent rates discussed in Sec.~\ref{sec_photon_rate}). 
However, one can 
see in Fig.~\ref{fig_thermocentroidal_time_window} the marked effects on the hierarchy of the time periods that are induced: high-energy photon production 
increases during late time periods in the first and second layer due to an anisotropy parameter of $\xi<$1, which in essence means a larger 
``longitudinal temperature" than in the transverse direction (recall the discussion in Sec.~\ref{sec_photon_rate}). 
In the second layer, late-time emission 
is now comparable to the early periods (even exceeding it at low photon energies), when a lot of centroid motion is present. The third layer sees a rearrangement of the hierarchy of time periods more in line with the evolution of temperature. The fourth layer did not see any appreciable change 
except that the last period became slightly harder. Overall, these features indicate that after the depletion of centroid energy, thermal anisotropies that 
``heat" the longitudinal direction can occur and play a significant role in photon production. However, the overall dominant source at photon energies 
above $\sim$60\,MeV remains the first layer during its early phases. Yet, at lower energies, the contributions from the second layer become very comparable (or even larger), including the late-time sources.

%\RR{end of modications for now -------- 04/03/25}

%%%%%%%%%%%%%%%%%%%%%%%%%%%%%%%%%
\section{Comparison to Measured Photon Energy Spectrum}
\label{sec_energy_spectrum}
\begin{figure}[ht!]
	\centering
	\includegraphics[width=\linewidth]{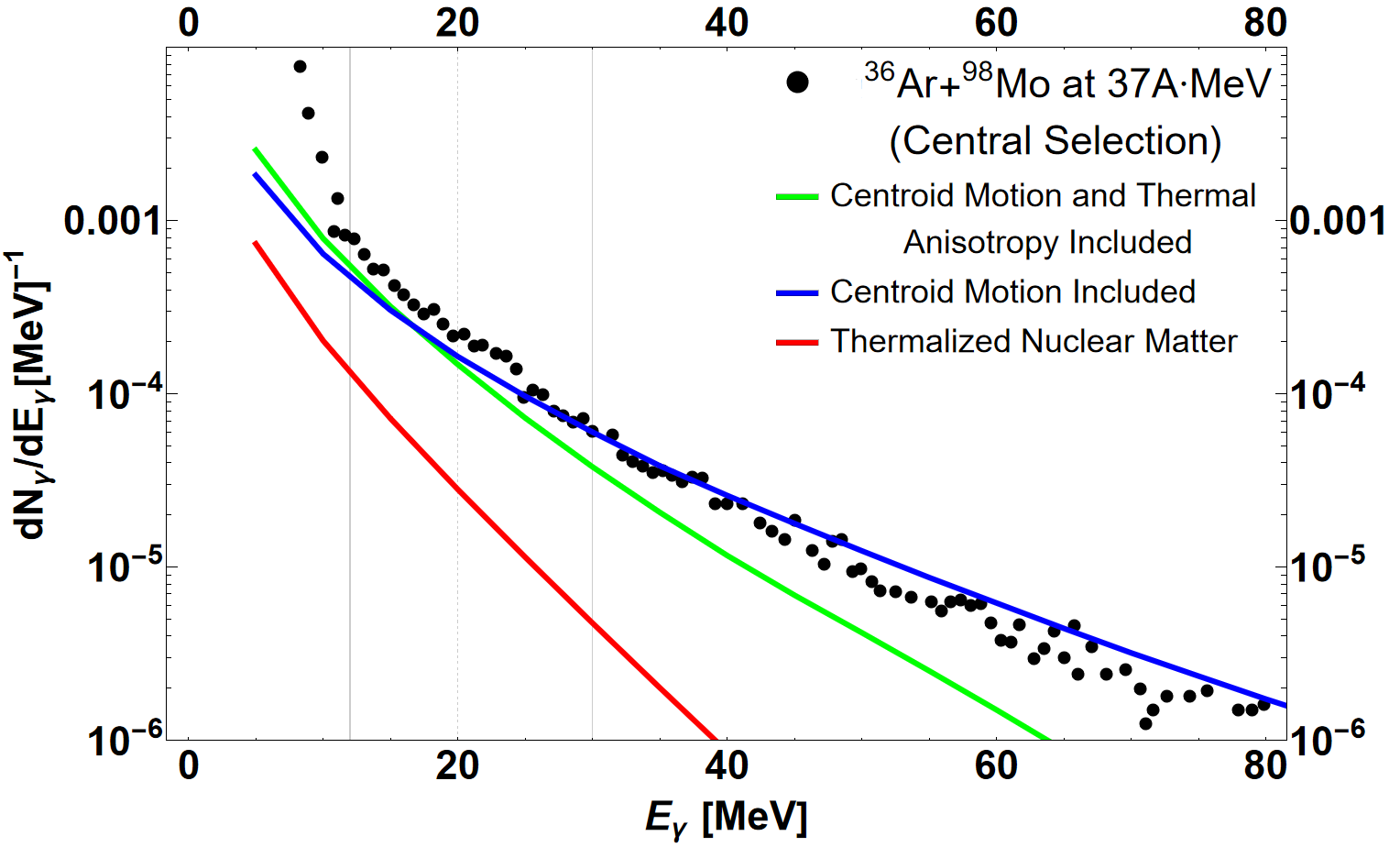}
	\caption{Comparison between the calculated energy spectrum of thermal photons produced from central $\ce{^{40}\text{Ca}}+\ce{^{40}\text{Ca}}$ at 35\,A$\cdot$MeV (red line), including centroid motion (blue line), and additionally including temperature anisotropy (green line), to the experimental data (black dots)~\cite{Piattelli:1996ppx,Santonocito:2002hy} measured in a central sample of $\ce{^{36}\text{Ar}}+\ce{^{98}\text{Mo}}$ at 37\,A$\cdot$MeV collisions. 
    The vertical solid lines denote the approximate bounds of the components of the photon energy spectrum. 
    %(see Fig.~\ref{fig_anatomy_of_photon_spectrum}). 
    }
	\label{fig_thermocentroidal_spectrum}
\end{figure}

%%%%%%%%%%%%%%%%%%%%%%%%%%%%%%%%%
The photon momentum-differential spectra were integrated according to 
\begin{equation}
    \frac{dN_\gamma}{dE_\gamma}=\int d\Omega_\gamma\bigg(E_\gamma\cdot E_\gamma\frac{d^3N_\gamma}{d^3k}\bigg)
    \label{eq_energy_spectrum}
\end{equation}
to obtain their corresponding energy spectra. The energy spectra were summed up over all time windows and layers to obtain the total radiation spectrum of the collision (including collision scaling and acceptance cuts). A comparison of the three different versions of our calculation (thermal, including centroid motion, and including thermal anisotropies) to experimental data is shown in Fig.~\ref{fig_thermocentroidal_spectrum}.
The energy spectrum calculated using the thermal parameters extracted from the transverse direction, represented by the red line in Fig.~\ref{fig_thermocentroidal_spectrum}, is well below the experimental data, but features a slope that is comparable to experimental data for photon energies around $E_\gamma\approx 15$~MeV. The lack of tapping into the energy stored in the initial collective motion of the nuclei is apparent. The calculation including this energy source through our implementation of centroid momenta in the nucleon distributions (blue line) shows a dramatic increase in the overall photon 
yield over the entire energy range, with the largest effect at high energies. The energy spectrum of this scenario (version) is generally harder than the slope of the
data, which eventually leads to a slight overestimate of the data for hard photons, $E_\gamma>50$~MeV. However, from our coarse-graining fits to the transport model results, we know that this version represents an overestimate of hard photons due to a ``longitudinal temperature" that overestimates the high-momentum tails of the longitudinal distributions. When accounting for this (green line), which represents our full result and best estimate within our current framework, the data for photon energies above $\sim$15\,MeV are slightly underestimated around 20~MeV, but more so, by up to a factor of $\sim$3-4 at energies above 
$\sim$40\,MeV. At the same time, the slope of this result is quite reminiscent of the data. We return to possible origins of this discrepancy below. 

%We also tested our scaling method by  applying it to an experimental data set gated on peripheral collisions using the method outlined in Sec.~\ref{subsec_scaling}), specifically by upslcaing the peripheral  for comparison with the data from central collisions to understand how photon emission changes with impact parameter. 
%Most of the hard photons were comparable, but for photons with energies $E_\gamma<40$ MeV, the spectrum from central collisions was much greater than the spectrum from peripheral collisions. This gap grew with decreasing photon energy. This may suggest that the 4/3-power collision-scaling is appropriate for hard photons which are produced during first-chance collisions and not appropriate for thermal photons. This is not a definitive conclusion, however. Thermal photons are proportional to the number of nucleons in the medium, suggesting that that region should be scaled by nucleon number to the first power. In addition to the number of $NN$ collisions occurring in the system, the number of thermal photons is proportional to the lifetime of the fireball \cite{Rapp:2014hha}. We posit that the number of collisions supplemented by the lifetime of the fireball validates using the scaling factor to study the photons of interest ($E_\gamma\gtrsim10$ MeV). There are still further considerations to the model detailed in this study that may provide the missing hard-photon yield.

%%%%%%%%%%%%%%%%%%%%%%%%%%%%%%%%%
\section{Comparison to Other Theoretical Calculations}
\label{sec_other_calculations}
%%%%%%%%%%%%%%%%%%%%%%%%%%%%%%%%%
The method for calculating photon emission from the medium produced in heavy-ion collisions in this work differs significantly from previous 
works~\cite{Kapusta:1977zb,Ko:1987zz,Guo:2021zcs} in a variety of ways. Photon production in low-energy HICs is coupled to nucleon-nucleon scattering, and, in those studies, it was calculated probabilistically from elastic scattering. The procedure used was to multiply the number of nucleon-nucleon 
collisions by the probability of producing a photon of a specific energy. Early works used a constant $NN$ cross section \cite{Ko:1987zz}, with a probability 
that was parameterized solely by the number of protons participating in the nuclear collision. In Refs.~\cite{Liu:2008as,Ma:2012zzb,Shi:2021far}, the cross 
section derived by Bauer et al. \cite{Bauer:1986zz} was used with a constant (``average") scattering radius as its only input parameter, which has limitations
in describing the energy dependence of the emission, which was, however, beyond the goal of early investigations to provide ballpark estimates.  
%Although the $NN$ cross section is known to increase with decreasing energy, quite rapidly for low kinetic energies \cite{RevModPhys.65.47,Stoks:1993tb}, the authors of Ref.~\cite{Bauer:1986zz} do state that it is not their goal to give a realistic and detailed microscopic formulation of free $NN$ scattering. 
Some previous theoretical calculations have accomplished this by fitting nucleon-nucleon dynamics to those photon spectra. This resulted in significantly larger cross sections as used in Refs.~\cite{Cassing:1986aya,Bauer:1986zz} and compared to the values for the $NN$ cross section used in this study. 
More microscopic methods for calculating the $NN\gamma$ cross section have applied quantum-field-theoretical approaches by utilizing
one-boson exchange interactions, specifically with Refs.~\cite{Gan:1994zz,Guo:2021zcs,Schaefer:1991rp,Cassing:1990dr}. 
Our remedy to the energy dependence of the cross section is implementing an empirical nucleon-nucleon (free) elastic cross section in a quantum-field-theoretical 
calculation of the photon production rate, following what has been done in calculations of axion emission from nucleon-nucleon 
Bremsstrahlung~\cite{Mahoney:2017jqk,Shin:2021bvz,Brinkmann:1988vi,Giannotti:2005tn}. 
However, our calculations are thus far limited to Bremsstrahlung from an external leg of the scattering nucleons, and a mechanism for photon emission from the exchange of internal bosons may prove crucial as the leading order term of those contributions is monopole; they are likely to produce more photons as the photon energy approaches the energies of the nucleons involved compared to calculations using the SPA shown by the completeness of the conservation of charge \cite{Low:1958sn,Nyman:1968jro,Nakayama:1989zza,Rrapaj:2015wgs}. This may, in fact, be the most promising avenue for improving the current discrepancy in our (thus far limited) comparisons to experimental data.

\begin{figure}[t!]
	\centering
	\includegraphics[width=\linewidth]{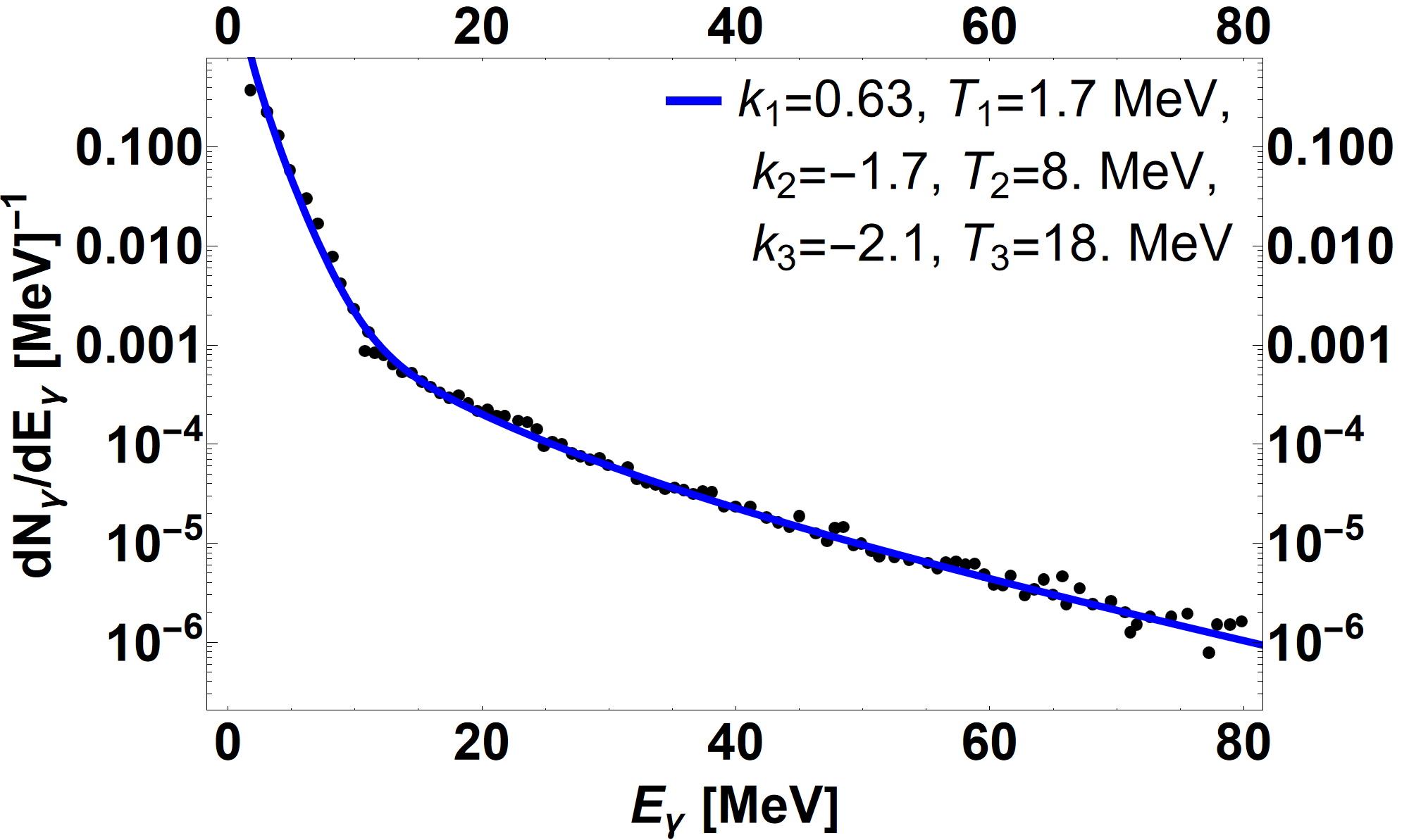}
	\caption{Schematic 3-component fit (blue lines) to experimental photon spectra (black dots)~\cite{Piattelli:1996ppx,Santonocito:2002hy} recorded for central collisions of $\ce{^{36}\text{Ar}}+\ce{^{98}\text{Mo}}$ at 37\,A$\cdot$MeV.}
	\label{fig_inverse_slope_spectrum}
\end{figure}

As a more qualitative exercise in thermometry, one may extract the inverse slopes of the photon spectra as a rough indicator of the emission sources or even temperature(s) reached in the nuclear fireballs. Clearly, the measured photon spectra discussed here are not based on a singular slope parameter, but by 
assuming that the components of the photon energy spectrum have an exponential dependence on temperature, one can still use a piece-wise fit ansatz to extract inverse-slope parameters. 
For the case at hand, a 3-component ansatz appears to be sufficient,  
$dN/dE = 10^{k_1}\cdot  exp(-E_\gamma/T_1)+10^{k_2}\cdot  exp(-E_\gamma/T_2)+10^{k_3}\cdot  exp(-E_\gamma/T_3)$, which indeed gives a  very good representation of the data, see Fig.~\ref{fig_inverse_slope_spectrum}.
The extracted ``temperatures" are about 2, 8 and 18\,MeV for the low- ($E_\gamma \lsim 15$~MeV), intermediate- ($15~{\rm MeV} \lsim E_\gamma \lsim 40$~MeV) and high-energy ($E_\gamma \gsim 40$~MeV) parts of the spectrum, respectively.
In light of our coarse-graining results (recall Sec.~\ref{sec_collision_evo}), the intermediate region approximately reflects the ``thermal radiation" part, leaving the high-energy part as the primordial component reflecting the initial hard production governed by centroid motion (with a scale of half the lab energy corresponding to the maximal available energy in the center of mass), and the low-energy part is commonly associated with final-state fragment decays. 
However, in reality, the situation is, of course, more complicated, as our simulations suggest that most of the spectrum above 15~MeV is still dominated by radiation from chiefly early, off-equilibrium collisions, while it is also known that there is an important contribution from giant-dipole-resonance decays in the intermediate region. The latter, as well as the final-state decays, is not part of our local radiation model and thus cause an expected discrepancy in comparison to our calculations.

%%%%%%%%%%%%%%%%%%%%%%%%%%%%%%%%%
\section{Conclusion}
\label{sec_concl}
%%%%%%%%%%%%%%%%%%%%%%%%%%%%%%%%%
In this paper, we have deployed a previously constructed coarse-graining method of transport simulations for Fermi-energy nuclear collisions to compute photon
emission spectra from the interaction zone formed in these reactions. This method, in particular, allows us to utilize field-theoretically calculated emission rates that are able to account for off-equilibrium effects in the early phases of the collision and thus provide a seamless treatment of the local system evolution from initial impact until kinetic freezeout (excluding final-state decays and long-range collective effects, such as giant resonance excitations). This approach puts the photon emission calculations on a rigorous quantum-mechanical footing and is amenable to systematic improvements in the future. In our current implementation, we have focused on Bremsstrahlung processes dominated by proton-neutron scatterings, employing a parametrized form of their total cross section. However, we did account for energy conservation including the emitted photon, thereby going beyond the soft-photon approximation, which turns out to be critical for quantitative applications in the Fermi energy regime where Pauli blocking plays an important role in limiting the final-state phase of the scattered nucleons. 

Using central $\ce{^{40}\text{Ca}}$ at 35A$\cdot$MeV collisions as a concrete example, we analyzed the space-time profile of the emitted 
radiation by dividing the collision zone in four spatial layers and 6 time windows. We have found that the two inner layers, essentially encompassing the initial 
nuclear size, generally provide the dominant contribution, but the time profile turned out to be more intricate. In a scenario where we only utilized the local 
thermodynamic parameters from the coarse-graining (\ie, neglecting any off-equilibrium effects), the final photon yields were dominated by late emission 
emanating from times of approximately 300-500~fm/$c$ after initial impact, driven by the highest temperature of circa 7\,MeV realized in the system. This dramatically changed when accounting for the initial collective motion of the incoming nuclei, where the now much higher available energies in individual 
$NN$ collisions cause an enhancement of the early emission from times of up to 100~fm/$c$ over the late emission rise by 2-4 orders of magnitude (increasing from low to high photon energies). An important mechanism in facilitating this enhancement is that the two-centroidal distribution in momentum space in the cells containing primordial collisions opens a significant phase space window around zero momentum in the cells' restframe, thus enabling the colliding nucleons to stop, 
thereby converting all their excess energy into photon emission. The overarching conclusion then is that the primordial collisions are the most important 
radiation source of hard photons throughout the fireball evolution.  

Finally, we applied our results to a comparison with experimental data available from the $\ce{^{36}\text{Ar}}+\ce{^{98}\text{Mo}}$ at 37A$\cdot$MeV system, which is reasonably close in energy and size to central $\ce{^{40}\text{Ca}}$ at 35A$\cdot$MeV. In carrying out this comparison, we estimated an appropriate 
scaling to match the centrality class of the asymmetric system and also made sure to transform our spectra back into the lab system where we then implemented the experimental acceptance cuts on the photon angles as given by the MEDEA detector. The experimental data are greatly underestimated by the ``thermal scenario", but upon including the centroid motion, the calculations are close to the experimental result. However, our most realistic scenario, which also includes a thermal stretch factor from the coarse graining necessary to accurately describe the somewhat reduced high-momentum tails in the $z$-direction compared to the transverse distributions, ultimately leads to an underestimate of the data by about a factor of $\sim$3-4 for large photon energies.
At intermediate photon energies, the discrepancy is smaller, and the slope of the calculated spectra is rather similar to the one observed in the data, with an 
inverse-slope parameter that is reminiscent of the highest temperatures extracted from the coarse-graining analysis of the system.

Future calculations will first and foremost have to address the current lack in emission yields compared to experiment, for which several possibilities are conceivable, such as additional contributions to the rate or a more accurate matching of the centrality selection. Depending on the outcome of these studies, further questions that could be addressed are the missing contributions of giant resonance decays in the 15-20\,MeV regime which in turn could help to increase the sensitivity to thermal emission, and/or isotopic dependencies as part of a beam energy and system scan, where the predominance of emission sources from $pn$ over $nn$ and $pp$ scattering provides a useful lever arm.
Work in some of these directions is already in progress.

\acknowledgments
We thank P.~Piattelli, M.~Colonna, and D.~Santonocito for their feedback on the acceptance of the MEDEA detector.
This work was supported by the Department of Energy and National Nuclear Security Administration under grant no. DE-NA 0003841 (C.E.N.T.A.U.R.), by the US National Science Foundation (NSF) under grant PHY-2209335 (RR), and was prepared by LLNL under Contract DE-AC52-07NA27344 (TO).

\bibliography{references}

%%%%%%%%%%%%%%%%%%%%%%%%%%%%%%%%%
\appendix
\section*{Appendices}
%%%%%%%%%%%%%%%%%%%%%%%%%%%%%%%%%
\section{Derivation of Nucleon-Nucleon Bremsstrahlung Rate}
\label{app_photon_rate}

The photon production rate can be derived from kinetic theory starting with the photon emissivity $\Dot{\epsilon}$, which is the emission rate of energy $E_\gamma$ by photons per unit four-volume $(\Dot{\epsilon}=E_\gamma\frac{d^4N_\gamma}{d^4x})$,
\bea
 \Dot{\epsilon}=\int{\frac{d^3k}{2 E_\gamma(2\pi)^3} E_\gamma}\int \frac{d^3p_1}{(2\pi)^3}\frac{d^3p_2}{(2\pi)^3}\frac{1}{2E_12E_2}f_1f_2
\nonumber\\
 \int \frac{d^3p_3}{(2\pi)^3}\frac{d^3p_4}{(2\pi)^3}\frac{1}{2E_32E_4}
  (1-f_3)(1-f_4)(2\pi)^4
\nonumber\\
  \times\delta^4(P_1+P_2-K-P_3-P_4)|\mathcal{M}|_{NN\gamma}^2.
\label{eq_emissivity}   
\eea
Equation (\ref{eq_emissivity}) has a similar functional form to the equations presented in Ref.~\cite{Weldon:1983jn}. In the present context, $|\mathcal{M}|_{NN\gamma}^2$ is the matrix element related to the emission of photons from nucleon-nucleon bremsstrahlung. In an expansion along the energy of the emitted photon $E_\gamma$, the matrix element for photon emission from external nucleon lines can be coupled to the matrix element for elastic nucleon-nucleon scattering $|\mathcal{M}|_{NN}^2$ through the relation \cite{Rrapaj:2015wgs,Chang:2016ntp,Nyman:1968jro,Bjorken:1965sts,Low:1958sn}
\begin{equation}
    \begin{split}
        |\mathcal{M}|_{NN\gamma}^2=4\pi\alpha(\epsilon^\mu \Tilde{J}_\mu)^2|\mathcal{M}|_{NN}^2,
    \end{split}
    \label{eq_photon_coupling}
\end{equation} 
where $\epsilon^\mu$ is the photon polarization vector and $\Tilde{J}_\mu$ is the appropriate nuclear electromagnetic current for the corresponding multipole contribution. The matrix element for elastic nucleon-nucleon scattering is connected to its differential cross section by the relation 
\begin{equation}
    \begin{split}
        \frac{d\sigma_{NN}}{d\Omega_{CM}}=\frac{|\mathcal{M}|_{NN}^{\ 2}}{64\pi^2 E_{CM}^{\ 2}}.
    \end{split}
    \
    \label{eq_cs_matrix_A}
\end{equation}
This relation can be derived in a similar manner to the steps presented in Section 4.5 of Ref.~\cite{Peskin:1995ev}. The cross section is defined in the nucleon-nucleon center-of-mass frame, so a transformation of the integration variables of Eq.~(\ref{eq_emissivity}) is done before substituting in the cross section for the matrix element. The specific transformation is that the incoming relative and total momenta are defined as $\overrightarrow{p}\equiv \frac{1}{2}(\overrightarrow{p_1}-\overrightarrow{p_2})$ and $\overrightarrow{q}\equiv \frac{1}{2}(\overrightarrow{p_1}+\overrightarrow{p_2})$, respectively. This transformation has a Jacobian of 1/8, resulting in an overall factor of 8 for the incoming phase space. The outgoing momenta have similar relations. After substituting in Eq.~(\ref{eq_photon_coupling}) and transforming the momentum variables, the photon rate evaluates to
\begin{equation}
\begin{split}
  E_\gamma\frac{d^7N_\gamma}{d^3kd^4x}=\frac{1}{4096\pi^{11}}8\int d^3pd^3q\frac{1}{2E_12E_2}f_1f_2 8\\
  \int d^3p'd^3q'\frac{1}{2E_32E_4}(1-f_3)(1-f_4)\\
  \times\delta^3\bigg(2\Vec{q}-2\Vec{q'}\bigg)\delta\bigg(\frac{p^2+q^2}{M}-E_\gamma-\frac{p'^2+q'^2}{M}\bigg)\\
    \times4\pi\alpha(\epsilon\cdot \Tilde{J})^2|\mathcal{M}|_{NN}^2
\end{split}
\label{eq_photon_rate_4-momentum}
\end{equation}
where $M$ is the nucleon mass. In the nonrelativistic approximation, we can say $E_3\approx E_4\approx M$. Photons produced from proton-neutron collisions are coupled to the nuclear electromagnetic current at the dipole level, with current
\begin{equation}
\begin{split}
  J_\mu^{(2)}=\bigg(\frac{P_1}{P_1\cdot K}-\frac{P_3}{P_3\cdot K}\bigg)_\mu=J_\mu,
\end{split}
\end{equation}
while the dominant contributions of the other nucleon-nucleon configurations are at the quadrupole level, with current
\begin{equation}
\begin{split}
  J_\mu^{(4)}=\bigg(\frac{P_1}{P_1\cdot K}+\frac{P_2}{P_2\cdot K}-\frac{P_3}{P_3\cdot K}-\frac{P_4}{P_4\cdot K}\bigg)_\mu=L_\mu.
\end{split}
\end{equation}
The evaluation of the electromagnetic factor $(\epsilon\cdot\Tilde{J})^2$ for the dipole level contribution is carried out in App.~\ref{app_em_factor}. A similar simplification is presented in Refs.~\cite{Rrapaj:2015wgs,Chang:2016ntp} with intermediate steps performed in Ref.~\cite{Haglin:1992fy} and Chapter 7 of Ref.~\cite{Bjorken:1965sts}. The result of the electromagnetic factor averaged over the photon angle is 
\begin{equation}
\begin{split}
    \int\frac{d\Omega_{\gamma}}{4\pi} (\epsilon\cdot J)^2=2\frac{T_{CM}}{M E_\gamma^{\ 2}}\frac{2}{3}(1-\cos{\theta_{CM}}).
\end{split}
\label{eq_real_electromagnetic_factor}
\end{equation}
This result agrees with the electromagnetic factors written in Refs.~\cite{Rrapaj:2015wgs,Chang:2016ntp} for the case of standard model photons. The electromagnetic factor used in those references is 
\begin{equation}
\begin{split}
    \frac{E_\gamma^{\ 2}}{4\pi}\int d\Omega_{\gamma}(\epsilon\cdot J)^2=2\frac{T_{CM}}{M}\bigg(1-\frac{k^2}{3\omega^2}\bigg)(1-\cos{\theta_{CM}}).
    \label{eq_electromagnetic factor}
\end{split}
\end{equation}
Returning to the calculation of the photon rate from Eq.~(\ref{eq_photon_rate_4-momentum}) after substituting in the electromagnetic factor, Eq.~(\ref{eq_real_electromagnetic_factor}), and the cross section, Eq.~(\ref{eq_cs_matrix_A}), the photon rate now has the form
\begin{equation}
\begin{split}
  E_\gamma\frac{d^7N_\gamma}{d^3kd^4x}=\frac{\alpha}{64\pi^{10}M^2}\int d^3pd^3q\frac{1}{2E_12E_2}f_1f_2 \\
  \int d^3p'd^3q'\delta^3\bigg(2\Vec{q}-2\Vec{q'}\bigg)   \delta\bigg(\frac{p^2+q^2}{M}-E_\gamma-\frac{p'^{\ 2}+q'^{\ 2}}{M}\bigg)\\
  \times(1-f_3)(1-f_4)\\
  \times\bigg[2\frac{T_{CM}}{M E_\gamma^{\ 2}}\frac{2}{3}(1-\cos\theta_{CM})\bigg]\bigg[64\pi^2E_{CM}^2\frac{d\sigma_{np}}{d\Omega_{CM}}\bigg]
\end{split}
\end{equation}
Using the properties of the Dirac delta function causes the photon rate to become
\begin{equation}
\begin{split}
  E_\gamma\frac{d^7N_\gamma}{d^3kd^4x}=\frac{\alpha}{64\pi^{10}M^2}\int d^3pd^3q\frac{1}{2E_12E_2}f_1f_2 \\
  \int d^3p'd^3q'\frac{1}{8}\delta^3\bigg(\Vec{q}-\Vec{q'}\bigg)
  M\\
  \delta(p^2+q^2-M E_\gamma-p'^2-q'^2)(1-f_3)(1-f_4)\\
   \bigg[2\frac{T_{CM}}{M E_\gamma^2}\frac{2}{3}(1-\cos\theta_{CM})\bigg]\bigg[64\pi^2E_{CM}^2\frac{d\sigma_{np}}{d\Omega_{CM}}\bigg].
\end{split}
\end{equation}
The energies of the incoming nucleons are approximated by $E_1\approx E_2\approx E_{CM}/2$. There is a distinction between the dynamics occurring in the transverse and longitudinal directions, so the rest of the derivation will be carried out in cylindrical coordinates. This causes the following relation for the scattering angle,
\begin{equation}
    \begin{split}
        \cos\theta_{CM}=\frac{p_\perp p'_\perp\cos(\phi-\phi')+p_z p'_z}{\sqrt{p_\perp^2+p_z^2}\sqrt{p_\perp^{'\ 2}+p_z^{'\ 2}}},
    \end{split}
\end{equation}
based on the dot product $\cos\theta_{CM}=\frac{\Vec{p}\cdot\Vec{p'}}{|\Vec{p}||\Vec{p'}|}$. The center-of-mass kinetic-energy is related to the incoming relative momentum through the relation $T_{CM}=(p_\perp^2+p_z^2)/M$. Substituting in the center-of-mass kinetic-energy
\begin{equation}
\begin{split}
  E_\gamma\frac{d^7N_\gamma}{d^3kd^4x}=\frac{\alpha}{8\pi^6 M^2}
  \int d^3pd^3q\frac{d\sigma_{np}}{d\Omega_{CM}}f_1f_2\\ \int d^3p'd^3q'(1-f_3)(1-f_4)\delta^3\bigg(\Vec{q}-\Vec{q'}\bigg)M\\
  \delta(p_\perp^2+p_z^2+q_\perp^2+q_z^2-M E_\gamma-p'_\perp{}^2-p'_z{}^2-q'_\perp{}^2-q'_z{}^2)\\
  \bigg[2\frac{p_\perp^2+p_z^2}{M^2 E_\gamma^2}\frac{2}{3}\bigg(1-\frac{p_\perp p'_\perp\cos(\phi-\phi')+p_zp'_z}{\sqrt{p_\perp^2+p_z^2}\sqrt{p_\perp^{'\ 2}+p_z^{'\ 2}}}\bigg)\bigg]
\end{split}
\end{equation}
\begin{equation}
\begin{split}
  E_\gamma\frac{d^7N_\gamma}{d^3kd^4x}=& \frac{\alpha}{6\pi^8 M^3 E_\gamma^2}\int d^3pd^3q\frac{d\sigma_{np}}{d\Omega_{CM}}(p_\perp^2+p_z^2)f_1f_2\\ 
  & \int d^3p' \delta(p_\perp^2+p_z^2-M E_\gamma-p'_\perp{}^2-p'_z{}^2) \\
  (1-f_3)& (1-f_4)\bigg(1-\frac{p_\perp p'_\perp\cos(\phi-\phi')+p_zp'_z}{\sqrt{p_\perp^2+p_z^2}\sqrt{p_\perp^{'\ 2}+p_z^{'\ 2}}}\bigg)
\end{split}
\end{equation}

The cross section implemented in this work is based on a parameterization of the data measured in Ref.~\cite{Stoks:1993tb} and presented in Ref.~\cite{Rrapaj:2015wgs}. For simplicity, the scattering of the nucleons is assumed to be angularly symmetric, which means only S wave contributions to the nucleon-nucleon scattering are included. The differential cross section is thus written as 
\begin{equation}
\begin{split}
  \frac{d\sigma_{np}}{d\Omega_{CM}}=\frac{1}{2\pi}\frac{d\sigma_{np}}{d(\cos{\theta_{CM}})}=\frac{\pi r_{np}^2}{4\pi}=\frac{r_{np}^2}{4}
\end{split}
\end{equation}
where $r_{np}$ is the hard-sphere scattering radius.
\begin{equation}
\begin{split}
  E_\gamma\frac{d^7N_\gamma}{d^3kd^4x}=& \frac{\alpha}{6\pi^8 M^3 E_\gamma^2}\int d^3pd^3q\frac{r_{np}^2}{4}(p_\perp^2+p_z^2)f_1f_2\\ & \int dp'_\perp d\phi' dp'_z\frac{p'_\perp}{2\sqrt{p_\perp^2+p_z^2-M E_\gamma-p'_z{}^2}}\\
  &\delta\bigg(p'_\perp-\sqrt{p_\perp^2+p_z^2-M E_\gamma-p'_z{}^2}\bigg)(1-f_3)\\ & 
  (1-f_4)\bigg(1-\frac{p_\perp p'_\perp\cos(\phi-\phi')+p_zp'_z}{\sqrt{p_\perp^2+p_z^2}\sqrt{p_\perp^{'\ 2}+p_z^{'\ 2}}}\bigg)
\end{split}
\end{equation}
Integrating over the transverse component of the outgoing relative momentum yields
\begin{equation}
\begin{split}
  E_\gamma\frac{d^7N_\gamma}{d^3kd^4x}=\frac{\alpha}{48\pi^8 M^3 E_\gamma^2}\int d^3pd^3q\ r_{np}^2(p_\perp^2+p_z^2)f_1f_2\\
  \int d\phi' dp'_z(1-f_3)(1-f_4)\\
  \bigg(1-\frac{p_\perp\sqrt{p_\perp^2+p_z^2-M E_\gamma-p_z^{'\ 2}}\cos(\phi-\phi')+p_zp'_z}{\sqrt{p_\perp^2+p_z^2}\sqrt{p_\perp^2+p_z^2-M E_\gamma}}\bigg).
\end{split}
\end{equation}
The integration over the angle associated with the total momentum can be evaluated trivially when all other vectors are measured in relation to the total momentum. (More clearly stated, the angles of the three vectors $\Vec{p}, \Vec{p'}, \Vec{q}$ always appear in addition to or subtraction from each other and never on their own. That means that the angles between them can be redefined in relation to the total momentum as opposed to an arbitrary coordinate frame. Furthermore, there is no loss of generality because the total momentum $\Vec{q}$ is still defined in relation to an arbitrary coordinate frame.)
\begin{equation}
\begin{split}
  E_\gamma\frac{d^7N_\gamma}{d^3kd^4x}=\frac{\alpha}{24\pi^7M^3 E_\gamma^2}\int{dp_\perp d\phi dp_z dq_\perp dq_z\ r_{np}^2 p_\perp}\\
  (p_\perp^2+p_z^2)q_\perp\ f_1f_2
   \int d\phi'dp'_z(1-f_3)(1-f_4)\\
   \bigg[1-\frac{p_\perp\sqrt{p_\perp^2+p_z^2-M E_\gamma-p_z'{ }^2}\cdot\cos{(\phi-\phi')}+p_z p_z'}{\sqrt{p_\perp^2+p_z^2}\sqrt{p_\perp^2+p_z^2-M E_\gamma}}\bigg]
 \label{eq_cylindrical_app}
\end{split}
\end{equation}
is the photon rate employed in this study when investigating photon emission in the presence of off-equilibrium effects. The photon rate was evaluated in the nucleon-nucleon center-of-mass frame after originally being defined in the thermal frame of the coarse-graining cell. So the substitutions of the momentum variables from the nucleon distribution functions are listed below:
\begin{equation}
\begin{split}
  p_1{}_\perp^2+\xi(p_1{}_z-p_0)^2=p_\perp^2+p_z^2+q_\perp^2+q_z^2+2q_\perp p_\perp\cos\phi\\
  +2q_z p_z  +(\xi-1)(q_z+p_z)^2-2\xi p_0(q_z+p_z)+\xi p_0^2
\end{split}
\end{equation}
\begin{equation}
\begin{split}
  p_1{}_\perp^2+\xi(p_1{}_z+p_0)^2=p_\perp^2+p_z^2+q_\perp^2+q_z^2+2q_\perp p_\perp\cos\phi\\
  +2q_z p_z  +(\xi-1)(q_z+p_z)^2+2\xi p_0(q_z+p_z)+\xi p_0^2
\end{split}
\end{equation}
\begin{equation}
\begin{split}
  p_2{}_\perp^2+\xi(p_2{}_z-p_0)^2=p_\perp^2+p_z^2+q_\perp^2+q_z^2-2q_\perp p_\perp\cos\phi\\
  -2q_z p_z  +(\xi-1)(q_z-p_z)^2-2\xi p_0(q_z-p_z)+\xi p_0^2
\end{split}
\end{equation}
\begin{equation}
\begin{split}
  p_2{}_\perp^2+\xi(p_2{}_z+p_0)^2=p_\perp^2+p_z^2+q_\perp^2+q_z^2-2q_\perp p_\perp\cos\phi\\
  -2q_z p_z  +(\xi-1)(q_z-p_z)^2+2\xi p_0(q_z-p_z)+\xi p_0^2
\end{split}
\end{equation}
\begin{equation}
\begin{split}
  p_3{}_\perp^2+\xi(p_3{}_z-p_0)^2=p_\perp^2+p_z^2-M E_\gamma+q_\perp^2+q_z^2\\
  +2q_\perp \sqrt{p_\perp^2+p_z^2-M E_\gamma-p'_z{}^2}\cos\phi'
  +2q_z p'_z+\\
  (\xi-1)(q_z+p'_z)^2  -2\xi p_0(q_z+p'_z)+\xi p_0^2
\end{split}
\end{equation}
\begin{equation}
\begin{split}
  p_3{}_\perp^2+\xi(p_3{}_z+p_0)^2=p_\perp^2+p_z^2-M E_\gamma+q_\perp^2+q_z^2\\
  +2q_\perp \sqrt{p_\perp^2+p_z^2-M E_\gamma-p'_z{}^2}\cos\phi'\\
  +2q_z p'_z\\
  +(\xi-1)(q_z+p'_z)^2+2\xi p_0(q_z+p'_z)+\xi p_0^2
\end{split}
\end{equation}
\begin{equation}
\begin{split}
  p_4{}_\perp^2+\xi(p_4{}_z-p_0)^2=p_\perp^2+p_z^2-M E_\gamma+q_\perp^2+q_z^2\\
  -2q_\perp \sqrt{p_\perp^2+p_z^2-M E_\gamma-p'_z{}^2}\cos\phi'  -2q_z p'_z\\
  +(\xi-1)(q_z-p'_z)^2-2\xi p_0(q_z-p'_z)+\xi p_0^2
\end{split}
\end{equation}
\begin{equation}
\begin{split}
  p_4{}_\perp^2+\xi(p_4{}_z+p_0)^2=p_\perp^2+p_z^2-M E_\gamma+q_\perp^2+q_z^2\\
  -2q_\perp \sqrt{p_\perp^2+p_z^2-M E_\gamma-p'_z{}^2}\cos\phi'  -2q_z p'_z\\
  +(\xi-1)(q_z-p'_z)^2+2\xi p_0(q_z-p'_z)+\xi p_0^2
\end{split}
\end{equation}

When off-equilibrium effects are not present, spherical distribution functions and a spherical energy-conserving delta function can be implemented. The integral of the photon emission rate can be reevaluated and simplified to the 4-dimensional integral of 
\begin{equation}
\begin{split}
  E_\gamma\frac{d^7N_\gamma}{d^3kd^4x}=&\frac{\alpha}{3\pi^5M^3 E_\gamma^2}\\
  &\int{dpd\theta dq\ \sin{\theta}\ r_{np}^2 p^4\sqrt{p^2-M E_\gamma}q^2 f_1f_2}\quad\\  
  &\int{d\theta'\ \sin{\theta'}(1-f_3)}(1-f_4)(1-\cos{\theta}\cos{\theta'}),
 \label{eq_spherical}
\end{split}
\end{equation}
as opposed to the 7-dimensional rate shown in Eq.~(\ref{eq_cylindrical_app}).

%%%%%%%%%%%%%%%%%%%%%%%%%%%%%%%%%
\section{Electromagnetic Current Coupling}
\label{app_em_factor}
%%\input{chapters/appendix}
%%\appendix{Appendix B: Electromagnetic Factor}
%%\label{app_em_factor}
%%%%%%%%%%%%%%%%%%%%%%%%%%%%%%
The electromagnetic factor $(\epsilon\cdot\Tilde{J})^2$ is responsible for coupling the photon to the movement of the nucleons. A similar simplification is presented in Refs.~\cite{Rrapaj:2015wgs,Chang:2016ntp} with intermediate steps performed in Ref.~\cite{Haglin:1992fy} and Ch. 7 of Ref.~\cite{Bjorken:1965sts}. Let's say that the photon is traveling in a way such that $K=(E_\gamma,E_\gamma,0,0)$ then the two polarization vectors are $\epsilon^{(1)}=(0,0,1,0)$ and $\epsilon^{(2)}=(0,0,0,1)$. Summing over the possible polarizations gives
\begin{equation}
\begin{split}
    (\epsilon\cdot \Tilde{J})^2=\sum\limits_{pols}\epsilon_\mu\epsilon_\nu \Tilde{J}^\mu \Tilde{J}^\nu.
\end{split}
\end{equation}
\begin{equation}
\begin{split}
    =-\Tilde{J}^2 \Tilde{J}^2-\Tilde{J}^3 \Tilde{J}^3
\end{split}
\end{equation}
\begin{equation}
\begin{split}
    =-\Tilde{J}^0 \Tilde{J}^0+\Tilde{J}^1 \Tilde{J}^1+\sum\limits_{\mu=0}^3 \Tilde{J}^\mu \Tilde{J}^\mu
\end{split}
\label{eq_penultimate_electromagnetic_factor}
\end{equation}
As a consequence of the conservation of nucleon current, we get 
\begin{equation}
\begin{split}
    K_\mu \Tilde{J}^\mu=0,
\end{split}
\end{equation}
which is the momentum space analog of $\frac{\partial\Tilde{J}^\mu(x)}{\partial x^\mu}$. This means that 
\begin{equation}
\begin{split}
    K_\mu \Tilde{J}^\mu \Tilde{J}^\nu=K_\nu \Tilde{J}^\mu \Tilde{J}^\nu=0.
\end{split}
\end{equation}
The components of $K$ are independent variables, which leads to
 \begin{equation}
\begin{split}
    \Tilde{J}^0 \Tilde{J}^\nu=\Tilde{J}^1 \Tilde{J}^\nu \text{ , } \Tilde{J}^\mu \Tilde{J}^0=\Tilde{J}^\mu \Tilde{J}^1\Longrightarrow \Tilde{J}^{00}=\Tilde{J}^{11}.
\end{split}
\end{equation}
Picking up after Eq.~(\ref{eq_penultimate_electromagnetic_factor}) we get
\begin{equation}
\begin{split}
    (\epsilon\cdot\Tilde{J})^2=-\Tilde{J}_\mu\Tilde{J}^\mu.
    \label{eq_simple_JJ}
\end{split}
\end{equation}
Equation~(\ref{eq_simple_JJ}) can be generalized to 
\begin{equation}
\begin{split}
    (\epsilon\cdot\Tilde{J})^2=-\bigg(g_{\mu\nu}-\frac{K_\mu}{E_\gamma}\frac{K_\nu}{E_\gamma}\bigg)\Tilde{J}^\mu\Tilde{J}^\nu
\end{split}
\end{equation}
for comparison with the electromagnetic factors in Refs.~\cite{Rrapaj:2015wgs,Chang:2016ntp}, which are for the case of dark photons with finite mass being emitted. Continuing from Eq.~(\ref{eq_simple_JJ}) for the case of proton-nucleon bremsstrahlung, which uses the dipole electromagnetic current, Eq.~(\ref{eq_dipole_current}), we get
\begin{equation}
\begin{split}
    (\epsilon\cdot J)^2=2\frac{P_1\cdot P_3}{(P_1\cdot K)(P_3\cdot K)}-\frac{M^2}{(P_1\cdot K)^2}-\frac{M^2}{(P_3\cdot K)^2}.
\end{split}
\end{equation}
Using the relation $\Vec{\beta}=\Vec{p}/E$, we get
\begin{equation}
\begin{split}
    (\epsilon\cdot J)^2=&2\frac{(1-\Vec{\beta_1}\cdot\Vec{\beta_3})E_1 E_3}{(1-\Vec{\beta_1}\cdot \hat{k})(1-\Vec{\beta_3}\cdot \hat{k})E_1E_3E_\gamma^{\ 2}}\\
    &-\frac{M^2}{(1-\Vec{\beta_1}\cdot \hat{k})^2E_1^2E_\gamma^{\ 2}} -\frac{M^2}{(1-\Vec{\beta_3}\cdot \hat{k})^2E_3^2E_\gamma^{\ 2}}
\end{split}
\end{equation}
By assuming that the photon is emitted isotropically from the nucleon-nucleon collision, the integral can then be averaged over the photon angle for simplicity.
\begin{equation}
\begin{split}
    E_\gamma^{\ 2}\int\frac{d\Omega_{\gamma}}{4\pi} (\epsilon\cdot J)^2=&2\int\frac{d\Omega_{\gamma}}{4\pi}\frac{1-\Vec{\beta_1}\cdot\Vec{\beta_3}}{(1-\Vec{\beta_1}\cdot \hat{k})(1-\Vec{\beta_3}\cdot \hat{k})}\\
    &-\int\frac{d\Omega_{\gamma}}{4\pi}\frac{M^2}{(1-\Vec{\beta_1}\cdot \hat{k})^2E_1^2}\\
    &-\int\frac{d\Omega_{\gamma}}{4\pi}\frac{M^2}{(1-\Vec{\beta_3}\cdot \hat{k})^2E_3^2}
\end{split}
\label{eq_em_factor_velocity}
\end{equation}
The last two integrals can be evaluated rather effortlessly using the relation
\begin{equation}
\begin{split}
    \frac{1}{4\pi}\int d\Omega_{\gamma}\frac{M^2}{(1-\Vec{\beta_3}\cdot \hat{k})^2E_3^2}=\frac{M^2}{E_3^2}\int\limits_{-1}^1 \frac{1}{2}\frac{dz}{(1-\beta_3z)^2}
\end{split}
\end{equation}
\begin{equation}
\begin{split}
    \frac{1}{4\pi}\int d\Omega_{\gamma}\frac{M^2}{(1-\Vec{\beta_3}\cdot \hat{k})^2E_3^2}=\frac{M^2}{E_3^2} \frac{1}{2}\frac{2}{1-\beta_3^{\ 2}}
\end{split}
\end{equation}
\begin{equation}
\begin{split}
    \frac{1}{4\pi}\int d\Omega_{\gamma}\frac{M^2}{(1-\Vec{\beta_3}\cdot \hat{k})^2E_3^2}=1
\end{split}
\end{equation}
where $z\equiv\cos\theta_\gamma$.
The first integral can be evaluated using a technique involving parameterized integrals 
\begin{equation}
\begin{split}
  \frac{1}{ab}=\int\limits_0^1 \frac{dx}{[ax+b(1-x)]^2}
\end{split}
\end{equation}
mentioned by Feynman \cite{Feynman:1949zx}. So the first integral
\begin{equation}
\begin{split}
  I_1\equiv2\frac{1}{4\pi}\int d\Omega_{\gamma}\frac{1-\overrightarrow{\beta_1}\cdot\overrightarrow{\beta_3}}{(1-\overrightarrow{\beta_1}\cdot \hat{k})(1-\overrightarrow{\beta_3}\cdot \hat{k})}
\end{split}
\end{equation}
can be rewritten as 
\begin{equation}
\begin{split}
  =&2(1-\overrightarrow{\beta_1}\cdot\overrightarrow{\beta_3})\\
  &\times\int\limits_0^1 dx\int \frac{d\Omega_{\gamma}}{4\pi}\frac{1}{[(1-\overrightarrow{\beta_1}\cdot \hat{k})x+(1-\overrightarrow{\beta_3}\cdot \hat{k})(1-x)]^2}.
\end{split}
\end{equation}
\begin{equation}
\begin{split}
  =&2(1-\overrightarrow{\beta_1}\cdot\overrightarrow{\beta_3})\\
  &\times\int\limits_0^1 dx\frac{1}{2}\int\limits_0^\pi d\theta_{\gamma}\frac{\sin{\theta_\gamma}}{\{1-\hat{k}\cdot[x\beta_1+(1-x)\beta_3]\}^2}.
\end{split}
\end{equation}
\begin{equation}
\begin{split}
  =2(1-\overrightarrow{\beta_1}\cdot\overrightarrow{\beta_3})\int\limits_0^1 dx\frac{1}{1-|x\beta_1+(1+x)\beta_3|^2}.
\end{split}
\end{equation}
\begin{equation}
\begin{split}
  =&2(1-\overrightarrow{\beta_1}\cdot\overrightarrow{\beta_3})\\
  &\times\int\limits_0^1 dx\frac{1}{1-x^2\beta_1^2-(1-x)^2\beta_3^2-2x(1-x)\overrightarrow{\beta_1}\cdot\overrightarrow{\beta_3}}.
\end{split}
\end{equation}
Using the soft-photon limit, we can say $|\overrightarrow{\beta_1}|=|\overrightarrow{\beta_3}|=|\Vec{\beta}|$.
\begin{equation}
\begin{split}
  =&2(1-\beta^2\cos\theta_{CM})\\
  &\times\int\limits_0^1 \frac{dx}{1-\beta^2+4\beta^2\sin^2{\big(\frac{\theta_{CM}}{2}\big)}x(1-x)}
\end{split}
\end{equation}
\begin{equation}
\begin{split}
  =&2(1-\beta^2\cos\theta_{CM})\\
  &4\frac{\arctan{\bigg(\frac{\sqrt{4\beta^2\sin^2{\big(\frac{\theta_{CM}}{2}\big)}}}{\sqrt{-4(1-\beta^2)-4\beta^2\sin^2{\big(\frac{\theta_{CM}}{2}}\big)}}\bigg)}}{-\sqrt{-4(1-\beta^2)-4\beta^2\sin^2{\big(\frac{\theta_{CM}}{2}}\big)}\sqrt{4\beta^2\sin^2{\big(\frac{\theta_{CM}}{2}\big)}}}
\end{split}
\end{equation}
To first order, the equation reduces to 
\begin{equation}
\begin{split}
  =&2(1-\beta^2\cos\theta_{CM})\frac{1}{1-\beta^2(1-\sin^2{\frac{\theta_{CM}}{2}})}.
\end{split}
\end{equation}
In the nonrelativistic limit $\beta\ll1$, the integral approximately evaluates to
\begin{equation}
\begin{split}
  I_1\approx2\bigg[1+\frac{2}{3}\beta^2(1-\cos\theta_{CM})\bigg].
\end{split}
\end{equation}
This result can be rewritten in terms of the center-of-mass energy $T_{CM}$,
\begin{equation}
\begin{split}
  I_1=2\bigg[1+\frac{2}{3}\frac{T_{CM}}{M}(1-\cos\theta_{CM})\bigg].
\end{split}
\end{equation}
Returning to the calculation of the total electromagnetic factor from Eq.~(\ref{eq_em_factor_velocity}), we get 
\begin{equation}
\begin{split}
    E_\gamma^{\ 2}\frac{1}{4\pi}\int d\Omega_{\gamma}(\epsilon\cdot J)^2=&2\bigg[1+\frac{2}{3}\frac{T_{CM}}{M}(1-\cos{\theta_{CM}})\bigg]\\
    &-1-1.
\end{split}
\end{equation}
\begin{equation}
\begin{split}
    \frac{E_\gamma^{\ 2}}{4\pi}\int d\Omega_{\gamma}(\epsilon\cdot J)^2=2\frac{T_{CM}}{M}\frac{2}{3}(1-\cos{\theta_{CM}})
\end{split}
\end{equation}
This result agrees with the electromagnetic factors written in Refs.~\cite{Rrapaj:2015wgs,Chang:2016ntp} for the case of standard model photons. The electromagnetic factor used in those references is 
\begin{equation}
\begin{split}
    \frac{E_\gamma^{\ 2}}{4\pi}\int d\Omega_{\gamma}(\epsilon\cdot J)^2=2\frac{T_{CM}}{M}\bigg(1-\frac{k^2}{3\omega^2}\bigg)(1-\cos{\theta_{CM}}).
\end{split}
\end{equation}
%%%%%%%%%%%%%%%%%%%%%%%%%%%%%%%%%

\end{document}